\documentclass[useAMS,usenatbib]{mn2e} 
\usepackage{graphicx}

\newcommand{\be}{\begin{equation}}
\newcommand{\ee}{\end{equation}}
\newcommand{\ba}{\begin{eqnarray}}
\newcommand{\ea}{\end{eqnarray}}

\newcommand {\smass} {$\mathcal{M} \;$}

\title[The different mechanisms that drive the SFHs of giants and dwarfs]{The Different
  Physical Mechanisms that Drive the Star-Formation Histories of Giant
  and Dwarf Galaxies}
\author[Haines et al.]{C. P. Haines,$^{1}$ A. Gargiulo,$^{1,2}$
 F. La Barbera,$^1$ A. Mercurio,$^1$ P. Merluzzi,$^1$
and 
\newauthor G. Busarello$^1$\\
$^{1}$INAF - Osservatorio Astronomico di Capodimonte,
via Moiariello 16, I-80131 Napoli, Italy; chris@na.astro.it \\
$^{2}$Department of Physics, Universit\`{a} Federico II, Napoli, Italy
}

\begin{document}

\maketitle
\label{firstpage}
\date{Accepted 2007 July 5. Received 2007 June 18; in original form
  2007 March 30}

\begin{abstract}
We present an analysis of star-formation and nuclear activity in
galaxies as a function of both luminosity and environment in the
fourth data release of the Sloan Digital Sky Survey (SDSS DR4). 
Using a sample of 27\,753 galaxies in the redshift range
\mbox{$0.005\!<\!z\!<\!0.037$} that is \mbox{$\ga\!9$0\%} complete to
\mbox{M$_{r}=-18.0$} we find that the EW(H$\alpha$) distribution is strongly 
bimodal, allowing galaxies to be robustly separated into passively-evolving 
and star-forming populations about a value \mbox{EW(H$\alpha)=2$\AA}. 
In high-density regions \mbox{$\sim\!7$0\%} of galaxies are passively-evolving 
independent of luminosity. In the rarefied field however, the fraction of 
passively-evolving galaxies is a strong function of luminosity, dropping 
from 50\% for \mbox{M$_{r}\!\la\!-21$} to zero by \mbox{M$_{r}\!\sim\!-18$}. 
Indeed for the lowest luminosity range covered 
\mbox{($-18\!<\!{\rm M}_r\!<\!-16$)} none of the \mbox{$\sim\!6$00} galaxies 
in the lowest density quartile are passively-evolving.
The few passively-evolving dwarf galaxies in field regions appear as
satellites to bright \mbox{($\ga\!L^{*}$)} galaxies.
We find a systematic reduction of $\sim\!3$0\% in the H$\alpha$
emission from dwarf \mbox{($-19\!<\!{\rm M}_{r}\!<\!-18$)} star-forming
galaxies in high-density regions with respect to field values,
implying that the bulk of star-forming  dwarf galaxies in groups and
clusters are currently in the process of being slowly transformed into
passive galaxies.  
The fraction of galaxies with optical AGN signatures decreases steadily from 
\mbox{$\sim\!5$0\%} at \mbox{M$_{r}\!\sim\!-21$} to \mbox{$\sim\!0$\%} by 
\mbox{M$_{r}\!\sim\!-18$} closely mirroring the luminosity-dependence
of the passive galaxy fraction in low-density environments. 
This result reflects the increasing importance of AGN feedback with
galaxy mass for their evolution, 
such that the star-formation histories of massive galaxies
are primarily determined by their past merger history.
In contrast, the complete absence of passively-evolving dwarf
galaxies more than \mbox{$\sim\!2$} virial radii from the nearest massive halo
(i.e. cluster, group or massive galaxy) indicates
that internal processes, such as merging, AGN feedback or gas consumption through star-formation, are not responsible for terminating
star-formation in dwarf galaxies. 
Instead the evolution of dwarf
galaxies is primarily driven by the mass of their host halo, probably
through the combined effects of tidal forces and ram-pressure stripping.  
\end{abstract}


\section{Introduction}
\label{intro}

It has been known for decades that local galaxies can be
broadly divided into two distinct populations \citep[e.g.][]{hubble1,hubble2,morgan,devaucouleurs}. The first are red, passively-evolving, bulge-dominated galaxies dominated
by old stellar populations that make up the red sequence; and the
second make up the ``blue
cloud'' of young, star-forming, disk-dominated galaxies \citep[e.g.][]{strateva,k03a,k03b,blanton03,baldry04,driver,mateus2}.

It has also long been known that the environment in which a galaxy
inhabits has a profound impact on its evolution in terms of defining
both its structural properties and star-formation histories \citep[e.g.][]{hubble3}. In
particular, passively-evolving spheroids dominate cluster
cores, whereas in field regions galaxies are typically both
star-forming and disk-dominated \citep{blanton05}. These differences have been
quantified through the classic morphology--density \citep{dressler80} and
star-formation (SF)--density relations \citep{lewis,gomez}. However,
despite much effort \citep*[e.g.][]{treu,balogh04,balogh04b,gray,kauffmann04,tanaka,christlein,rines05,baldry06,blanton06,boselli,haines,paper1,mercurio,sorrentinovoids,weinmann06a,weinmann06b,mateus4}, it
still remains unclear whether these environmental trends are:
(i) the direct result of the initial conditions in which the galaxy forms,
whereby massive galaxies are formed earlier preferentially in the
highest overdensities in the primordial density field, and have a more
active merger history, than galaxies that form in the smoother
low-density regions;
or (ii) produced later by the direct interaction of the galaxy with
one or more aspects of its environment, whether that be other
galaxies, the intracluster medium (ICM), or the underlying dark-matter
distribution (e.g. tidal stripping).
Several physical mechanisms have been proposed that could cause the
transformation of galaxies through interactions with their environment
such as ram-pressure stripping \citep{gunn}, galaxy harassment
\citep{moore}, and suffocation (also known as starvation or
strangulation) in which the diffuse gas in the outer galaxy halo
is stripped preventing further accretion onto the galaxy before the
remaining cold gas in the disk is slowly consumed through star-formation \citep*{larson}.

The morphologies and star-formation histories of galaxies are also
strongly dependent on their masses, with high-mass galaxies
predominately passively-evolving spheroids, and low-mass galaxies
generally star-forming disks. A sharp transition between
these two populations is found about a characteristic stellar mass of
\mbox{$\sim\!3\times10^{10}\,{\rm M}_{\odot}$}, corresponding to
\mbox{$\sim\!{\rm M}^{\star}\!+1$} \citep{k03a,k03b}. This bimodality
implies fundamental
differences in the formation and evolution of high- and low-mass
galaxies. The primary mechanism behind this transition appears to be
the increasing efficiency and rapidity with which gas is converted
into stars for more massive galaxies according to the
Kennicutt-Schmidt law \citep{kennicutt,schmidt}. 
This results in massive galaxies with their deep potential wells consuming
their gas in a short burst \mbox{($\la\!2$\,Gyr)} of star-formation at \mbox{$z\!>\!2$} \citep{chiosi},
while dwarf galaxies have much more extended star-formation histories
and gas consumption time-scales longer than the Hubble time \citep{vanzee}.

In the monolithic collapse model for the formation of elliptical
 galaxies this naturally produces the effect known as ``cosmic
 downsizing'' whereby
the major epoch of star-formation occurs earlier and over a shorter
period in the most massive galaxies and progressively later and over
more extended time-scales towards lower mass galaxies. 
This has been confirmed observationally both in terms of the global
 decline of star-formation rates in galaxies since $z\!\sim\!1$
 \citep{noeske07a,noeske07b} and the fossil records of low-redshift
 galaxy spectra \citep{heavens,panter}. Finally in analyses of the
 absorption lines of local quiescent galaxies, the most massive galaxies are found to have higher mean stellar ages and abundance
ratios than their lower mass counterparts, indicating that they formed
stars earlier and over shorter time-scales \citep{thomas,nelan}.  
In this scenario, the mass-scale at which a galaxy becomes quiescent
should decrease with time, with the most massive galaxies becoming
quiescent earliest, resulting in the red sequence of
passively-evolving galaxies being built up earliest at the bright end \citep{tanaka05},
and the evolving mass-scale of E+A (post-starburst) galaxies
\citep{poggianti}. 

However the standard paradigm for the growth of structure and the
evolution of massive galaxies within a CDM universe is the
hierarchical merging scenario \citep[e.g.][]{white,kauffmann93,lacey}
in which massive elliptical galaxies are assembled through the merging
of disk galaxies as first proposed by \citet{toomre} \citep[for a full
  historical review of this subject see][]{struck}.
Although downsizing appears at first sight to be at odds with the
standard hierarchical 
model for the formation and evolution of galaxies, \citet{merlin} are
able to reproduce the same downsizing as seen in the earlier ``monolithic''
models in a hierarchical cosmological context, resulting in what they
describe as a revised monolithic scheme whereby the merging of
substructures occurs early in the galaxy life \mbox{($z>2$)}. Further
contributions to cosmic downsizing and the observed bimodality in
galaxy properties could come from the way gas from the halo cools and
flows onto the galaxy \citep{dekel,keres} and which affects its
ability to maintain star-formation over many Gyr, in conjunction with feedback
effects from supernovae and AGN \citep[e.g.][]{springel,croton}.
These mechanisms which can 
shut down star-formation in massive galaxies allow the hierarchical
CDM model to reproduce very well the rapid early formation and quenching of
stars in massive galaxies \citep*[e.g.][]{bower,hopkins06b,hopkins07b,birnboim}. 
In particular, the transition from cold to hot accretion modes of gas
when galaxy halos reach a mass \mbox{$\sim\!10^{12}{\rm
  M}_{\odot}$} \citep{dekel} could be responsible for the observed sharp transition in galaxy
properties with mass.

If the evolution of galaxies due to internal processes is effectuated
earlier and more rapidly with increasing mass, then this would give
less opportunity for external environmental processes to act on
massive galaxies. Moreover, low-mass galaxies
having shallower potential wells are more susceptible to disruption
and the loss of gas due to external processes such as ram-pressure
stripping and tidal interactions. This suggests that the relative
importance of internal and external factors on galaxy evolution and on
the formation of the SF-, age- and morphology-density relations could
be mass-dependent, in particular the relations should be stronger for
lower mass galaxies. Such a trend has been observed by \citet{smith}
who find that radial age gradients (out to \mbox{1\,R$_{\rm vir}$}) are more
pronounced for lower mass \mbox{($\sigma\!<\!175\,$km\,s$^{-1}$)}
cluster red sequence galaxies than their higher mass subsample.

The environmental trends of fainter galaxies
\mbox{(M$_{r}\!\ga{\rm M}^{*}\!+1$)} have generally been examined
using galaxy colours as a measure of their star formation
history. Whereas the colour of massive galaxies becomes steadily redder
with increasing density, a sharp break in the mean colour of faint
galaxies is observed at a critical density corresponding to
\mbox{$\sim\!{\rm R}_{\rm vir}$} \citep{gray,tanaka}. In a photometric
study of galaxies in the environment of the Shapley supercluster core,
we found that the fraction of faint \mbox{(${\rm M}^{*}\!+2<{\rm
  M}_{r}<{\rm M}^{*}\!+6$)} red galaxies dropped from
\mbox{$\sim\!9$0\%} in the cluster cores to \mbox{$\sim\!2$0\%} by the virial radius
  \citep{haines}, while the shape of the faint end of the red galaxy
  luminosity function changes dramatically with density inside the
  virial radius \citep{mercurio}. 

In \citet[hereafter Paper I]{paper1} we investigated the possible
mass-dependency of the age-density and SF-density relations by
comparing the global trends with environment for giant \mbox{(M$_{r}\!<\!-20$)}
and dwarf \mbox{($-19\!<\!{\rm M}_{r}\!<\!-17.8$)} galaxies in the vicinity of the
\mbox{$z=0.03$} supercluster centred on the rich cluster A\,2199, the
richest low-redshift \mbox{($z\!<\!0.05$)} structure covered by the Sloan Digital
Sky Survey (SDSS) DR4 spectroscopic dataset.

A strong bimodality was
seen in the mean stellar age-M$_{r}$ distribution about a mean stellar
age of 7\,Gyr, with a population
of bright ($L^*$) galaxies \mbox{$\sim\!1$0\,Gyr} old, and a second population
of fainter galaxies dominated by young \mbox{($\la\!3$\,Gyr)} stars,
while a clear
age-density distribution was identified for both giant and dwarf
subsamples. We confirmed the findings of 
\citet{smith} that the age-density relation is stronger for dwarf
galaxies, while the critical density at which the ages increase
markedly is higher for dwarf galaxies, occurring at values typical of
the cluster virial radius. 
In the highest-density regions we found that \mbox{$\ga\!8$0\%} of both giant and dwarf
subsamples were old \mbox{($>\!7$\,Gyr)}. However, whereas the fraction of old giant
galaxies declines gradually with decreasing density to the global
field value of \mbox{$\sim\!5$0\%,} that of dwarf galaxies drops rapidly, tending to {\em zero} for the lowest density
bins. Identical trends with density were independently observed when
passive galaxies were identified from their lack of H$\alpha$ emission. 

Looking directly at the spatial distribution of galaxies in the
vicinity of the supercluster, in field regions the giant population
shows a completely interspersed mixture of both young and old
populations, indicating that their evolution is driven
primarily by their merger history rather than direct interactions with
their environment. In contrast, the mean stellar ages of dwarf
galaxies were strongly correlated with their immediate environment:
those passively-evolving or old dwarf galaxies found outside of the rich clusters were always found within poor groups or as a satellite to an  old,
giant \mbox{($\ga\!L^{*}$)} galaxy. No isolated old or passively-evolving dwarf
galaxies were found. 

In this article we extend the study to cover the entire SDSS DR4
footprint, creating a volume-limited sample of $\sim28\,000$ galaxies
with \mbox{$0.005\!<\!z\!<\!0.037$} that is \mbox{$\ga\!9$0\%} complete to
\mbox{M$_{r}=-18$} \mbox{(M$^{*}\!+3.2$)}. 
We reexamine the arguments of Paper I taking advantage of this
much larger dataset to provide quantitative measures of the
environmental dependencies on star-formation in galaxies and in
particular how these vary with the galaxy mass/luminosity. We attempt to
disentangle the different contributions to the SF-density relation
caused by physical mechanisms internal to the galaxy (e.g. AGN
feedback) and those caused by the direct interaction of the galaxy
with its surroundings \citep[for a review of how the diverse
  mechanisms leave different imprints on the environmental trends see e.g.][]{treu}. 

In $\S$~\ref{data} we describe the dataset
used, the classification of galaxies,
and the measures used to remove any biases due to aperture effects and
the complex
survey geometry, while in $\S$~\ref{environment} we describe the
adaptive kernel method used to define the local galaxy density.  
In $\S$~\ref{bimodality} we quantify how the fraction of
passively-evolving galaxies depends on both environment and
mass/luminosity, while in $\S$~\ref{sfr} we examine the ongoing effects of
environment on galaxies that are still forming stars. In
$\S$~\ref{neighbours} we examine which aspects of environment are most
important for defining the star-formation history of galaxies in field
regions, in particular whether the presence of a nearby
galaxy of a particular mass has any role in star-formation being
truncated in a galaxy. 
In $\S$~\ref{agn} we quantify the fraction of
galaxies having AGN signatures as a function of both luminosity and
environment, and examine the possible connection between AGN feedback
and the shut-down of star-formation in galaxies. In $\S$~\ref{semi} we
compare our results with predictions from the semi-analytic models of
\citet{croton}. 
In $\S$~\ref{discussion} we discuss
the possible physical mechanisms that can affect star-formation in
galaxies and produce the observed environmental trends, and finally in
$\S$~\ref{summary} we present a summary of our results and conclusions.
Throughout we assume a concordance $\Lambda$CDM cosmology
with \mbox{$\Omega_{M}=0.3$}, \mbox{$\Omega_{\Lambda}=0.7$} and
\mbox{H$_{0}=70\,$km\,s$^{-1}$Mpc$^{-1}$}. 

\section{The Data}
\label{data}

The sample of galaxies used in this work is taken from the fourth
release of Sloan Digital Sky Survey \citep[SDSS DR4;][]{sdssdr4} an imaging and spectroscopic survey covering an area of
$\pi$ sr principally located in the North Galactic Cap \citep{york}. 
The area covered by the imaging survey has a total extension of 6670
square degrees, from which $ugriz$ broad band imaging data 
have been acquired for 180 million objects. 

Instead of the
spectroscopic catalogue obtained with the SDSS reduction and calibration
procedures, we refer to the low redshift catalogue (LRC) taken from
the New York University Value Added Galaxy Catalogue (NYU-VAGC) of
\citet{nyuvagc}. The LRC is a catalogue of SDSS galaxies with
\mbox{$0.0033\!<\!z\!<\!0.05$}, \mbox{$r_{Petro}\!<\!18$} 
and with \mbox{$\mu_{50}\!<\!24.5$\,mag\,arcsec$^{-2}$} in the $r$-band (where
$\mu_{50}$ is the surface brightness within a circular aperture
containing half of Petrosian flux). 
\citet{nyuvagc} also perform some further quality control checks on
the catalogues, including: a
procedure for dealing with large, complex galaxies that were
incorrectly deblended by the SDSS pipeline, and correctly associating
spectra from fibres that were offset from the actual centres of the
objects; bringing into the LRC a number of galaxies with redshifts but
morphologically classified as stars; and finally performing a number
of visual checks on objects in the catalogue.
In addition, \citet{nyuvagc} computed the $K$-correction for each
object of the NYU-VAGC using the version 3.2 of the software
 \textsc{k-correct} \citep{kcorrect} providing in this way, the
 absolute magnitudes of the objects.

The region covered by the SDSS survey in this release includes two
wide contiguous regions in the North Galactic Cap, one centred
roughly on the Celestial Equator and the other around \mbox{$\delta =
+40^{\circ}$}, both with \mbox{$120^{\circ}\!<\!\alpha\!<\!240^{\circ}$}, and
three narrower stripes, one centred on the Celestial
Equator again, and two at \mbox{$\delta =+15^{\circ}$} and \mbox{$\delta=
-10^{\circ}$} with \mbox{$-60^{\circ}\!<\!\alpha\!<\!60^{\circ}$}. We have
excluded the LRC galaxies belonging to the three stripes as the
computation of local galaxy density could be biased due to the narrow
dimension of these regions (no point is more than 3.4\,Mpc from the
boundary).

In this paper we focus our attention particularly on the environmental
impact on the dwarf galaxy population. Hence we  
 have only selected from the LRC the galaxies with
\mbox{$0.005\!<\!z\!<\!0.037$} in order to obtain a catalogue which is
\mbox{$\ga\!9$0\%} complete to \mbox{M$_{r}\!<\!-18$} 
\mbox{(M$_{r}\!\la{\rm M}^{\star}\!+3$)}
yielding a final catalogue of 27\,753 objects. 
The lower limit in redshift at $z=0.005$
is due both to the peculiar velocities which can seriously influence the
distance estimates and to the great problems arising from the
deblending of the large and resolved objects.  

As dwarf galaxies tend to
have low surface brightnesses, it is important to consider whether
significant numbers of dwarf galaxies are missing from the SDSS
spectroscopic catalogue (and hence ours) due to surface brightness
selection effects, which are three-fold: (i) photometric
incompleteness; (ii) galaxies not being targetted due to being
shredded by the deblending algorithm; (iii) targetted galaxies which
did not yield reliable redshifts.
\citet{blanton05a}
analysed the surface brightness completeness of the LRC up to
\mbox{$\mu_{50}\!<\!24.5$} and found that for \mbox{M$_{r}\!<\!-18.0$}
the LRC does not suffer from significant ($\ga\!1$0\%) incompleteness due to surface brightness effects. 
At fainter magnitudes they find that although low-surface
brightness dwarf galaxies are clearly detectable in the SDSS images,
the photometric pipeline tended to mistakenly deblend them or
overestimate the background sky levels.

\subsection{Spectral indices of the galaxies}
The stellar indices used are taken from MPA/JHU SDSS DR4 catalogues
\citep[][hereafter K03]{k03a}, in which a continuum fitting code was adopted that was
optimized to work with SDSS data in order to recover also the weak
features of the spectra and to account for the Balmer absorption
\citep{tremonti}. The library of spectra templates are composed
of single stellar population models following the assumption that
the star-formation history of a galaxy is made up of a set of discrete
bursts. The models are based on new population synthesis code of
\citet{bc03} which incorporates a spectral library covering
the 3200-9300\AA\, range and with high resolution (3\AA) matching
the SDSS data. The templates span a wide set of ages and
metallicities. After a Gaussian convolution of the templates in order
to match the stellar velocity dispersion of each galaxy, the best
fitting model is constructed by performing a non-negative least-squares
fit with the dust extinction values A$_{z}$ of K03 and the
\mbox{$\lambda$$^{-0.7}$} attenuation law of \citet{charlot}.
K03 use the amplitude of the 4000\AA\, break \citep[as defined
  in][]{balogh99} and the strength of the H$\delta_{A}$ absorption
line as diagnostics of the stellar populations of the galaxies, from
which maximum-likelihood estimates of the $z$-band mass-to-light
ratios are made. These in conjunction with the $z$-band absolute
magnitude and the dust attenuation A$_{z}$ yield the stellar mass
\smass of each galaxy.

The stellar mass estimates of K03 are only available for galaxies in the range \mbox{$14.5\!<\!r\!<\!17.77$}. For
the remainder of the galaxies we use the same technique as \citet{baldry06} who
estimate the stellar mass-to-light ratio of each galaxy from its $u-r$
colour, using the analysis of \citet{belldejong} who show that for
models of star-forming disk galaxies with reasonable metallicities and
star-formation histories, the
stellar mass-to-light ratios correlate strongly with the colours of the
integrated stellar populations. 
Briefly, for each 0.05\,mag bin in $u-r$ we determine the
median value of ($\mathcal{M}/L_{r}$) for those galaxies with stellar
mass estimates by K03, and linearly interpolating between
bins, create a relation between the $u-r$ colour and its stellar
mass-to-light ratio. This relation is then used to estimate stellar
masses for the remaining galaxies from their $r$-band luminosity and
$u-r$ colour.     

\subsection{Aperture biases}
\label{aperture}

Throughout this article we quantify the current star-formation and
nuclear activity in our sample of galaxies from their spectral
indices, in particular their H$\alpha$ emission.  
One possible cause of bias in estimating the star-formation rate of
galaxies in our sample is due to the galaxy
spectrum being obtained through a 3\,arcsec diameter aperture rather than
over the full extent of the galaxy. Significant radial star-formation
gradients are possible within galaxies, particularly those undergoing
nuclear star-bursts or spiral galaxies with a prominent
passively-evolving bulge, that can result in the ``global''
star-formation rate being significantly over or underestimated based
upon spectra containing flux dominated by the galaxy
nucleus. \citet*{kewley05} indicate that star-formation rates based on
spectra obtained through apertures covering less than $\sim$20\% of the
integrated galaxy flux can be over or underestimated by a factor
$\sim$2, and to ensure the SDSS fibres sample more than this 20\%
require galaxies to be at \mbox{$z\!>\!0.04$}. Clearly in order to use
the SDSS dataset to study star-formation in \mbox{M$_{r}\!\sim\!-18$}
galaxies this is not possible, as at $z=0.04$ they are already too
faint to be included in the SDSS spectroscopic
sample. \citet{brinchmann} quantify the effects of aperture bias on
their estimates of star-formation rates in SDSS galaxies, and find
that indeed in the case of galaxies with
\mbox{$\mathcal{M}\!\ga\!10^{10.5}{\rm M}_{\odot}$} strong trends are
apparent when plotting SFR/$\mathcal{M}$ as a function of redshift
(their Fig. 13), the star-formation rate being systematically
underestimated for galaxies at the lowest redshifts by as much as a
factor three. However they also find that for lower-mass galaxies,
which cover the same redshift range as our dataset, the
aperture biases are considerably smaller ($\la\!2$0\%) and a simple
scaling of the fiber SFR by the $r$-band flux, as done by
\citet{hopkins}, is perfectly acceptable.         

In a subsequent paper \citep[][Paper III]{paper3}
we perform a complementary analysis of the same volume-limited sample
combining the SDSS $r$-band photometry with {\em GALEX} NUV imaging to
obtain {\em integrated} measures of recent star-formation in
\mbox{$\sim\!1$5} per cent of our galaxies. A comparison of the
integrated \mbox{${\rm NUV}-r$}
colours with the SDSS fiber spectral indices allows us to quantify the
effects of aperture bias within our sample. We find that for
M$_{r}\!<\!-20$ galaxies aperture biases are significant with
$\sim\!7$ per cent of galaxies classified as passive
(EW[H$\alpha]\!<\!2$\AA) from their spectra yet also having blue colours \mbox{(${\rm NUV}-r\!\la\!4$)} indicative of recent
star-formation. 
As expected, many of these galaxies appear as
face-on spiral galaxies with prominant bulges. This fraction drops
to zero at fainter magnitudes \mbox{(M$_{r}\!\ga\!-19.5$)} as the
SDSS fibres cover a greater fraction of the galaxies, while lower
luminosity 
galaxies tend to be either late-type spirals or dwarf ellipticals.
The luminosity function of 
early-type spirals (Sa+b) for which aperture biases are by far the
most important has a Gaussian distribution centred at
\mbox{M$_{r}\!\sim\!-21.7$} and width
\mbox{$\sigma\!\sim\!0.9$\,mag} \citep{delapparent}, and hence are
rare at \mbox{M$_{r}\!\ga\!-20$}.   
 
We thus indicate that star-formation rate estimates for \mbox{M$_{r}\!>\!-20$}
galaxies in our sample based on H$\alpha$ fluxes obtained
through the SDSS fibres should be reasonably robust against aperture
biases. In the case of \mbox{M$_{r}\!<\!-20$} galaxies where aperture
biases are important, we limit our analysis to that based on the
simple separation of passive and star-forming galaxies, and where
possible refer to comparable studies performed using samples limited
to redshifts where aperture effects are much reduced
\citep[e.g.][]{gomez,balogh04,tanaka}. Throughout this article we
indicate the possible effects of aperture biases on our results.   

\subsection{The completeness of the catalogue}
\label{completeness}

The completeness (i.e. the fraction of galaxies brighter than the SDSS 
spectroscopic magnitude limit of $r=17.77$ that have been
spectroscopically observed resulting in good redshifts) of our catalogue
is strictly influenced by three factors:\\ 
i) The dimension of the fibres which prevents two objects closer than
55$''$ from being observed. Roughly 6\% of the objects are not
spectroscopically observed for this reason \citep{blanton03a}.\\ 
ii) The blending of bright galaxies with
saturated stars. Bright galaxies
which overlap saturated stars are flagged themselves as {\sc saturated}
and hence will not be targeted spectroscopically. As one goes to
fainter magnitudes the blending goes down as the area covered by
the galaxy decreases.
The fraction of galaxies not targeted for spectroscopy 
for this reason rises from 1\% overall to 5\% at the bright end of
galaxy sample \mbox{($r\!<\!15$)} \citep{strauss}. \\
iii) The selection criteria set by
LRC are broader than those of the selection algorithm used to target
galaxies for the spectroscopic SDSS survey \citep{strauss}. 

The pronounced incompleteness of the spectroscopic catalogue at the
bright end may bias the detection and characterization of low-$z$ groups of galaxies, since
the most luminous objects of these structures are not included. 
To cope with this deficiency, we have matched the photometric catalogue
of SDSS with the NASA/IPAC
Extragalactic Database (NED). For all
the objects with a positive match we have associated the SDSS $ugriz$
photometry with the corresponding redshift from NED and calculated the
absolute magnitudes. There are a total of 803 galaxies added this way
to our catalogue with \mbox{$0.005\!<\!z\!<\!0.037$} and
\mbox{$r\!<\!17.77$}. Their contribution is largest at bright
magntitudes where the 202 \mbox{$r\!<\!14.5$} galaxies from NED make
up $\sim\!8$\% of the catalogue. We do not have the spectral indices
for the galaxies taken from NED, and so they are only used here in
defining the local environment of the LRC galaxies.

Despite the contribution from NED, our improved spectroscopic
catalogue is still incomplete. To compute the completeness, assuming
that in the catalogue all the objects with spectra are correctly
classified, it is firstly necessary to check the classification of the
objects without spectra.
From a first visual check on the limited sample of bright galaxies
($r\!<\!14.5$) with no spectra we found many objects such as saturated stars
and satellite tracks classified as galaxies. To remove these
objects from our photometric catalogue in the most automatic way, we
have compared 
their flags with those of known galaxies (i.e. with redshifts) looking
for some peculiar differences. From this comparison we have noticed
that, differently from galaxies, all the saturated stars have the
flags {\sc saturated},
{\sc satur\_center} and the great part of those due to satellite tracks have the flag
{\sc edge} \citep{stoughton}.
After removing the objects with these flags, we have performed a
visual inspection of a subsample of galaxies in the range
\mbox{$14.5\!<\! r\!<\!17.7$} which were not targetted for
spectroscopy. In this subsample we found that the photometric pipeline
sometimes fails the detection, recognizing non-existent objects. Real $r\!<17.7$
galaxies should be clearly detected also in at least the $g$, $i$ and
$z$ images, whereas this should not be the case for non-existent
objects, and hence to exclude
these objects we have only selected galaxies as having \mbox{$g,i,z\!<\!21$}
and \mbox{$r\!<\!17.77$} the last limit due to the selection criteria of the SDSS spectroscopic survey.
Finally, in the photometric catalogue we also found a small percentage \mbox{($\sim\!1$\%)} of stars classified as galaxies and of badly deblended
objects. Since no particular flag characterizes them and it being
impossible to reject these by hand we have left these objects in the
catalogue their influence on determining the completeness being negligible.

To compute the completeness of this {\em cleaned} catalogue we have followed
the prescription of \citet{blanton03a} based on the algorithm
used by SDSS to locate the plates and to assign the
fibres. 
This procedure places on the area covered by the survey a set of
1$^{\circ}49'$ radius circles (defined tiles) such to maximize the number
of available fibres. The intersection between the tiles and the
survey region defines a set of spherical polygons. The union of all the
polygons that could have been observed by a unique set of tiles is
called ``sector''. These sectors are the regions over which we have
computed the completeness, $\mathcal{C}$, as the fraction of galaxies
in the cleaned photometric catalogue that have good redshifts.

\setcounter{figure}{0}
\begin{figure*}
\label{densitymap}
  \begin{center}
    \includegraphics[width=17.0cm]{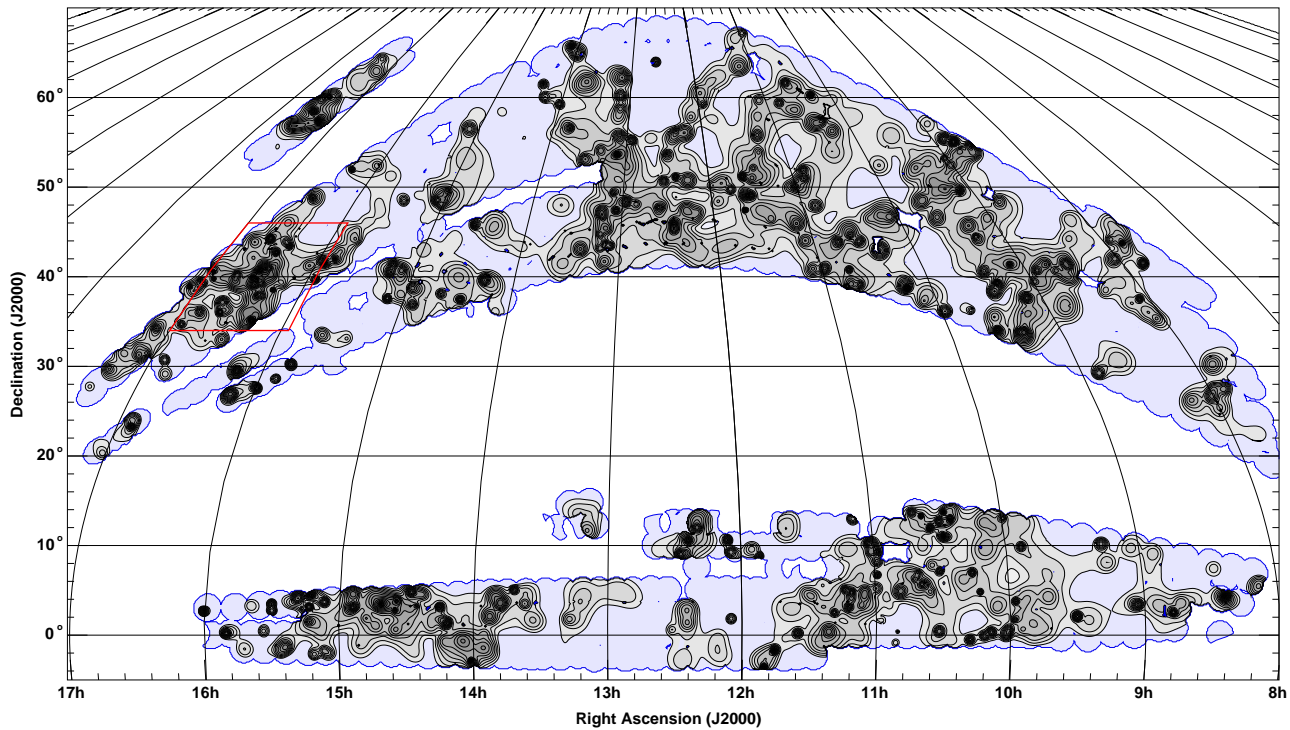}
    \caption{Luminosity-weighted density map of \mbox{M$_{r}\!<\!-18$} galaxies over the redshift range \mbox{$0.023\!<\!z\!<\!0.037$} over the entire SDSS DR4
      North Galactic Cap region. The isodensity contours are
      logarithmically spaced, the spacing between each contours indicating
      a factor $\sqrt{2}$ increase in the $r$-band luminosity-weighted
      local density. The red box indicates the
      \mbox{$12^{\circ}\times12^{\circ}$} region containing the $z=0.03$ A2199
      supercluster analysed in Paper I.}
  \end{center}
\end{figure*}

\section{Definition of Environment}
\label{environment}

To study how the evolution of galaxies is related to their local
environment, we firstly need to define the environment by means
of the local number density of \mbox{M$_{r}\!<\!-18$} galaxies.

To compute the local number density $\rho$($\textbf{x}$, $z$) we use a
variant of the adaptive kernel estimator \citep{silverman,pisani93,pisani96} where
each galaxy $i$ with \mbox{M$_{r}\!<\!-18$} is represented with an adaptive
Gaussian kernel $\kappa_{i}(\textbf{x}, z$) in redshift
space. 
Differently from \citet{silverman} and its previous
applications to astronomical data
\citep[e.g.][]{haines04a,haines04b,haines} in which the kernel width
$\sigma_{i}$ is iteratively set to be proportional to $\rho_{i}^{-1/2}$, we fix the radial width to
500\,km\,s$^{-1}$ and the transverse width 
$\sigma_{i}$ to $(8/3)^{1/2}\,D_{3}$, where $D_{3}$ is the distance of the third nearest neighbour within 500\,km\,s$^{-1}$, a limit
which includes $\sim9$9\% of physical neighbours (as determined from
the Millenium simulation considering the three nearest galaxies in
real-space) and minimizes the
contamination of background galaxies. The choice of $D_{3}$ was made
to maximize the sensitivity of the density estimator to poor groups
containing as few as four galaxies, while the $(8/3)^{1/2}$ smoothing
factor was added to reduce the noise of the estimator, so that in
``field'' regions $\sigma_{i}\approx D_{8}$ which, as shown in
Appendix~\ref{tests} and Table~\ref{estimators}, appears the optimal value
for the smoothing length-scale. 

The
choice both of the method and of the kernel dimensions is designed
to resolve the galaxy's environment on the scale of its host dark
matter halo, as it is the mass of its host halo and whether the galaxy is the central or a 
satellite galaxy, that is believed to be the dominant factor in 
defining its global properties such as star-formation history or 
morphology \citep[e.g.][]{lemson,kauffmann04,yang,blanton06}. 
In the case of galaxies within groups or clusters, the local
environment is measured on the scale of their host halo (0.1--1\,Mpc),
while for galaxies in field regions the local density is estimated by
smoothing over its 5--10 nearest neighbours or scales of 1--5\,Mpc. 

For each galaxy $i$ the local galaxy density is
defined as:
\begin{equation}
\rho_{i}({\mathbf x},z)\propto\sum_{j} \eta_{j} \exp \left[-\,\frac{1}{2} \left\{ \!\!\left( \frac{
  D_{ij}}{\sigma_{j}} \right)^{\!\!2}\!\!+\left(
  \frac{\nu_{i}-\nu_{j}}{500\,{\rm km\,s}^{-1}}
  \right)^{\!\!2} \right\} \right]
\end{equation}
 where \mbox{$\eta_{j}=\mathcal{C}_{j}^{-1}(2\pi)^{-3/2}\sigma_{j}^{-2}$} is the
 normalization factor, D$_{ij}$ is the projected 
distance between the galaxies $i$ and $j$, $\nu_{i}$ is the recession
 velocity of galaxy $i$, and the sum is over all
 galaxies with \mbox{M$_{r}\!<\!-18$}. Note that we also calculate the
 local galaxy density for galaxies fainter than \mbox{M$_{r}=-18$}.

We have performed a number of tests of the efficiency of this density
estimator, in particular with regard to identifying group and isolated
field galaxies, by applying the estimator to the public galaxy catalogues from
the Millennium simulation \citep{springelsim}, and comparing it to
other estimators based on the nearest-neighbour algorithm as applied by
\citet[hereafter BB04]{balogh04b} and \citet[hereafter
  BB06]{baldry06}. These tests are described in detail in the
Appendix, and confirm that the estimator is at least as efficient as
any variant of the nearest-neighbour algorithm for the same dataset. In
particular the estimator is very sensitive to the presence of even
poor groups containing as few as four galaxies, the result being that
selecting galaxies with \mbox{$\rho\!<\!0.5\,{\rm Mpc}^{-2}(500\,{\rm
    km\,s}^{-1})^{-1}$} a pure field sample is produced, with {\em no}
contamination from group members. In contrast 90\% of $\rho>4$
galaxies lie within the virial radius of a galaxy group or cluster,
while those galaxies in the transition regions between groups and
field environments ($r\!\sim\!{\rm R}_{vir}$) have densities in the
range $1\!\la\!\rho\!\la\!4$. 

Figure~1 shows the resultant $r$-band luminosity-weighted density
map for galaxies with \mbox{M$_{r}\!<\!-18$} over the redshift range
\mbox{$0.023\!<\!z\!<0.037$} for the whole SDSS DR4 North Galactic Cap region. For
comparison, the red box indicates the \mbox{$12^{\circ}\times12^{\circ}$}
region containing the A2199 supercluster that was analysed in Paper
I. The adaptive kernel estimator used
has the advantage of being able to be used as a group-finder
\citep[e.g.][]{bardelli,haines04a}, by identifying groups and clusters as local maxima in the galaxy
density function $\rho(\textbf{x}, z)$, and as demonstrated in the
Appendix all groups and clusters having four or more
\mbox{M$_{r}\!<\!-18$} galaxies in the SDSS DR4 catalogue will be marked
by local maxima in the density map of Figure 1. To put this in
perspective, we are sensitive to environments comparable to the Local
Group (which contains four \mbox{M$_{r}\!<\!-18$}
galaxies: Milky Way, LMC, M\,31 and M\,33) and the other nearby groups
\citep[the M81, Cen\,A/M\,83 and Maffei groups;][]{karachentsev05}. 
Such poor groups represent the preferential major-merger mass scale 
\mbox{(M$_{\rm halo}\!\sim\!10^{12}{\rm M}_{\odot}$} for galaxies of stellar mass
\mbox{$\sim\!10^{10}$--1$0^{11}\,{\rm M}_{\odot}$} \citep{hopkins07a}.

\section{The bimodality in EW(H$\bmath{\alpha}$) and its dependence on luminosity and environment}
\label{bimodality}

\begin{figure}
\centerline{\includegraphics[width=70mm]{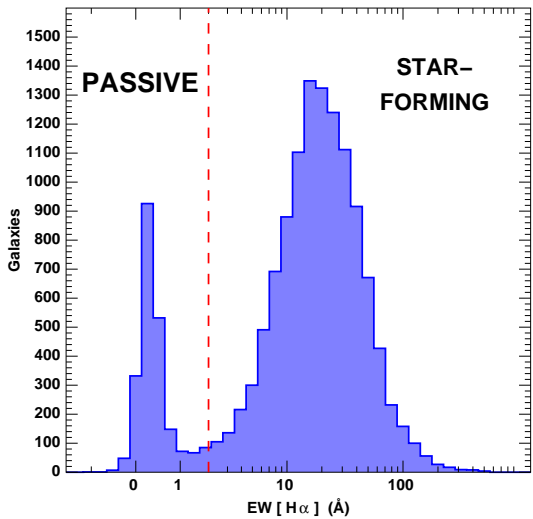}}
\caption{The EW(H$\alpha$) distribution of \mbox{M$_{r}<-18.0$}
  galaxies (AGN excluded) in
  the redshift range \mbox{$0.005<z<0.037$}.}
\label{Ha_hist}
\end{figure}

One of the best understood and calibrated indicators of the
star-formation rate (SFR) in galaxies is the H$\alpha$ nebular
emission-line, whose luminosity is directly proportional to the ionizing radiation from massive \mbox{$(>\!10\,{\rm M}_{\odot})$} short-lived
\mbox{($<\!2$0\,Myr)} stars, and hence the H$\alpha$ emission provides
  a near-instantaneous measure of the current star-formation rate
  \citep{kennicutt}. 
Figure~\ref{Ha_hist} shows the EW(H$\alpha$) distribution of
\mbox{M$_{r}\!<\!-18.0$} galaxies in the redshift range \mbox{$0.005\!<\!z\!<\!0.037$} from the
SDSS DR4 spectroscopic dataset. The x-axis is scaled as \mbox{$\sinh^{-1}{\rm EW(H}\alpha)$}: this results
in the scale being linear at \mbox{EW(H$\alpha)\!\approx\!0$\AA} where measurement
  errors dominate, and logarithmic for \mbox{EW(H$\alpha)\!\ga\!10$\AA}  
  allowing the lognormal distribution of equivalent widths for star-forming
  galaxies to be conveniently displayed.

We exclude galaxies showing an AGN signature, as their H$\alpha$ emission may be dominated by emission
from the AGN rather than star-formation. AGN are defined using the
[N\,{\sc ii}]$\lambda6584$\,/\,H$\alpha$ versus
[O\,{\sc iii}]$\lambda5007$\,/\,H$\beta$ diagnostics of \citet*{baldwin}
as
lying above the 1$\sigma$ lower limit of the models defined by
\citet*{kewley}. When either the [O\,{\sc iii}]$\lambda5007$ or H$\beta$ lines are
unavailable \mbox{(S/N$<3$)}, the two-line method of \citet{miller} is used,
with AGN identified as having
\mbox{log([N\,{\sc ii}]$\lambda6584$\,/\,H$\beta)\!>\!-0.2$}. 
We also
exclude those galaxies without an H$\alpha$ measurement.

The distribution is clearly bimodal, with two approximately Gaussian
distributions: one that is narrow and centred at
\mbox{EW(H$\alpha)\!\sim\!0.2$\AA}, corresponding to passively-evolving galaxies
with little or no ongoing star-formation; and another that is wider and
centred at \mbox{EW(H$\alpha)\!\sim\!20$\AA}, corresponding to galaxies currently
actively star-forming. Midway between these two distributions there
are relatively-speaking very few galaxies, and we identify the dividing
line between passive and star-forming galaxies as being
\mbox{EW(H$\alpha)=2$\AA}, that corresponds approximately to the minimum in
the distribution between the two peaks. Note that this value is
different to that used in the studies of \citet[hereafter
  B04]{balogh04} and \citet[hereafter T04]{tanaka} who use
\mbox{EW(H$\alpha)=4$\AA} to separate passive and star-forming
galaxies, but is the same as used by \citet[hereafter R05]{rines05}.
The lower value however appears justified empirically from
Fig.~\ref{Ha_hist}, and is sufficiently large that even for the
faintest galaxies ($r\!\sim\!17.77$) the limit still represents a
4$\sigma$ detection in H$\alpha$, the median uncertainty in EW(H$\alpha$) only
reaching 0.5\AA\, by $r=17.77$. The inclusion of galaxies with optical
AGN signatures would tend to fill in the gap in the bimodal
distribution, their H$\alpha$ equivalent widths typically in the range
0.5--10\AA\, (median=1.56\AA). 

\begin{figure*}
\includegraphics[width=170mm]{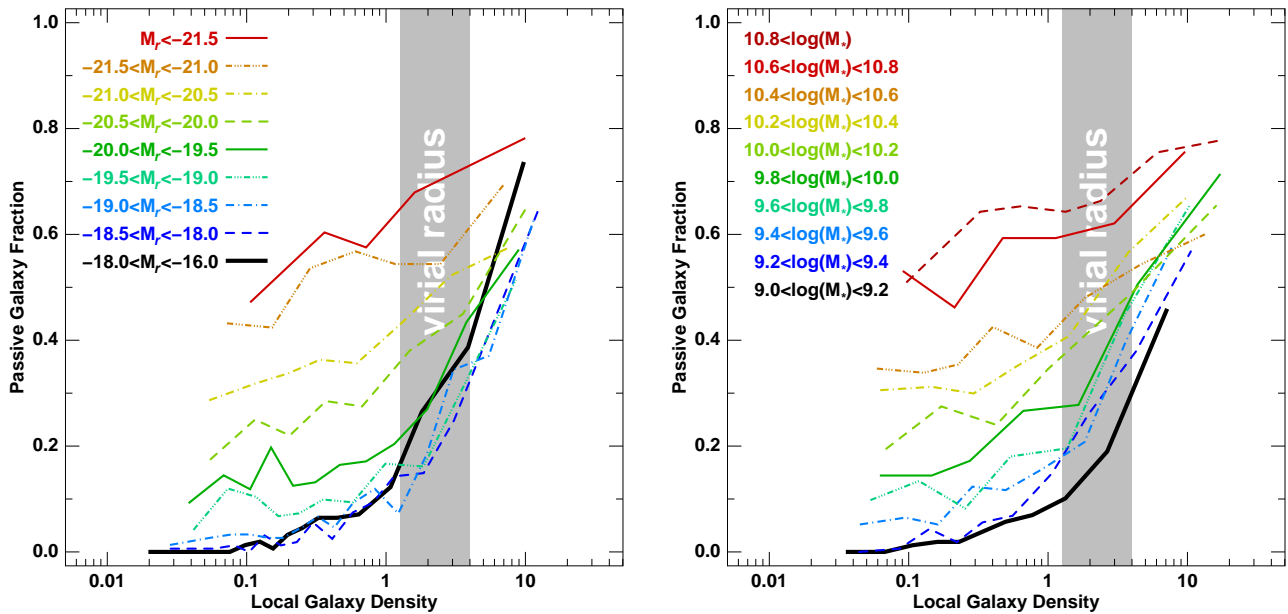}
\caption{The fraction of passively-evolving galaxies
  \mbox{(EW[H$\alpha]<2$\,\AA)} as a function of both local density and
  luminosity (left panel) or stellar mass (right panel). Each coloured
  curve corresponds to a different luminosity / stellar mass bin as
  indicated. Each density bin contains 150 galaxies. The grey shaded
  region indicates the typical densities found for galaxies near the virial
  radius \mbox{($0.8<(r/{\rm R}_{vir})<1.2$)} of groups or clusters in the Millennium simulation (see Fig.~\ref{frac_rho}).}
\label{passive_rho}
\end{figure*}

The left panel of Fig.~\ref{passive_rho} shows how the bimodality in EW(H$\alpha$), and hence
the ongoing star-formation rate of galaxies, depends on both luminosity and
environment. Each coloured curve shows the fraction of
passively-evolving galaxies \mbox{(EW[H$\alpha]\!<\!2$\AA)} as a function of
local density for a particular luminosity range as indicated. 
The lowest luminosity bin \mbox{($-18<{\rm M}_r<-16$)} is far from complete,
and is biased heavily towards galaxies close to the bright magnitude
limit, but the environmental trends should be representative of those
galaxies slightly fainter than \mbox{M$_{r}=-18$.} 
Galaxies that lie very close to the edge of the SDSS DR4 footprint
are likely to have biased density estimates, and so
galaxies that are within 2\,Mpc or $\sigma_{i}$\,Mpc, whichever is
smaller, of the survey boundary are excluded from all further
analyses. This results in a final sample of 22\,113 galaxies.  

At the highest densities \mbox{($\rho\ga\!5$)}, corresponding to the centres of
galaxy clusters or groups, passive galaxies dominate for
the entire luminosity range studied, with \mbox{$\sim\!7$0\%} of galaxies being
passive independent of luminosity. At lower densities in contrast
the fraction of passive galaxies depends strongly on luminosity. Even
at densities comparable to those seen at the virial radius of
groups/clusters, the fraction of \mbox{M$_r\!\ga\!-20$} galaxies that are passive
has dropped to \mbox{$\sim\!2$0\%} or lower, while that of brighter galaxies
has dropped only slightly. The luminosity dependence is
greatest for the lowest density regions corresponding to field
environments well beyond the environmental influence of galaxy
clusters or groups. Here the fraction of passive galaxies drops from
\mbox{$\sim\!5$0\%} for \mbox{M$_r\!\la\!-21$} galaxies to \mbox{$\sim\!0$\%} for \mbox{M$_r\!\ga\!-19$}  
In the lowest luminosity bin \mbox{($-18\!<\!{\rm M}_r\!<\!-16$)} the passive galaxy
fraction has dropped to precisely zero in the lowest density
regions. In fact, there are no passive galaxies in the lowest density quartile, corresponding to \mbox{$\simeq6$00} galaxies in total. 

These results can be compared with the analysis of BB04 who show in
their Fig.~2 the fraction of red sequence galaxies as a function of
both environment and $r$-band luminosity using data from SDSS
DR1. As here, BB04 find that \mbox{$\sim\!7$0\%} of galaxies in their highest
density bin belong to the red sequence. However, in their lowest
density bin, the luminosity dependence is somewhat less than presented
here, dropping from \mbox{$\sim\!3$5\%} for \mbox{$-22\!<\!{\rm M}_{r}\!<\!-21$} to \mbox{$\sim\!8$\%}
for \mbox{$-19\!<\!{\rm M}_{r}\!<\!-18$}.

The right panel of Fig.~\ref{passive_rho} repeats the analysis using
stellar mass ($\mathcal{M}$) instead of $r$-band luminosity. 
Essentially the same
results are obtained, with passive galaxies dominating in high-density
regions independent of stellar mass, while in low-density regions the
fraction of passively-evolving galaxies depends strongly on stellar
mass, dropping from \mbox{$\sim\!5$0\%} at \mbox{$\mathcal{M}\!\sim\!10^{10.8}\,{\rm
  M}_{\odot}$} to zero by  \mbox{$\mathcal{M}\!\sim\!10^{9.2}\,{\rm M}_{\odot}$}. 
We note that for stellar masses below \mbox{$10^{9.2}\,{\rm M}_{\odot}$} we
are no longer volume-limited introducing a selection bias, whereby
passively-evolving galaxies are more likely to
be missed by the \mbox{$r=17.77$} magnitude limit than star-forming galaxies
of the same mass and at the same distance. 

BB06 have performed a very similar analysis of the same SDSS DR4 dataset,
examining how the fraction of {\em red sequence} galaxies varies as a
function of both environment and stellar mass (their Fig.~11a). They consider a much
larger volume than our analysis, resulting in a significantly larger sample,
particularly at the high-mass end, allowing them to follow the
environmental trends for stellar mass bins to \mbox{$\log\mathcal{M}=11.6$}.
 BB06 use a different approach to K03 to calculate the mass-to-light
 ratios of the galaxies based on the $u-r$ colour only, but they use
 the same IMF, and as shown in Fig.~5 of BB06 obtain stellar
 masses that on average are within 0.1 dex of one another. 
The global trends are qualitatively the same, with red sequence galaxies
 dominating in high-density environments independently of stellar
 mass, while in the lowest density environments the fraction of red
 sequence galaxies is a strong function of stellar mass. This latter
 trend extends to the higher stellar masses studied by BB06, falling from
 $\sim\!1$00\% at \mbox{$\log\mathcal{M}\!\sim\!11.6$} to 5\% by \mbox{$\log\mathcal{M}\!\sim\!9.0$}.   
However for the same stellar mass bin, the red sequence fractions of
BB06 in low-density regions are systematically \mbox{$\sim1$0\%} higher than
the passive galaxy fraction from our analysis. 

Although the trends shown here in Fig.~\ref{passive_rho}
are similar to those of BB04 and BB06, as discussed above there are
some important differences. In particular, we find that for
\mbox{M$_{r}\!\ga\!-18.0$} or \mbox{$\mathcal{M}\la\!10^{9.2}{\rm M}_{\odot}$} there
are no passively-evolving galaxies in the lowest-density bins, whereas
for the same stellar mass / luminosity ranges 
both BB04 and BB06 find that 5--10\% of the galaxies belong to the red
sequence in their lowest density bin. 
This difference has important consequences for the conclusions that
can be drawn from the data (see $\S$~\ref{discussion}). 
What is the cause of this remnant
population of faint red galaxies in low-density environments, that
disappears in our analysis\,? Firstly, as discussed previously, the
local density estimator used in BB04 and BB06 is not completely able
to separate group and field galaxies, so that even for the lowest
density bin considered \mbox{$\sim5$\%} of the galaxies are group members,
the majority of which lie on the red sequence at all luminosities. 
Secondly, not all red sequence galaxies are passively-evolving: a
significant fraction are known to be star-forming, and appear red due
to high levels of dust extinction. In an analysis of the SDSS main
sample galaxies covered by infrared imaging from the SWIRE survey,
\citet{davoodi} find that 17\% of red sequence galaxies are 
dusty star-forming galaxies (identified by their high 24$\mu$m to
3.6$\mu$m flux ratios and H$\alpha$ emission), while \citet{wolf05} find that
 dusty star-forming galaxies constitute more than one-third of the
red sequence population in the A901/2 supercluster region. 

Conversely, due to the SDSS spectra being obtained through
3$^{\prime\prime}$ diameter fibres, the region covered may only cover
the central bulge region of nearby large galaxies, resulting in
galaxies 
appearing passive despite having normal star-forming disks. As
discussed earlier ($\S$~\ref{aperture}) 
based on a comparison of the SDSS and {\em GALEX} NUV photometry of
\mbox{$\sim\!1$5\%} (4065 galaxies) of our low-redshift sample we find that \mbox{$\sim\!8$\%} (20 out
of 246) of bright
\mbox{(M$_{r}\!<\!-21$)} galaxies are classified as passive yet have blue UV-optical colours
\mbox{(${\rm NUV}-r<4$)} indicative of normal star-forming
galaxies (Paper III). This fraction drops steadily with magnitude
(being 2.5\% for \mbox{$-21\!<\!{\rm M}_{r}\!<\!-20$} galaxies),
falling to zero (0 out of 1375) for galaxies at \mbox{M$_{r}\!\ga\!-19.5$}. 
We find no significant variation of these fractions with environment.
We thus indicate that the passive galaxy fractions obtained for the
higher luminosity/mass bins are overestimated due to aperture effects,
but that those for the lower luminosity galaxies
\mbox{(M$_{r}\!\ga\!-20$)} are robust against aperture biases.

\section{Star-forming Galaxies}
\label{sfr}

\begin{figure}
\centerline{\includegraphics[width=80mm]{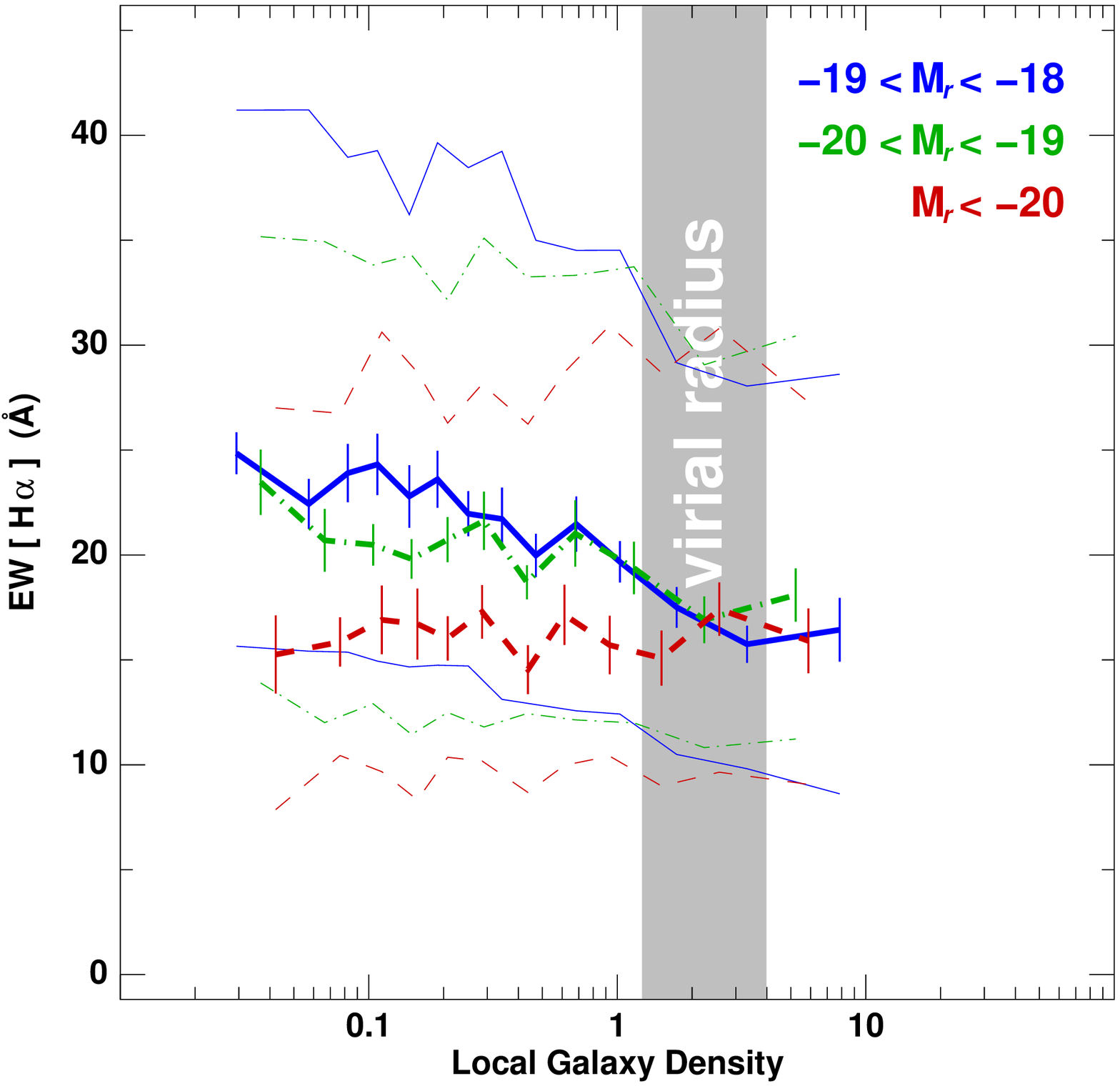}}
\caption{The dependence of EW(H$\alpha$) on local density for star-forming
  galaxies with EW(H$\alpha)>2$\AA. The red dashed lines
  represent giant galaxies \mbox{(M$_r<-20$)}, the green dot-dashed lines represent
  galaxies with \mbox{$-20<{\rm M}_r<-19$}, while the blue solid lines represent
  dwarf galaxies \mbox{(M$_r>-19$)}. The thick and thin lines show
  respectively the median and interquartile values of the
  distribution. The median lines are accompanied by $1\sigma$ error
  limits estimated by bootstrap resampling including the measured
  error in EW(H$\alpha$). Each bin contains 300 galaxies.}
\label{ha_rho}
\end{figure}

If star-forming galaxies at the present day are affected by
environmental mechanisms when they move from low- to high-density
regions, we should see a signature of this transformation which
depends on the relevant time-scale. In particular, if the dominant
environmental mechanism produces a gradual ($\ga\!1$\,Gyr) decline in
star-formation when galaxies become bound to groups or clusters
(e.g. suffocation), then star-forming galaxies in dense regions should
show systematically lower star-formation rates or EW(H$\alpha$). On
the other hand, if the dominant environmental mechanism suppresses
star-formation in galaxies on a very short timescale, then we should
not expect any significant changes in the EW(H$\alpha$) distribution
of star-forming galaxies, since the galaxies will quickly become
classed as passive and hence not contribute to the EW(H$\alpha$)
distribution. In the previous studies of B04 and T04 the distributions of EW(H$\alpha$) of giant
\mbox{(${\rm M}_{r}\!<\!{\rm M}^{\star}\!+1$)} star-forming \mbox{(EW[H$\alpha]\!>\!4$\AA)}
galaxies show no
dependence on local density, while R05 found no difference in the
EW(H$\alpha$) distributions of star-forming galaxies inside the virial
radius, in infall regions \mbox{($1<(r/{\rm R}_{200})<5$)} or in field
regions.
From these results they imply that few giant
galaxies can be currently undergoing a gradual decline in
star-formation due to environmental 
processes. However, when considering fainter galaxies with \mbox{${\rm
  M}^{\star}\!+1\!<\!{\rm M}\!<\!{\rm M}^{\star}\!+2$} T04 found the
EW(H$\alpha$) of star-forming galaxies to be slightly smaller in dense
regions, a result taken to be a signature of the slow truncation of
star-formation in faint galaxies. 

\begin{figure*}
\includegraphics[width=170mm]{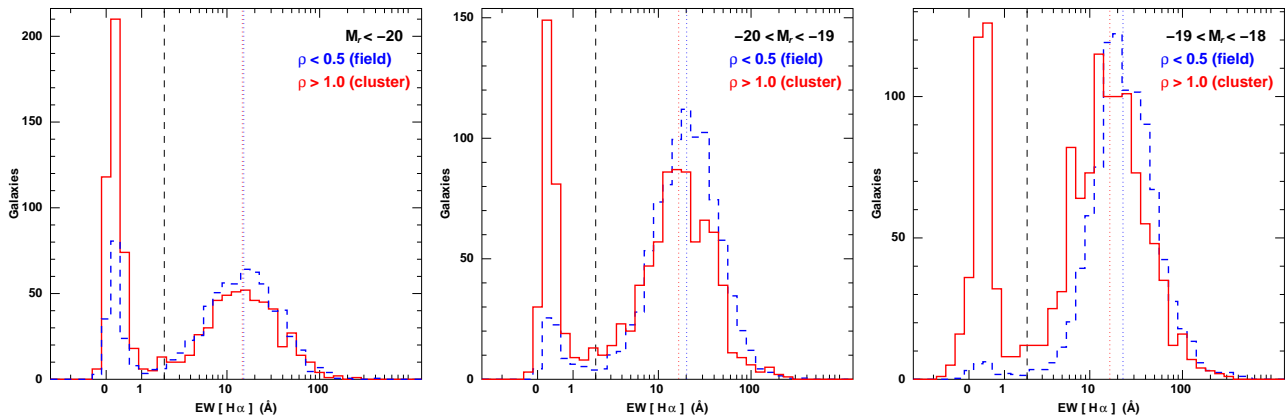}
\caption{A comparison of the EW(H$\alpha$) distributions for
  galaxies in high- ($\rho>1.0$; red histogram) and low-density
  ($\rho<0.5$; blue dashed histogram) regions for three luminosity
  ranges, corresponding to \mbox{M$_r<-20$} (left panel), \mbox{$-20<{\rm M}_r<-19$}
  (middle) and \mbox{$-19<{\rm M}_r<-18$} (right). The vertical scale
  corresponds to the number of galaxies per bin in the high-density
  histogram, while the low-density histogram has been scaled to allow
  comparison of the distribution of star-forming galaxies. The
  red and blue dotted lines indicate the median values of
  star-forming galaxies \mbox{(EW[H$\alpha]\!>2$\,\AA)} in the high- (red) and
  low-density (blue) regions.} 
\label{ha_dist}
\end{figure*}

\subsection{H$\balpha$-density relation for star-forming galaxies}
   
Following B04 and T04 we show in
Fig.~\ref{ha_rho} the EW(H$\alpha$) distribution of star-forming galaxies as a
function of local density for three luminosity ranges: \mbox{M$_r\!<\!-20$} (red
dashed lines) which can be compared with the results of B04
\mbox{(M$_B\!<\!-20.2$; M$_r\!<\!-21.3$)} or the bright sample of T04
\mbox{(M$_r\!<\!-20.3$)}; \mbox{$-19\!<\!{\rm M}_r\!<\!-20$} (green
dot-dashed lines) which is comparable to the
faint sample of T04; and \mbox{$-18\!<\!{\rm M}_r\!<\!-19$} (blue
solid lines). In
each case, star-forming galaxies are defined as having
EW(H$\alpha)>2$\AA\, as throughout this article. Note that both
B04 and T04 use EW(H$\alpha)>4$\AA, but using
this value instead makes no noticeable difference to the results.
We also exclude here those galaxies classified as AGN.
As observed in previous studies, there is no apparent
dependence on density for the EW(H$\alpha$) distribution for galaxies
with \mbox{M$_r\!<\!-20$}. In contrast, the trends for lower luminosity galaxies
show a significant drop in EW(H$\alpha$) with increasing density, most
of the drop occuring within the range \mbox{$0.5\!\la\!\rho\!\la\!2$} which
represents the transition between galaxies inside and outside bound
structures. The significance of the trends are measured using the
Spearman rank correlation test and reported in
Table~\ref{ha_table}. Whereas the EW(H$\alpha$) distribution of \mbox{M$_r\!<\!-20$} star-forming
galaxies shows no correlation with local density $\rho$, significant
anti-correlations are found for the \mbox{$-20\!<\!{\rm M}_r\!<\!-19$} and
\mbox{$-19\!<\!{\rm M}_r\!<\!-18$} star-forming galaxy populations at the 5$\sigma$ and
10$\sigma$ level respectively.   

We do not expect aperture biases to have any significant effects on
the environmental trends presented here, as we observe no dependencies
on local galaxy density for the distribution of SDSS fiber aperture
covering fractions in any of the luminosity ranges. Similarly we find
no environmental trends for the fraction of the early-type spiral
galaxies classified as passive from their SDSS spectra yet having blue
NUV$-r$ colours. The only possible effect could be a systematic
underestimation of the H$\alpha$ emission in the M$_{r}<-20$
luminosity bin, but our results for this bin 
are fully consistent with the comparable trends obtained by
T04 and B04 based upon galaxy samples at
$0.03<z<0.065$ and $0.05<z<0.095$ respectively, where
aperture effects should not be important \citep{kewley05}.

\begin{table*}
\begin{center}
\begin{tabular}{cccccc} \hline
Magnitude & \multicolumn{2}{c}{Median EW[H$\alpha$](\AA)} & Probability & U
test & Spearman rank \\
range & \mbox{$\rho<0.5$} & \mbox{$\rho>1.0$} & (Kolmogorov-Smirnov)
& ($\sigma$) & correlation $\rho$\\ \hline
\mbox{M$_r<-20$} & 15.19 & 14.66 & 0.464 & 0.07 & 0.0102$\pm$0.0242\\
\mbox{$-20<{\rm M}_r<-19$} & 20.23 & 16.58 & \mbox{$2\times10^{-6}$} & 5.26 & -0.0753$\pm$0.0179\\
\mbox{$-19<{\rm M}_r<-18$} & 22.85 & 16.48 & \mbox{$6\times10^{-25}$} & 11.27 & -0.1580$\pm$0.0156\\ \hline
\end{tabular}
\label{ha_table}
\caption{Comparison of the EW(H$\alpha$) distributions in high
  \mbox{($\rho>1.0$)} and low \mbox{($\rho<0.5$)} density environments.} 
\end{center} 
\end{table*}

To see exactly how the distribution of EW(H$\alpha$) changes with
environment, Fig.~\ref{ha_dist} shows the EW(H$\alpha$) distribution
of galaxies in high (\mbox{$\rho\!>\!1.0$}; red histogram) and low (\mbox{$\rho\!<\!0.5$};
blue dashed histogram) density envrironments for three luminosity ranges,
corresponding to \mbox{M$_r\!<\!-20$} (left panel), \mbox{$-20<\!{\rm M}_r\!<\!-19$} (middle)
and \mbox{$-19<{\rm M}_r\!<\!-18$} (right). The vertical red and blue dotted
lines indicate the median values of star-forming galaxies (EW[H$\alpha]>2$\AA).

The bimodal character of the EW(H$\alpha$) distribution is apparent
in both the high- and low-density environments for each of the
luminosity ranges studied. The two environmental dependencies
described in Figs.~\ref{passive_rho} and~\ref{ha_rho} can both be seen
when comparing the EW(H$\alpha$) distributions for the high- and
low-density environments. 

Firstly, a global shift in the relative
fractions of star-forming and passively-evolving galaxies is
apparent. The two histograms have been normalized so that
distributions of the star-forming galaxies appear to have approximately the
same height. As a result, the relative increase in the fraction of
passively-evolving galaxies from low- to high-density environments is
clear. This relative increase is strongly dependent on luminosity,
rising from about a factor 2.5--3 for luminous (M$_r<-20$) galaxies
to a factor $\sim20$ for the dwarf ($-19<{\rm M}_r<-18$) galaxy
population.  

The
second effect can be seen as a global shift in the EW(H$\alpha$)
distribution of the star-forming galaxies from high- to low-density
environments. In each environment and luminosity range, the EW(H$\alpha$)
distribution of the star-forming galaxies can be well described as being
log-normal (and hence appearing as a Gaussian distribution in the figure).   
However, whereas there is no apparent difference in the high- and
low-density distributions for luminous \mbox{(M$_r\!<\!-20$)} star-forming
galaxies, at lower luminosities, the high-density EW(H$\alpha$)
distributions are {\em systematically} shifted to lower valaues than their
low-density counterparts. The level of this shift is quantified by
comparison of the median values of the distribution, while the
significance of the differences between the two distributions are
estimated through application of the non-parametric Kolmogorov-Smirnov and
Wilcoxon-Mann-Whitney U tests, the results of which are shown in
Table~1. 
These results confirm that while the
EW(H$\alpha$) distribution of the high- and low-density \mbox{M$_r\!<\!-20$}
galaxy populations are fully consistent with one another, for the
lower luminosity samples the null hypothesis that the high- and
low-density star-forming populations have the same EW(H$\alpha$)
distribution is rejected at very high significance levels. For the
\mbox{$-19\!<\!{\rm M}_r\!<\!-18$} sample, the H$\alpha$ emission from star-forming
galaxies in high-density environments is systematically lower by
\mbox{$\sim\!3$0\%} with respect to their low-density counterparts. 

The H$\alpha$ emission (and hence star-formation) must be
suppressed in a significant fraction of galaxies when they fall into
a cluster or group for the first time.  However, for these galaxies to
remain classed as star-forming, this suppression must act over a long
period of time, to allow a significant fraction of galaxies to be seen
in the process of transformation into passively-evolving galaxies. 
If we assume that the H$\alpha$ emission of galaxies declines
exponentially with time as they are being transformed, and that the rate at
which galaxies are transformed remains constant, the EW(H$\alpha$) of
galaxies which are 
{\em currently} in the process of being transformed but are still classed as
star-forming, will drop from $\sim2$0\AA\, to 2\AA, with an average of
$\sim8$\AA. Hence 
star-forming galaxies in the process of transformation will have {\em
  on average} $\sim4$0\% of their emission prior to their being transformed. 
To
produce a global systematic reduction of $\sim3$0\% in the H$\alpha$
emission would then require $\sim5$0\% of the dwarf star-forming
galaxies in high-density regions to be in the process
of being transformed into passive galaxies. Given that, as discussed
previously, as many as 30--40\% of galaxies in the high-density do not
lie within the virialized regions of a cluster or group, this suggests
that the vast majority of dwarf star-forming galaxies in groups or
clusters are {\em currently} in the process of being transformed into
passive galaxies.  

\subsection{SFR-density relation for star-forming galaxies}

\begin{figure}
\includegraphics[width=84mm]{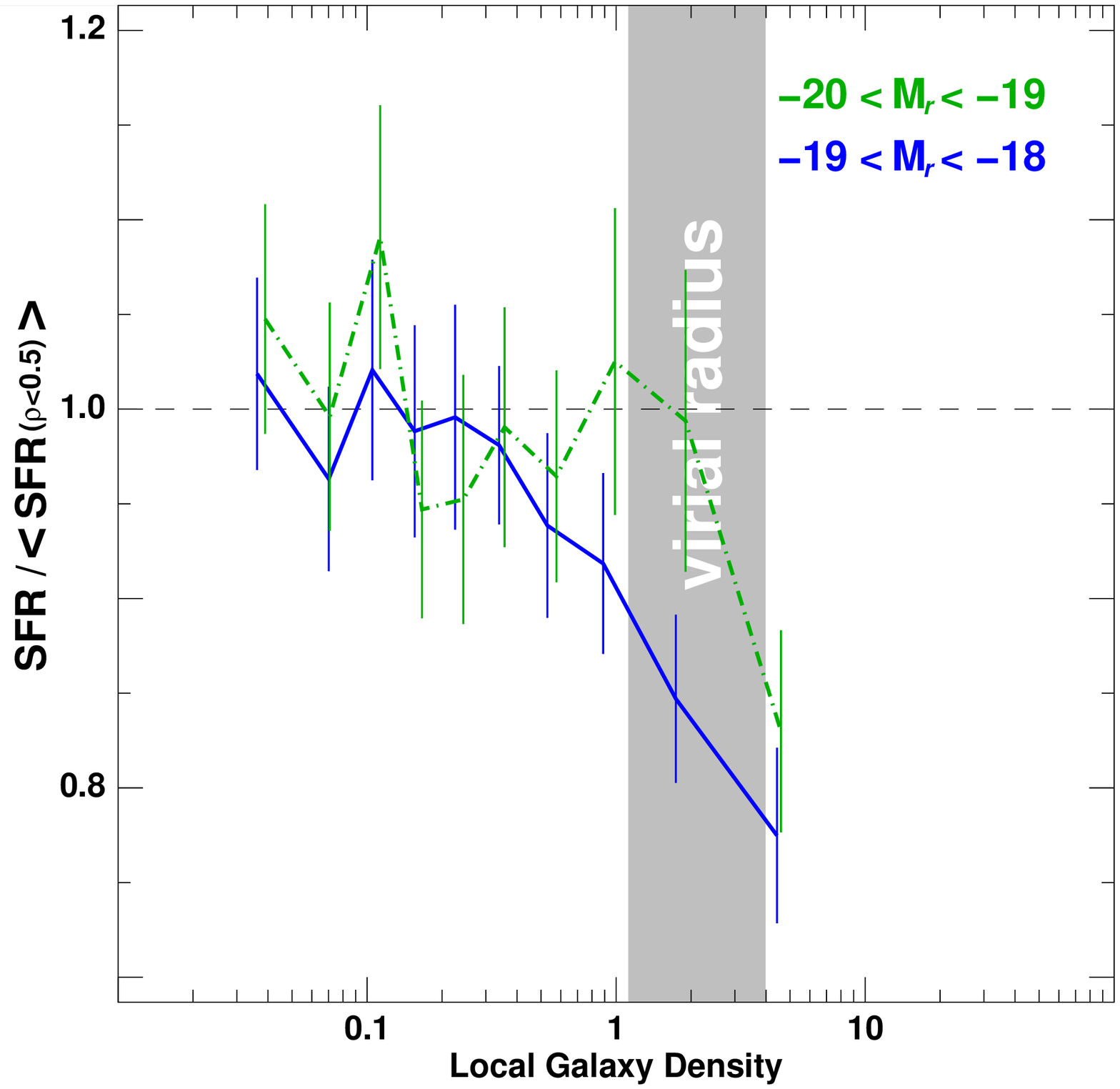}
\caption{The dependence of the median star-formation rate on local
  density for star-forming  galaxies with
  \mbox{EW(H$\alpha)\!>\!2$\AA}. The green
  dot-dashed lines represent
  galaxies with \mbox{$-20\!<\!{\rm M}_r\!<\!-19$}, while the blue
  solid lines represent
  dwarf galaxies \mbox{(M$_r\!>\!-19$)}. The lines are accompanied by $1\sigma$ error
  limits estimated by bootstrap resampling and include the
  measured error in EW(H$\alpha$) and uncertainties in the level of
  dust obscuration. Each bin contains 300 galaxies.}
\label{sfr_rho}
\end{figure}

\begin{table}
\begin{center}
\begin{tabular}{ccccc} \hline
Magnitude & \multicolumn{2}{c}{Median SFR} & & \\
range & \multicolumn{2}{c}{(M$_\odot$yr$^{-1}$)} & P(K-S) & U test \\
of galaxies & $\!\rho<0.5\!$ & $\!\rho>1.0\!$ & & ($\sigma$) \\ \hline
\mbox{$-20\!<\!{\rm M}_r\!<\!-19$} & 0.496 & 0.443 & 0.0079 & 2.30 \\
\mbox{$-19\!<\!{\rm M}_r\!<\!-18$} & 0.191 & 0.157 & \mbox{$\!7\!\times\!10^{-9}\!$} & 6.52 \\ \hline
\end{tabular}
\caption{Comparison of the SFRs of star-forming galaxies in high
  ($\rho>1.0$) and low ($\rho<0.5$) density environments.} 
\label{sfr_table}
\end{center} 
\end{table}

The current star-formation rate of a galaxy can be estimated from its
H$\alpha$ flux through the calibration given by
\citet{kennicutt}:
\begin{equation}
{\rm SFR} \left( {\rm M}_\odot{\rm yr}^{-1}\right) = \frac{L({\rm
 H}\alpha)}{1.27\times10^{34}{\rm W}}
\end{equation} 
 Before applying this calibration, it is necessary to correct for the
 effects of dust obscuration and account for the effects of emission
 lost by virtue of the spectra being obtained through a fibre whose
 aperture may be significantly smaller than the galaxy. The
 obscuration correction is measured by the Balmer decrement, estimated
 by measuring the ratio of the stellar absorption-corrected H$\alpha$
 and H$\beta$ line fluxes, and assuming case B recombination and the
 obscuration curve of \citet*{cardelli}. The aperture correction is
 quantified as the ratio of the observed $r$-band Petrosian flux and the
 continuum flux at the wavelength of H$\alpha$ within the fibre
 aperture. A full discussion of these corrections and the use of
 H$\alpha$ line emission as a SFR indicator in SDSS data is given in \citet{hopkins} where the explicit calculation used is given as
 equation B2. \citet*{moustakas} compare the integrated SFRs estimated from the H$\alpha$ flux using the above procedure
 with those estimated from IRAS infra-red data and find the two
 estimates consistent with a precision of $\pm$70\% and no systematic
 offset, confirming that the extinction-corrected H$\alpha$ luminosity
 can be used as a reliable SFR tracer, even for the most dust-obscured systems.         
Using the above calibration and corrections, we plot in
 Fig.~\ref{sfr_rho} the median SFR of star-forming galaxies as a
 function of local density for galaxies in the luminosity range
 \mbox{$-20\!<\!{\rm M}_{r}\!<\!-19$} (green dot-dashed line) and
 \mbox{$-19\!<\!{\rm M}_{r}\!<\!-18$} (blue solid line). 
As discussed in $\S$\ref{aperture} aperture effects will strongly bias
 the estimates of star-formation rates made using the method of
 \citet{hopkins} for the most massive galaxies in our sample and so we
 do not plot the results for \mbox{M$_{r}\!<\!-20$}  galaxies.
 To allow the effect of
 high-density environments on star-formation to be measured, each
 curve is normalized to the median SFR of ``field''
 \mbox{($\rho\!<\!0.5$)} star-forming galaxies in the same luminosity range. 

The environmental trends in SFR broadly match those shown earlier in
Fig.~\ref{ha_rho} for the EW(H$\alpha$) distribution of star-forming
galaxies, confirming that those trends do indeed reflect changes in
the global SFR with environment, and are not due to
variations in dust obscuration or aperture biases. These trends are
quantified in Table~\ref{sfr_table} which compares the median SFRs of
star-forming galaxies in high- and low-density environments for both luminosity ranges, as well as estimates the significance
of any differences.  The most significant result \mbox{($\sim\!6\sigma$)} is the observed
systematic drop of \mbox{$\sim\!2$0\%} in the median SFR of dwarf
\mbox{($-19\!<\!{\rm M}_r\!<\!-18$)} star-forming galaxies in high-density regions with respect to
field galaxies. In both luminosity bins there appears a
systematic drop in SFR for densities greater than 1\,Mpc$^{-2}$, which
suggests that star-formation is suppressed in a significant fraction
of galaxies when they infall for the first time into a cluster or
group.

As discussed earlier we do not expect aperture biases to be important
for galaxies in these luminosity bins. Moreover we find no
dependencies on local galaxy density for the fraction of galaxy flux
covered by the SDSS fiber apertures in either luminosity bin.
 
\begin{figure}
\includegraphics[width=80mm]{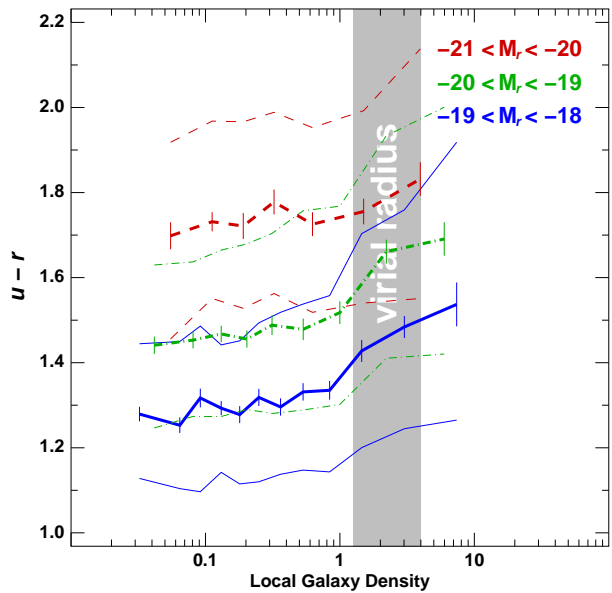}
\caption{The dependence of the $u-r$ galaxy colour on local density for star-forming
  galaxies with EW(H$\alpha)>2$\AA. The red dashed lines
  represent giant galaxies (M$_r<-20$), the green dot-dashed lines represent
  galaxies with $-20<{\rm M}_r<-19$, while the blue solid lines represent
  dwarf galaxies (M$_r>-19$). The thick and thin lines show
  respectively the median and interquartile values of the
  distribution. The median lines are accompanied by $1\sigma$ error
  limits estimated by bootstrap resampling and include the measured
  errors in $u-r$. Each bin contains 300 galaxies.}
\label{u_r_rho}
\end{figure}

As a final check to confirm that aperture effects are not behind the
observed environmental trends in EW(H$\alpha$) and star-formation
rates, we repeat the analyses using galaxy $u-r$ colours as a measure
of their current/recent star-formation. The $u-r$ colours are determined
over apertures defined by the Petrosian radius, and hence represent an
integrated measure of a galaxy's star-formation history. The resultant
trends in the median $u-r$ colour with local density for each of the
luminosity ranges are presented in Fig.~\ref{u_r_rho}.

For each of the three
luminosity ranges star-forming galaxies become increasingly redder
with density. The strength of the trend increases with decreasing
luminosity from 0.07\,mag for \mbox{$-21<{\rm M}_r<-20$} galaxies to
\mbox{$\sim0.2$\,mag} for \mbox{M$_r>-20$} galaxies. 
 Almost identical trends were observed by
\citet{balogh04b} for galaxies selected as star-forming by their $u-r$ colour.
In the case of the two lower luminosity ranges \mbox{(M$_r>-20$)} the bulk of
the change in $u-r$ colour with density occurs at \mbox{$\rho>1$}, as seen
for the trend in SFR of Fig.~\ref{sfr_rho}. 
These trends are fully consistent with those seen in the H$\alpha$
emission and SFR, confirming that the previous trends are not the
result of aperture effects.

\section{Which aspects of environment define the SF-density relation
  ?}
\label{neighbours}

To this point we have examined the environmental dependence on
star-formation in galaxies using densities measured by smoothing over
the nearest 5--10 galaxies. This has allowed us to describe the
effects of the group and cluster environments on the galaxies. It is
also possible that galaxies are affected by the presence of individual
neighbouring galaxies, for example through disturbance from tidal forces.

\begin{figure*}
  \begin{center}
  \includegraphics[width=17cm]{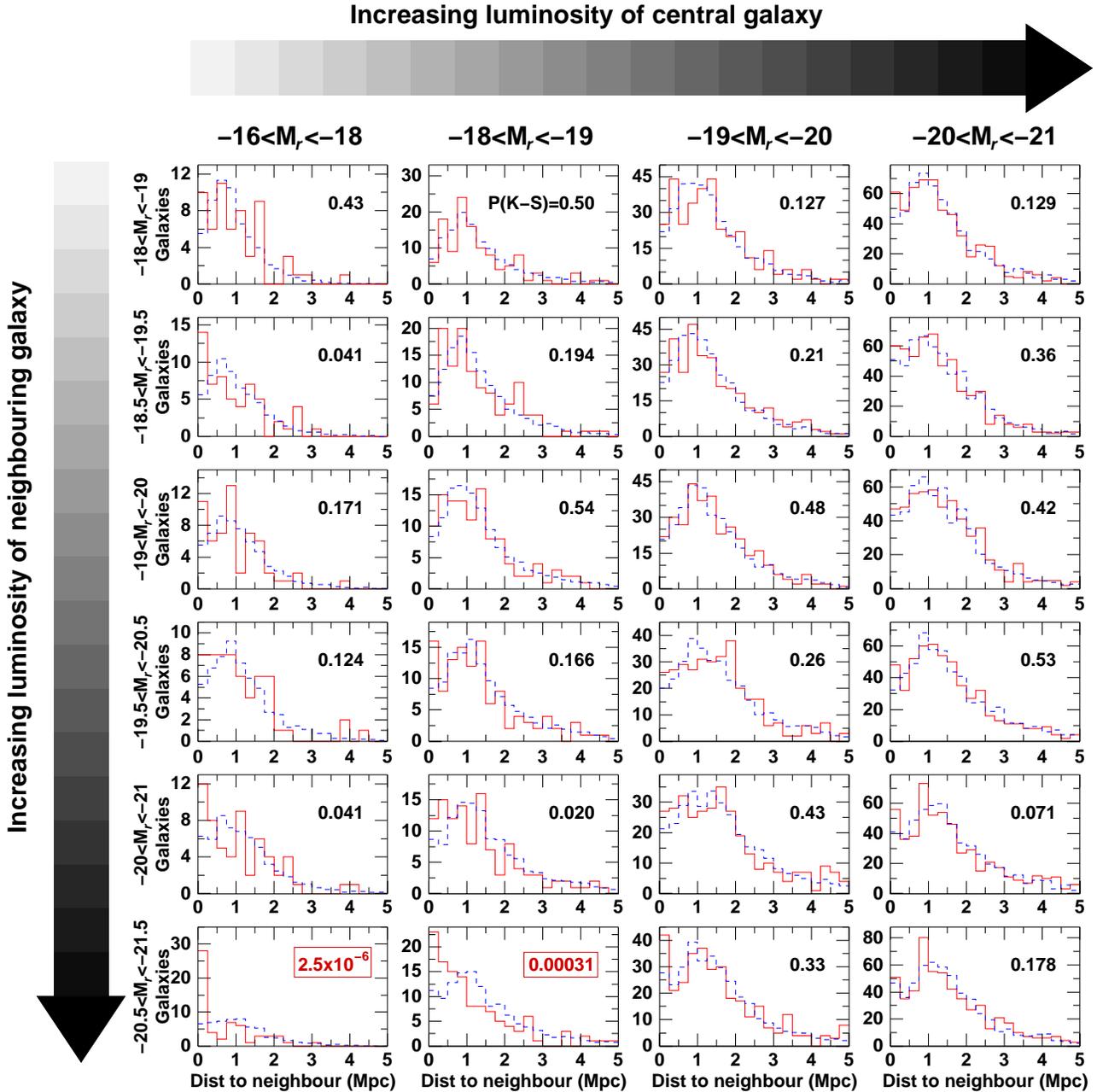}\\
  \caption{The distributions of distances to the nearest neighbouring
  galaxy within a specific luminosity range (in order of increasing
  luminosity from top to
  bottom as indicated) for \mbox{$\rho\!<\!0.5$} passive
  (red line) and star-forming (blue dashed line) {\em central}
  galaxies also within 
  a specific luminosity range (in order of increasing luminosity from left to right as indicated).
The probabilities that the two histograms
  are taken from the same distribution according to the
  Kolmogorov-Smirnov test are indicated in the top-right of each panel.}  

\label{histo1}
  \end{center}
\end{figure*}

In particular, we wish to reexamine for the much larger volume covered
by the SDSS dataset the long noted morphological
segregation of dwarf galaxies in the local \mbox{($\la\!3$0\,Mpc)}
neighbourhood, whereby dwarf ellipticals are confined to groups,
clusters and satellites to massive galaxies, while dwarf irregulars
tend to follow the overall large-scale structure without being bound
to any of the massive galaxies \citep*{binggeli90,ferguson}. \citet{einasto} first noted this segregation when
comparing the spatial distribution of dwarf companions to our Galaxy
and the nearby massive spirals M\,31, M\,81, and M\,101. He found a
striking separation of the regions populated by dE and dIrr galaxies
with dEs confined to being close satellites to the primary galaxies,
and dIrr found at larger distances. 
In Paper I we 
found {\em no} isolated passively-evolving dwarf galaxy, always
finding them gravitationally
bound to clusters/groups or as satellites of $\ga\!L^{\star}$ field
galaxies, differently from star-forming field dwarfs which appeared randomly
distributed throughout the region. 

\begin{figure*}
  \begin{center}
  \includegraphics[width=170mm]{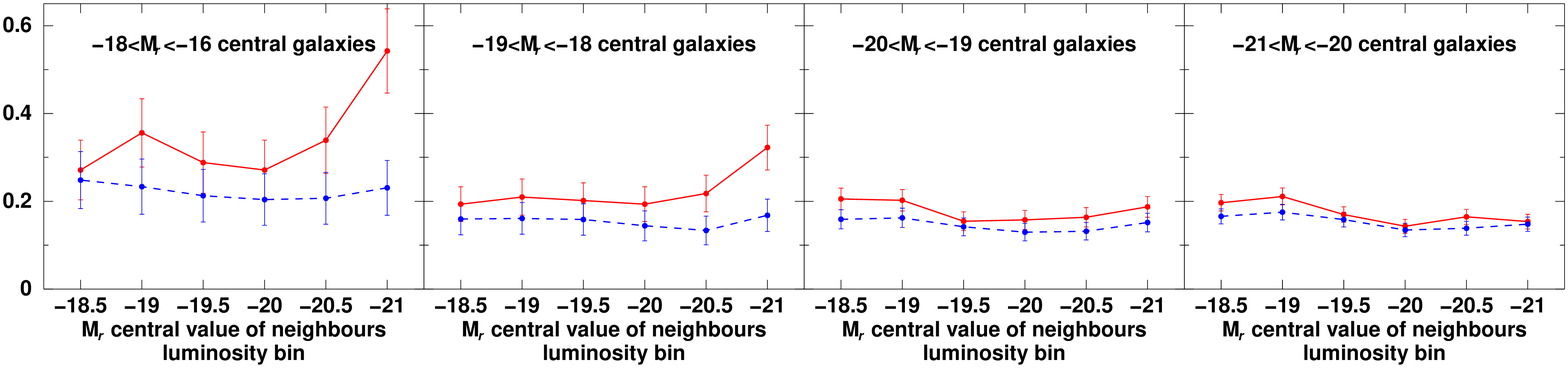}
  \caption{The fraction of passive (red lines) and star-forming (blue
    dashed lines) {\em central} galaxies which have within 0.5
    Mpc one or more galaxies belonging to a fixed range
    of magnitude. Each panel corresponds to {\em central} galaxies
    within a specific magnitude range in order of increasing
    luminosity from left to right as indicated. The top left panel shows that dwarf passive field
    galaxies are much more likely to be found in the proximity of
    massive galaxies than star-forming dwarfs while the giant passive
    field galaxies show no preference as to their neighbours.}
\label{fraction}
  \end{center}
\end{figure*}
 
In this context we wish to look at the effects of neighbouring galaxies
on the star-formation histories of field galaxies ($\rho<0.5$),
i.e. those not in 
groups or clusters for which other processes may well dominate. 
There are two main physical mechanisms whereby a neighbouring galaxy
could affect star-formation in another galaxy, tidal interactions, and
ram-pressure stripping caused by the passage of the galaxy through the
gaseous halo of its neighbour. In both of these mechanisms, the
mass/luminosity of both the {\em central} galaxy (i.e. that which is
being acted on) and the neighbouring galaxy are important for defining
the strength of the effect on the {\em central} galaxy, in particular
the greater mass-ratio between the neighbouring galaxy and the central galaxy, the stronger the effect is likely to be. 
To measure the effect of both the central and neighbouring galaxy
masses, we split both the central and neighbouring galaxies into bins
of luminosity. 
It is also important that we take out the effect of the
large-scale ($\ga\!1$\,Mpc) galaxy density from the equation, as galaxies
in higher density regions will naturally have closer neighbours than
lower density regions. 
To measure the effect of the presence of a
neighbouring galaxy on the
star-formation history of the central galaxy we compare the distances
to the nearest neighbour (within a certain luminosity range) for
passively-evolving and star-forming central galaxies that have the
same mass/luminosity and the same large-scale environment (i.e. their
local densities are the same). If the presence of a neighbouring galaxy is important for causing the central galaxy
to become passive, we would expect passively-evolving central galaxies
to have nearer neighbours (within a certain luminosity range) than
star-forming galaxies of the same luminosity and large-scale
environment.

In Figure~\ref{histo1} the distribution of distances between
passively-evolving \mbox{(EW[H$\alpha]\!<\!2$\AA;} red histograms) and
star-forming \mbox{(EW[H$\alpha]\!>\!2$\AA;} blue dashed
histograms) central galaxies in field regions \mbox{($\rho\!<\!0.5$)} and their nearest
neighbours are compared for both different magnitude ranges of central
galaxies (in order of increasing luminosity from left to right as indicated)
and different magnitude ranges of neighbouring galaxies (in order of
increasing luminosity from top to
bottom as indicated)

As a
consequence of the SF-density relation, even in field regions
passively-evolving galaxies will on average be in higher density
regions than star-forming galaxies of the same luminosity, and hence
on this basis alone would be expected to have closer neighbours on
average. To remove this bias, we normalize the density distribution of
the star-forming galaxies to that of the passively-evolving
galaxies. This is done by splitting the galaxies into ten density
bins
of equal logarithmic width (0.2\,dex) in the range
\mbox{$0.01\!<\!\rho\!<\!1$} and for each bin $j$ identify a weight
\mbox{$\omega_{j}={\rm N}^{j}_{\rm passive}/{\rm N}^{j}_{\rm SF}$} where
\mbox{N$^{j}_{\rm passive}$} and
\mbox{N$^{j}_{\rm SF}$} are the total number of passive and star-forming galaxies in
that density bin. Each star-forming galaxy belonging to density bin
$j$ is then given the corresponding weight $\omega_{j}$. The resulting
weighted population of star-forming galaxies has the same density
distribution as their passive counterparts. The blue-dashed histograms
then represent the distribution of distances to the nearest neighbour
for the star-forming galaxies where each galaxy $i$ is represented by its
corresponding weight $\omega_{i}$.

For each of the panels in Fig.~\ref{histo1} corresponding to a
particular luminosity range for central and neighbouring galaxies, we
estimate the significance of any differences between the distributions
of the distance to the nearest neighbours of passive and star-forming
galaxies (the red and blue histograms) using the Kolmogorov-Smirnov test. The
results of these are indicated in the top-right of each panel, with
significant differences \mbox{(P$_{\rm KS}\!<\!0.01$)} highlighted by red boxes.
 
Looking at the histograms in the last two columns (corresponding to
central galaxies with \mbox{M$_{r}\!<\!-19$}) we see that the
distributions of distances to the nearest neighbours of any luminosity
range are the same for the passively-evolving and star-forming
galaxies. This implies that the star-formation histories of \mbox{M$_{r}\!<\!-19$}
galaxies are not significantly affected by the presence of individual
galaxies in their immediate neighbourhoods, and instead it is only the
global large-scale environment (as measured here by $\rho$) to which
their SFRs are correlated. Equally if we look at the histograms in the
top four rows (corresponding to neighbouring galaxies with
\mbox{M$_{r}\!>\!-20.5$}) the distance distribution to the nearest
neighbours are the same for passively-evolving and star-forming central
galaxies of any luminosity range. Hence star-formation in galaxies (at
least for M$_{r}\!\la\!-16$) is not significantly affected by the presence
of neighbouring galaxies with \mbox{M$_{r}\!\ga\!-20$}.   

The only luminosity combinations of central and neighbouring galaxies
that show any significant difference \mbox{(P$_{\rm KS}\!<\!0.01$)}
between the distance distribution to the nearest neighbours of the
passive and star-forming central galaxies are the two lower-left
panels corresponding to
low-luminosity central galaxies \mbox{(M$_{r}\!>\!-19$)} that have 
bright \mbox{(M$_{r}\!<\!-20.5$)} neighbours. In both these panels we
see that passively-evolving dwarf galaxies \mbox{(M$_{r}\!>\!-19$)} are
  much more likely to have a nearby bright \mbox{(M$_{r}\!<\!-20.5$)}
  neighbour within $\sim\!5$00\,kpc than would a star-forming galaxy of
  the same luminosity and having the same global environment (as
  measured with $\rho$). This implies that massive galaxies can
  influence the star-formation history of neighbouring dwarf galaxies,
  presumably orbiting as satellites, by causing them to become stop
  forming stars.

\begin{table*}
\begin{center}
\label{passivedwarfs}
\begin{tabular}{ccccccccccccc} \hline
RA, Dec & $z$ & M$_{r}$ &$d_{cl}$& Group & Refs & N$_{gal}$ &
$\mu_{nu}$ & $\sigma_{z}$ & R$_{vir}$ & T$_{X}$ &log($L_{X}$)\\
(J2000) & & & $(Mpc)$ & Name & & & 
&(km/s)&(Mpc)&(keV)&(erg/s)\\ \hline
14:37:13.70, +02:28:35.9 & 0.0255 & -17.67 & 2.56 & MKW\,8 & 1,2 &
147 & 0.0267 & 492 & 1.56 & 3.03 & 41.98\\ 
12:00:26.60, +01:40:07.6 & 0.0209 & -17.32 & 1.46 & MKW\,4 & 1,2 &
99 & 0.0199 & 428 & 1.25 & 1.83 & 41.96\\ 
12:00:37.72, +02:08:47.9 & 0.0201 & -17.61 & 1.41 & MKW\,4 & 1,2 &
99 & 0.0199 & 428 & 1.25 & 1.83 & 41.96\\
15:14:04.35, +03:24:04.9 & 0.0297 & -17.82 & 0.76 & & & 
6 & 0.0292 & 221 & 0.52 & &\\
09:47:15.48, +37:03:07.1 & 0.0223 & -17.62 & 1.83 & WBL\,236 & 3,4 & 
24 & 0.0219 & 324 & 0.96 & & 41.76\\ \hline
\end{tabular}
\end{center}
References: (1) \citet{rines}, (2) \citet{popesso}, (3) \citet{white99}, (4) \citet{mahdavi}.
\caption{Candidate isolated passively-evolving dwarfs and possible
  associated groups}
\end{table*}

We reillustrate these effects in
Figure~\ref{fraction} where we plot the fraction of {\em central}
passive (red solid line) and star-forming (blue dashed line) galaxies 
that have one or more neighbours within 0.5\,Mpc belonging to a fixed
magnitude range. Each panel corresponds
to a different magnitude range of central galaxies as before, while
each point corresponds to one of the six magnitude ranges of
neighbouring galaxies of Figure~\ref{histo1}, the central value of
which is indicated along the x-axis.     
The fractions of passive and
star-forming giant \mbox{($-21\!<\!{\rm M}_{r}\!<\!-20$)} galaxies with a
neighbour within 0.5\,Mpc are quite similar for every magnitude range
of surrounding systems; not only, but the common fractions are quite
constant independently of the luminosity of neighbouring
galaxies. These observations show that both passive and star-forming
massive galaxies have no preferences about their neighbours suggesting
a uniform distribution for these systems. Different results are found
for dwarf \mbox{($-18\!<\!{\rm M}_{r}\!<\!-16$)} field galaxies. In fact, the
fractions of passive dwarfs with neighbours within 0.5\,Mpc are
different from those found for their star-forming counterparts. The
latter show a similar behavior with the field giants having, on
average, no preference for neighbours of a particular luminosity. On
the contrary the fraction of passive field dwarfs with a close-by
galaxy strongly increases with the luminosity of the neighbour, underling
that these systems, as was firstly pointed out by \citet{einasto}, are not uniformly distributed but are commonly found close
to massive galaxies. This trend is also present, even if in a less
strong way, for \mbox{$-19\!<\!{\rm M}_{r}\!<\!-18$} {\em central} galaxies and
disappears at brighter magnitudes.

These results suggest that the mechanisms transforming giants and
dwarfs from star-forming to passive systems are different.  The quite
uniform spatial distribution of passive field giant galaxies
underlines the negligible influence of any environmental interactions
in stopping star-formation for these systems, while the frequent
presence of nearby massive galaxies to passive field dwarf galaxies is
a clear indication of the fundamental  
impact of massive galaxies on star-formation in nearby dwarf systems.

Out of the 252 passively-evolving dwarf galaxies in the lowest luminosity
bin \mbox{($-18\!<\!{\rm M}_{r}\!<\!-16$)} 48 are in regions with
\mbox{$\rho\!<\!0.5$}. Of these 48, 34 were found to have bright
\mbox{(M$_{r}\!<\!-20$)} galaxies within $\sim5$00\,kpc and
\mbox{$\sim\!5$00\,km\,s$^{-1}$}, 24 of which were within \mbox{$\sim\!2$00\,kpc} and
\mbox{$\sim\!2$00\,km\,s$^{-1}$}. No further neighbours are identified if the
magnitude limit is extended from \mbox{M$_{r}\!<\!-20$} to \mbox{M$_{r}\!<\!-19$}. 
A further nine galaxies were identified as
not actually being passively-evolving dwarfs, either having apparent
H$\alpha$ emission not identified by the MPA/JHU pipeline, appearing
blue, or having bad photometry which made the galaxy appear much
fainter than it actually was. 

Only five passively-evolving dwarf
galaxies appear to be isolated, being 0.8--1.2\,Mpc from the nearest
bright galaxy. However, looking a little further out we find that all five appear to lie in the infall regions of galaxy groups
at around 1.5--2 virial radii from the group centres, as indicated in
Table~3. The centres and redshifts of each of the groups were
identified as maxima in the luminosity-weighted galaxy density
distribution, and the cluster velocity dispersions and virial radii
determined as in \citet{girardi} based on the galaxy radial velocities within
2\,Mpc and $3\sigma_{\nu}$ of the cluster centre. 

Although the first galaxy lies some 2.5\,Mpc from MKW\,8,
this cluster is part of a larger structure which extends for
$\sim7$\,Mpc around the ``isolated'' passive dwarf galaxy. It
seems reasonable to assume this structure is still in the process of
assembly, and hence the dwarf galaxy may have been left behind or
thrown out by a previous interaction between the structures. The
remaining nearby clusters are rather more isolated and regular, and so
it seems less likely that the other four dwarf galaxies were thrown out by
cluster interactions. The most likely mechanism for these galaxies to
have become passive is that in the past their orbits took them through
their neighbouring group/cluster, whereupon they became passive through
ram-pressure stripping and/or tidal interactions. From cosmological
N-body simulations \citet{mamon} find that infalling galaxies on
radial orbits can bounce out of the clusters, reaching maximum
clustercentric distances of between 1 and 2.5 virial radii. The main
difficulty is to understand why these galaxies haven't been able to
start forming stars again once they are no longer affected by the
cluster environment. In particular, while these galaxies may have been
completely stripped of gas while in the cluster, outside gas recycling
from stellar mass loss should be able to produce enough gas to be
detectable \citep[e.g.][]{jungwiert} and subsequently allow
star-formation to restart after a few Gyr, although \citet{grebel}
suggest that ram-pressure from the passage of the galaxy through the
low-density IGM may be sufficient to strip the stellar mass loss as it
is recycled. It would be
interesting to confirm whether these isolated dwarfs are truly
passively-evolving or whether they have any detectable H\,{\sc i}.   

\section{The connection with AGN}
\label{agn}

In recent years observations have shown that at the heart of most if
not all massive galaxies is a supermassive black hole \citep[for a
  review see][]{ferrarese05}, and it has become increasingly clear that the evolution of the galaxy and the
central black hole are strongly interdependent. This is manifested
most clearly by the tight correlations between the mass of the
central supermassive black hole (SMBH) and the global properties of their
host galaxies, such as the stellar mass of the bulge component
\citep[M$_{\rm BH}=0.0014\pm0.0004\,{\rm M}_{\rm Bulge}$;][]{haring}, the
stellar velocity dispersion
\citep{gebhardt,ferrarese05}, and the host bulge Sersic index \citep{graham07}. The scatter in the black hole masses
around these relations are only of the order 0.3\,dex. 

\citet{silkrees} suggested that these correlations arise naturally through the
self-regulated growth of SMBHs through accretion
triggered by the merger of gas rich galaxies. 
Tidal torques produced
by the merger channel large amounts of gas onto the central nucleus
fuelling a powerful starburst and rapid black hole growth, until
feedback from accretion is able to drive quasar winds and expel the
remaining gas from the remnant galaxy. Hydrodynamical simulations of gas rich
galaxy mergers incorporating star-formation, supernova feedback and
black hole growth \citep*{springel} confirm this picture, reproducing
the observed M$_{\rm BH}-\sigma$ relation
\citep*{dimatteo}. \citet{springel} show that the presence of the
central SMBH has a strong impact on the remnant galaxy, producing
passively-evolving spheroidal galaxies \citep{springel05b}
consistent with the observed scaling relations \citep{robertson} and
whose gas is heated by the quasar winds forming the observed X-ray
emitting halos \citep{cox}, whereas the remnant galaxies in models
without SMBHs continued to form stars at a non-negligible rate.

Given the tight correlation between the mass of the central SMBH
and that of the host galaxy, we should expect the effects of AGN
feedback to be strongly dependent on galaxy mass, being reduced or
even negligable for low-mass galaxies.
\citet{springel05b} find that for merging galaxies with
$\sigma=8$0\,km\,s$^{-1}$ the effects of black hole growth on the
remnant are negligable, the spheroids that form remaining gas rich
with ongoing star-formation. Indeed in low-mass galaxies SMBHs may not
form at all during mergers. The rapidity of the gas accretion onto the
central object depends on the depth of the potential well of the
host galaxy, and in low-mass galaxies the accreting gas may have time
to fragment and form stars, preventing further dissipation and
collapse of the gas \citep{haehnelt}. 
Indeed most (50--80\%) dwarf galaxies 
(M$_{B}\ga-18$) appear to host central compact stellar nuclei, regardless of
their morphological class \citep{cote,rossa}, the masses of which scale
with the mass of their host galaxy, following the same correlation as
that observed for SMBHs \citep{ferrarese,wehner}.

In recent years there have been significant advances in the
theoretical framework in understanding galaxy evolution, in particular
the ability of semi-analytic models to reproduce the global properties
observed in the current large scale surveys such as the SDSS. One
large problem that the theoreticians have been facing is reproducing
the observed break and exponential cut off in the luminosity function at the
bright end along with the observation that most massive galaxies are
passively-evolving, and have been for many Gyr. These massive galaxies have
halos of X-ray emitting hot gas, which without constant energy
injection should cool through radiative losses onto the galaxy,
fuelling further star-formation \citep{mathews}. However, no evidence
of this cooling gas is observed, and AGN feedback has often been put forward as a means of
supplying energy to the hot gas, and preventing it from cooling.
\citet{croton} have developed semi-analytic models which include AGN
feedback to prevent this gas cooling onto massive galaxies, which have been able to successfully reproduce the exponential
cut-off in the bright end of the luminosity function and the fact that
most massive galaxies are quiescent and dominated by old stars. It is
important to note that the feedback considered here is low-level AGN
activity from the quasi-continuous accretion from a static atmosphere
of hot gas that is in thermal equilibrium with the DM halo.  

\begin{figure}
\includegraphics[width=80mm]{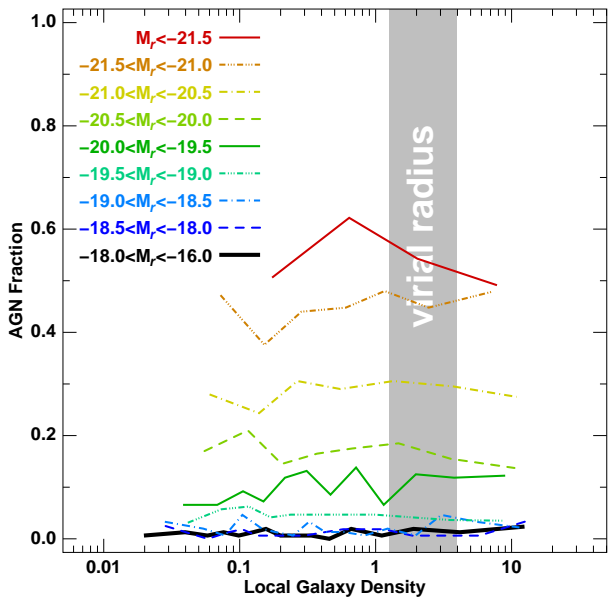}
\caption{The fraction of galaxies classed as AGN from their emission
  line ratios as a function of local density. Each coloured curve corresponds to a different luminosity bin as indicated. Each density bin contains 150 galaxies.}
\label{agn_rho}
\end{figure}

To examine the possible role of AGN feedback in terminating
star-formation in galaxies Fig.~\ref{agn_rho} shows how the AGN
fraction depends on both luminosity and environment. Each coloured
curve shows the fraction of galaxies classed as AGN from their
location in the [N{\sc ii}]$\lambda6584$\,/\,H$\alpha$ versus [O\,{\sc iii}]$\lambda5007$\,/\,H$\beta$
diagnostic diagram \citep{baldwin} as a function of local density for a particular
luminosity range as indicated. 
We note
that these diagnostics are not sensitive to type 1 AGN with broad-line
emission, however these are mostly limited to massive galaxy hosts and
at these redshifts they constitute only $\la\!1$\% of the \mbox{M$_{r}\!\la20$}
galaxy population \citep[e.g.][]{hao,sorrentino}.
For ease of comparison, the colours and
curves correspond to the same luminosity ranges as
Fig.~\ref{passive_rho}. 

The fraction of galaxies with an AGN remains constant with local
density from the cores of clusters to the rarefied field, for each of
the luminosity ranges covered. This is consistent with the result obtained by
\citet{miller} for bright \mbox{(M$_r\!<\!-20$)} galaxies.
Whereas the AGN fraction is independent of density, it
varies by more than an order of magnitude with the galaxy's
luminosity, from \mbox{$\sim\!5$0\%} for the brightest galaxies \mbox{(M$_r<-21$)} to
0--3\% by \mbox{M$_r\!\sim\!-18$} \citep[see also][]{k03c}. Hence, the processes that power an AGN must be
internal to the galaxy and dependent only on the mass/luminosity of
the galaxy, while the galaxy's local environment has little or no effect on its
ability to host or power an AGN. Most of the AGN detected here are
low-luminosity AGN powered by low-level, quasi-continuous accretion of
gas onto a SMBH. 
High-luminosity \mbox{($L$[O\,{\sc iii}$]\!>\!10^{7}L_{\odot}$)}
AGN and quasars are instead preferentially found in the outskirts of
clusters or in low-density regions, being relatively absent in cluster
cores \citep{kauffmann04,sochting,ruderman}.
 These results are consistent with the model for
high-luminosity AGN/quasars being triggered by the merging of two
gas-rich galaxies, as in the centres of clusters most galaxies have
already lost most of their gas.

\citet{martini} find that only \mbox{$\sim\!1$0\%} of their X-ray selected AGN in
clusters had obvious optical spectroscopic AGN signatures, although in
total they found only 5\% of \mbox{M$_{R}\!<\!-20$} cluster members had X-ray emission
characteristic of AGN, which is much smaller than the \mbox{$\sim\!3$0\%} of
\mbox{M$_{r}\!<\!-20$} galaxies observed here with the optical
signatures of AGN. These optically dull AGN are found to have higher
inclination angles than those showing optical emission, indicating
that extranuclear dust in the host galaxy hides the emission lines of
optically dull AGN \citep{rigby}.
It should be noted that AGN detection rates are subject to strong
aperture biases, because as the physical aperture subtended by the
SDSS fibre becomes larger, the nuclear emission from the AGN is
increasingly diluted by emission from the surrounding host
galaxy. This effect is particularly great for low-luminosity AGN and
LINERs where the fraction detected in massive galaxies from their SDSS
spectra drops from $\sim6$0\% to $\sim2$0\% with increasing redshift
\citep{k03a,kewley06}. For our sample, being at low redshifts the SDSS fibres cover only the
nuclear region ($\la\!1$kpc) of massive galaxies and so we should not
lose a significant fraction of AGN due to dilution of their emission.   
Another possible bias to the results presented here could be
signal-to-noise effects in the classification of AGN. In particular, a
$3\sigma$ detection of H$\alpha$ emission is required, yet the median
H$\alpha$ equivalent width of galaxies classed as AGN is just
1.56\,\AA. This could produce a dependency on the $r$-band
magnitude of the galaxy, as the median uncertainty in EW(H$\alpha$)
rises from 0.1\,\AA\, at $r=14$ to 0.5\,\AA\, by $r=17.77$. Thus a
significant fraction of faint galaxies \mbox{($r\!\ga\!17$)} may not be
classified as AGN simply due to their low signal-to-noise spectra.
However, although this could produce in theory a small bias against
detecting AGN in the dwarf galaxies,  
it should be negligible in comparison to the
observed trend with luminosity presented in Fig.~\ref{agn_rho}. 

The most striking result in Fig.~\ref{agn_rho} is the lack of AGN in
dwarf galaxies.  As shown in Fig.~\ref{passive_rho} thse galaxies are
star-forming and many have strong emission lines due to this
star-formation, which could in theory affect the ability to detect any
low-luminosity AGN from the BPT diagnostic method. However, as
\citet{k03c} show that the presence of even a low-luminosity AGN with
\mbox{$10^{5}\!<L$[O{\sc iii}$]<10^{6}\,L_{\odot}$} would perturb the emission-line
ratios enough to be detected in 93\% of dwarf galaxies classifed as
star-forming. 

Although the biases due to signal-to-noise effects, dust-obscuration,
aperture effects, emission from star-forming regions etc are likely to
affect the ability to detect AGN in galaxies, the observed trend with
luminosity is very strong, and appear broadly consistent with
expectations from the tight correlation between the mass of the
central SMBH and the host bulge component. At the
lowest luminosities, the predicted mass of the central black hole (if
there is one at all) is too low to be able to power an AGN detectable
in the optical spectrum. With increasing galaxy luminosity, the
typical mass of the central SMBH increases \citep{ferrarese05}, becoming
increasingly capable of powering an AGN detectable in the optical
spectrum, resulting in the AGN fraction increasing with luminosity. 

A comparison between Figs.~\ref{passive_rho} and~\ref{agn_rho} finds a
remarkable match between the fraction of passive galaxies in
low-density environments and the AGN fraction for all of the
luminosity bins covered. In these regions, environment-related
processes such as galaxy harassment or ram-pressure stripping are not
effective, and their star-formation histories are dependent only on
mechanisms {\em internal} to the galaxy, such as their merger history
and feedback processes. The strong correlation between the fraction of
AGN and passive galaxies in these regions suggests a direct
connection, with galaxies becoming passive through AGN feedback, and
appears consistent with the current model of the coevolution of AGN
and galaxies \citep[e.g.][]{springel,hopkins06a,hopkins06b,hopkins07b}.

\section{Comparison with semi-analytic models}
\label{semi}

To better understand the physical mechanisms that contribute to the
observed different environmental trends of galaxies with luminosity,
we compare our results with those produced by the semi-analytic models
of \citet[hereafter C06]{croton}. These are implemented on top of the Millennium Run
N-body simulation, currently the largest dark matter simulation of the
concordance $\Lambda$CDM cosmology with $\sim10^{10}$ particles in a
periodic box $500h^{-1}$\,Mpc on a side, giving a mass resolution of
$8.6\times10^{8}h^{-1}$\,Mpc \citep{springelsim}. Dark matter halos
and subhalos are identified as having 20 or more bound particles,
their merging trees constructed, which are subsequently used to populate
the halos and subhalos with galaxies according to the prescriptions
described in C06. For each galaxy there are four
components: stars, cold gas in the disk, hot gas in
the halo, and the central supermassive black hole. The two novel
features of the C06 model are: (i) the modeling of gas
infall from the halo onto the cold disk, which occurs either through
rapid cooling primarily in low-mass galaxies, or cooling from a static
halo of hot gas heated by accretion shocks, the dominant process in
massive galaxies; and (ii) its inclusion of the growth of
black holes, and their subsequent effects on the cold and halo gas in
the galaxy. These effects are two-fold, during galaxy mergers a
certain fraction of the cold gas is accreted by the black hole,
although any resultant energy released such as quasar winds are not
modeled, while instead low-level accretion of the hot gas in the halo
on the black hole is also described and results in energetic 'radio
mode' feedback which can prevent the further cooling of gas from the halo. 

As described in Appendix~\ref{tests} the resultant galaxy
properties are used to create a mock SDSS redshift catalogue, and the
local density for each galaxy estimated in the same manner. As in the
mock catalogues we do
not have information regarding the H$\alpha$ emission from each
galaxy, we instead define galaxies as being passive if they are both
red \mbox{($u-r\!>\!2.2$)} and have a current specific star-formation rate
(SFR/$\mathcal M$) less than 10\% that of the median value of
star-forming galaxies \mbox{($\sim 10^{-10}h^{-1}{\rm M}_{\odot}$yr$^{-1}$/\,M$_{\odot}$)}.  

\begin{figure}
\includegraphics[width=80mm]{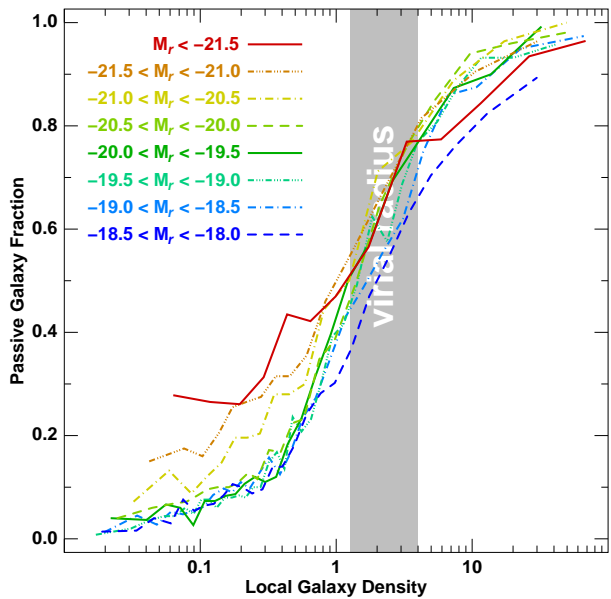}
\caption{The fraction of passively-evolving galaxies as a function of
  local density in the C06 semi-analytic model. Each coloured
  curve corresponds to a different luminosity bin as indicated and are
  analogous to those in Fig.~\ref{passive_rho}. Each
  density bin contains 150 galaxies.}
\label{croton}
\end{figure}

Figure~\ref{croton} shows the resultant fraction of passively-evolving
galaxies as a function of both local density and luminosity. Each
coloured curve corresponds to a different luminosity bin as indicated,
and are analogous to those for the SDSS dataset shown in
Fig.~\ref{passive_rho} allowing a direct comparison.
There are several important discrepancies between
 the model predictions and the SDSS data, which indicate areas where
 the models do not accurately describe the physical processes which
 define whether galaxies are still forming stars or not. 

The most notable difference is that the much smaller apparent luminosity
 dependence in the fractions of passively-evolving galaxies in
 low-density regions from the C06 models in comparison to those observed
 in the SDSS data. Whereas the passive galaxy fraction increases from
 \mbox{$\sim\!2$--8\%} to \mbox{$\sim\!3$0\%} from the lowest to the highest luminosity bins in the models, in the SDSS dataset the increase over the same
 luminosity range is from 0\% to \mbox{$\sim\!5$0\%}. As in these low-density
 regions environmental processes should not be important, the
 differences must be due to the models treatment of internal feedback
 processes that truncate and regulate star-formation in galaxies. The
 discrepancies appear greatest for the most massive galaxies where AGN
 feedback should be the most important mechanism for truncating
 star-formation in galaxies, and so this suggests that the
 prescription for AGN feedback in the C06 model is not efficient
 enough, the most likely cause of this being the neglect of feedback
 from quasar winds.

There is also a discrepancy at the
low-luminosity end, with fewer passively-evolving dwarf galaxies
observed than predicted by the models. While in the SDSS data the
passive galaxy fraction continues to fall with luminosity until it
reaches zero at \mbox{M$_{r}\!\sim\!-18$}, in the models there appears no
luminosity dependence for galaxies fainter than \mbox{M$_{r}\!\ga\!-20.5$}, in any
environment. The C06 model predicts that \mbox{$6.1\pm0.3$}\% of
\mbox{$-18\!<\!{\rm M}_{r}\!<\!-19$} galaxies in field regions
\mbox{($\rho\!<\!0.2$)} should be passively-evolving, a factor \mbox{$\sim\!3$}
greater than the \mbox{$1.9\pm0.3$}\% observed in the SDSS dataset.     
Three-quarters of these galaxies in the C06 model are satellites to
\mbox{M$_{r}\!<\!-20$} galaxies, indicating that this excess most
likely related to how the C06 model deals with the evolution of satellite galaxies.
In the C06 model, as in most semi-analytic models, when a galaxy becomes
a satellite within a more massive halo, it instantly loses the gas
from its halo to that of its host, ``suffocating the galaxy'', and
consequently uses up the remainder of its cold gas until it is no
longer able to continue forming stars \citep{larson}. This finding of too many
passively-evolving dwarf galaxies in low-density environments suggests
that the incorporation of ``suffocation'' into the model is too
efficient, terminating star-formation in galaxies too rapidly once
they become satellites. This is confirmed by \citet{weinmann06b} who
split galaxies in the SDSS and C06 catalogues into satellite and
central galaxies, and find that in the C06 model catalogues just
\mbox{$\sim\!2$0\%} \mbox{M$_{r}\!\sim\!-18$} satellite galaxies are blue, much
lower than the \mbox{$\sim\!6$0\%} observed in the SDSS data. 

\citet*{bekki01,bekki02} performed N-body and hydrodynamical
simulations following the effects of the cluster environment on the
gaseous halo of an infalling $L^{*}$ spiral galaxy, and found that for
a cluster of mass \mbox{$\sim\!10^{14}{\rm M}_{\odot}$} the
combination of the cluster tidal field and ram-pressure were able to
efficiently strip $\sim8$0\% of the halo gas over a period of 1--2\,Gyr. 
The stripping should be more rapid for lower mass galaxies. 
\citet*{balogh00} elaborated the model of \citet{larson}, indicating that
the gradual decline of star-formation on the $\sim3$\,Gyr timescales predicted by suffocation could reproduce the Butcher-Oemler effect and the observed
gradual star-formation density relations extending well beyond the
cluster virial radius \citep[e.g.][]{lewis,treu}. 
By simply assuming the Schmidt-Kennicutt law, \citet{balogh00} obtain
a relation for the decline in star-formation where no further gas
accretion is possible as, 
\begin{equation}
{\rm SFR}(t)={\rm SFR}(0) \left( 1+0.33 \frac{t}{t_{e}}
  \right)^{-3.5} {\rm M}_{\odot}\,{\rm yr}^{-1},
\label{gasconsume}
\end{equation}
where SFR(0) is the initial star-formation rate, and
$t_{e}\approx2.2\,[{\rm SFR}(0)/{\rm M}_{\odot}\,{\rm
    yr}^{-1}]^{\,-0.29}$\,Gyr is the characteristic gas consumption
time-scale, including the effects of gas recycling. 
For a typical $L^{*}$
spiral galaxy with SFR$(0)\sim5\,{\rm M}_{\odot}\,{\rm yr}^{-1}$ we obtain 
$t_{e}\sim1.4$\,Gyr, resulting in the galaxy taking $\sim4$\,Gyr to
become passive \mbox{(SFR(t)/SFR$(0)\sim0.1$)}, a time-scale
consistent with the cluster spiral population in $z\sim0.4$ clusters
becoming passive by $z=0$.

Applying Eq.~\ref{gasconsume} instead to a typical field dwarf galaxy with
\mbox{$M_{r}\!\sim\!-18$} and \mbox{SFR$(0)\sim0.2\,{\rm M}_{\odot}\,{\rm
 yr}^{-1}$} we obtain \mbox{$t_{e}\!\sim\!3.5$\,Gyr}, resulting in
the galaxy taking \mbox{$\sim\!1$0\,Gyr} to 
become passive, consistent with their observed gas depletion
time-scales of \mbox{$\sim\!20$\,Gyr} \citep{vanzee}. 
Hence, even if deprived of further gas
accretion through suffocation, star-formation in dwarf galaxies
occurs at a sufficiently low rate that they are unlikely to have
consumed all their gas and become passive by the present day if they
are acted on only by the mild stripping envisaged in suffocation.   

If suffocation alone is unable to terminate star-formation in the
dwarf satellites of massive galaxies, then a stronger form of gas
stripping is required. \citet{mayer,mayer06} show that dwarf spiral
galaxies orbiting a Milky Way type galaxy can suffer significant mass
loss and have their entire gas content removed or used up in a
star-burst, transforming them into passive dwarf ellipticals over a
period of $\sim5$\,Gyr through the combinination of tidal shocks and
ram-pressure stripping if their orbits take them within $\sim 5$0\,kpc
of the primary. Such behaviour is difficult to model within a
cosmological simulation such as that of C06, in particular as the
tidal forces acting on the dwarf satellites change rapidly along their
orbits, such that the star-formation histories and evolutions (in terms
of mass loss) of dwarf satellites could be
very strongly affected by even small variations in their orbits \citep{kravtsov,sales}.

Looking at the possible effect of the Millenium simulation neglecting
tidal stripping on the satellite population, we find an overabundance of
faint satellite galaxies around massive field galaxies in the models
in comparison to the SDSS data. In particular for each \mbox{M$_{r}\!<\!-20$}
field galaxy we observe on average 
\mbox{$0.125\pm0.007$} \mbox{$-19\!<\!{\rm M}_{r}\!<\!-18$} galaxies that lie within
250\,kpc and 500\,km\,s$^{-1}$ of it (i.e. that are probable
satellites), whereas in the C06 model we find on average
\mbox{$0.195\pm0.004$}. A certain fraction of
these galaxies will in fact be interlopers, and should be subtracted
from this analysis. We estimate the number of interlopers as the
number of dwarf \mbox{($-19\!<\!{\rm M}_{r}\!<\!-18$)} galaxies found around a
dwarf galaxy which has no M$_{r}<-20$ galaxy within 1\,Mpc and
1\,000\,km\,s$^{-1}$, so that neither of the dwarf galaxies are
satellites, obtaining a value of $0.083\pm0.003$ per galaxy. This
takes into account the natural clustering of dwarf galaxies,
independent of the presence of nearby giant galaxies.
 An
alternate estimate for the number of interlopers can be made by
assuming interlopers are distributed evenly over the volume covered,
and so in this case we would expect 0.019 dwarf galaxies within the
volume around each \mbox{M$_{r}\!<\!-20$} galaxy. Considering the contamination
from interlopers to be between these two values, we estimate the
excess dwarf satellites in the Millennium simulation to be 65--167\%.

\section{Discussion}
\label{discussion}

Using a volume-limited sample of galaxies from the SDSS DR4
spectroscopic dataset we have examined the star-formation activity of galaxies
as a function of both luminosity and environment, the main results of
which are summarized in Figure~\ref{passive_rho}. In high-density
regions the bulk of galaxies are passively-evolving independent of mass. 
Many processes could contribute to terminating the
star-formation in these galaxies, either internal to the galaxy or
related to the hostile cluster environment, making it
difficult to identify those which are most important. To gain insights
into the relative importance of these different mechanisms we instead
focus on galaxies in low-density regions for which we can be sure that their
star-formation histories have not been influenced by cluster-related
processes (thanks to our robust separation of cluster and field
populations). Hence the star-formation histories of these galaxies can
only be influenced by internal mechanisms such as merging, gas
consumption through star-formation and AGN
feedback.

We find that the fraction of passively-evolving galaxies in
field regions drops steadily from $\sim5$0\% at \mbox{M$_{r}\!\la\!-21$} to
zero by \mbox{M$_{r}\!\sim\!-18$}. This implies that internal
mechanisms are not responsible for the formation of passively-evolving
dwarf galaxies, while they become increasingly important with galaxy mass for
driving their star-formation histories. This would be consistent with
the increasingly early and rapid conversion of gas into stars for more
massive galaxies resulting from the Kennicutt-Schmidt law
\citep[$\Sigma_{SFR}\propto\Sigma_{gas}^{1.4\pm0.1}$;][]{kennicutt,schmidt},
which has been shown to hold over several orders of magnitude of gas
density, in conjunction with the appearance of a critical gas density
below which star-formation does not occur
\citep{martin}. Hydrodynamical simulations of the formation of massive
$\sim\!M^{*}$ field galaxies show that passively-evolving elliptical
galaxies can be produced without recourse to major mergers or feedback
mechanisms consuming their gas in a short burst \mbox{($\la\!2$\,Gyr)}
of star-formation at \mbox{$z\!\ga\!2$} \citep{chiosi,naab}. In dwarf
galaxies instead star-formation occurs very inefficiently resulting in
gas consumption time-scales much longer than the age of the Universe
\citep{vanzee} and global star-formation rates that have remained
approximately constant since \mbox{$z\!\sim\!3$} \citep{panter}. These
simulations and the observed studies of the global decline of
star-formation since \mbox{$z\!\sim\!1$} through downsizing \citep{heavens,noeske07a,noeske07b} suggest that gas exhaustion
through star-formation may play a dominant role in the {\em global}
evolution of star-formation.

This does not appear to be the complete answer, in particular it does
not explain the wide variety of star-formation histories seen in field
\mbox{$\sim\!M^{*}$} galaxies \citep{paper1}, where completely
interspersed mixtures of young and old galaxies (in terms of their
stellar populations) are found. This instead suggests that some stochastic
processes play a role in determining the
star-formation histories of massive galaxies, the most natural being their hierarchical assembly
through mergers and the resultant growth of SMBHs and subsequent AGN
feedback effects \citep{hopkins06a,hopkins06b,hopkins07b}.
Moreover we find a close
mirroring of the increase in AGN fraction with galaxy luminosity to
that observed for the passive galaxies in field regions, suggesting a
direct connection between nuclear activity and galaxies becoming
passive, reflecting the increasing importance of AGN feedback with
galaxy mass for their evolution.

\subsection{Massive galaxies affected by merging}
\label{giants}

In the case of massive galaxies
(\mbox{$\mathcal{M}\!\ga\!10^{10.3}{\rm M}_{\odot}$};
\mbox{M$_{r}\!\la\!-20$}) we confirm the recent results on the
environmental dependence of star-formation \citep{lewis,gomez,tanaka,rines05}, finding gradual trends for
the fraction of passively-evolving bright galaxies with environment
that extend to the lowest densities studied
($\S$~\ref{bimodality}). Furthermore we find that this correlation
with the local density is independent of the presence or absence of
{\em individual} neighbouring galaxies ($\S$\ref{neighbours}). 
\citet{rines05} find that
the fraction of passively-evolving galaxies depends on the local
density, and this dependency is the same for galaxies in different
types of global environment (within the cluster virial radius, in
infall regions, and the field), while \citet{tanaka} find no
dependence on system richness for galaxies residing in groups with
\mbox{$\sigma\!>\!200$\,km\,s$^{-1}$}.
In Section~\ref{sfr} we also find that the
distribution of EW(H$\alpha$) of massive galaxies with {\em current}
star-formation shows no dependence on local density in agreement
with the results obtained by \citet{balogh04}, \citet{tanaka} and
\citet{rines05}. These results together indicate a limited role for
mechanisms specific to cluster environments on the termination of
star-formation in massive galaxies, in particular excluding mechanisms which
result in a gradual decline in the SFR of infalling
galaxies.
Instead star-formation in massive galaxies is terminated
rapidly (on time-scales \mbox{$\la$1\,Gyr}) through processes that
depend only marginally on the local density 
and/or occur at high-redshifts such that ongoing
transformations are rare.   

This early cessation of star-formation in massive passively-evolving
galaxies and the independence of the star-formation histories on the
global cluster environment, point towards the evolution of massive
galaxies being driven primarily by {\em internal} mechanisms and their
merger history \citep{hopkins07b}. The environmental trends are then defined by the
initial conditions in which galaxies form, whereby massive galaxies are
formed earlier preferentially in the highest overdensities in the
primordial density field \citep{maulbetsch}, and have a more active merger history \citep{gottlober}, than
those that form in the smoother low-density regions. This implies the
environmental trends should be imprinted early on in the massive galaxy
population, as is observed with the morphology--density
and colour--density relations being already largely in place
by \mbox{$z\!\sim\!1$} \citep{smithg,cooper}, and possibly apparent even at \mbox{$z\!>\!2$} \citep{quadri}.

This rapid formation of massive galaxies and the early shutdown of
star-formation is produced naturally in the monolithic collapse model,
where the deep potential wells allow for rapid and efficient
conversion of the gas into stars according to the Kennicutt-Schmidt
law \citep{chiosi} in a short burst \mbox{($\la\!2$\,Gyr)} of star-formation
at \mbox{$z\!\ga\!2$}. 

While the monolithic collapse model appears at odds
with the hierarchical growth of structures produced in $\Lambda$CDM
models, \citet{merlin} are able to reproduce the same qualitative
star-formation histories for massive galaxies within a hierarchical
cosmological context, most of the merging of substructures occuring
early in the galaxy life ($z>2$).
This so-called ``revised monolithic'' scheme is consistent with the
observed rapid evolution and growth of massive galaxies through
merging at \mbox{$z>1$} \citep{conselice}, and the mild evolution in the number
density of massive \mbox{($\mathcal{M}\!\ga\!10^{10.5}{\rm M}_{\odot}$)} red galaxies
to \mbox{$z\!\sim\!1$} \citep{bell,willmer} and beyond
\citep{glazebrook,cirasulo,renzini}. 
Equally the observed
bimodality in the colour distribution of galaxies at \mbox{$z\!\sim\!1.2$}
\citep{bell,willmer} implies that star-formation must be truncated rapidly and
completely in massive galaxies at \mbox{$z\ga\!1.5$} or even earlier
\citep{kriek}. Given the short gas-consumption time-scales of massive
galaxies, to maintain star-formation to the
present day would require massive galaxies to continuously accrete of
fresh gas from their surroundings \citep{larson}. However, massive galaxies in halos
of mass \mbox{$\ga\!6\times10^{11}{\rm M}_{\odot}$} stable
virial shocks form which heat infalling gas to the virial temperature,
signficantly reducing the accretion rate onto the galaxy, and making
the gas vulnerable to feedback effects such as AGN
\citep{dekel}. Indeed in many massive galaxies further accretion of
gas may be completely shutdown, resulting in the subsequent
termination of star-formation without the need for further quenching
mechanisms such as mergers or AGN feedback \citep*{birnboim}. 

In Section~\ref{agn} we determined the fraction of galaxies with
optical AGN signatures as a function of both luminosity and
environment. We find the AGN fraction to be independent of local
density for galaxies of a given luminosity for
\mbox{M$_{r}\!<\!-18$}. In contrast the AGN fraction is strongly
luminosity dependent, increasing from \mbox{$\sim\!0$\%} at
\mbox{M$_{r}\!\sim\!-18$} to \mbox{$\sim\!5$0\%} by
\mbox{M$_{r}\!\sim\!-21$} in a way that closely mirrors the
luminosity-dependence of the passive galaxy fraction in low-density
environments. These results suggest that the ability of a galaxy to
host an AGN depends only on its mass, as would be expected from the
tight \mbox{M$_{\rm BH}-\sigma$} relation \citep{ferrarese05}, while the strong
correlation between the fractions of galaxies hosting AGN and being
passive suggests a direct connection between nuclear activity and a
galaxy becoming passive. This appears consistent with the current
models of the coevolution of AGN and galaxies
\citep[e.g.][]{springel,hopkins06a,hopkins06b,hopkins07b}, and reflects the
increasing importance of AGN feedback with galaxy mass for their
evolution, and its increasing efficiency in permanently shutting down
star-formation, by expelling gas from the galaxy through quasar winds \citep{dimatteo}
and/or preventing further cooling of gas from the surrounding halo \citep{croton}. 

It seems impossible for environmental processes to be
able to shut off star-formation in massive galaxies at such an early
epoch as required: ram-pressure stripping for example is unlikely to
be efficient until later epochs \mbox{($z\!\la\!1$)} once the dense cluster ICM has had time to build
up \citep{kapferer}, while both galaxy harassment and suffocation take several Gyr to
terminate star-formation in galaxies, and simply haven't had time to
act. Moreover, even in field regions where these processes are not
efficient, massive, passively-evolving galaxies are seen to be in
place at $z\!>\!1$, with only slight
differences are observed in the mean stellar ages and mass-to-light
ratios of galaxies in diverse environments \citep{thomas,smith,vandokkum}.

Although the formation of passively-evolving early-type galaxies
through merging largely took place at \mbox{$z\ga\!1$}, there is
evidence indicating this process is continuing at lower redshifts.
Early-type galaxies that are currently passively-evolving, but with
remnant young (1--4\,Gyr) stellar populations or E+A spectra indicating
\mbox{$<\!1$\,Gyr} old stars, are preferentially located in low-density
environments or poor groups, where low-velocity
interactions should be most frequent \citep*{goto,nolan}. 
The fraction of blue, spiral galaxies in
clusters is also observed to drop rapidly from \mbox{$z\!\sim\!0.5$} to \mbox{$z\!\sim\!0$} ---
the Butcher-Oemler effect --- while many spiral galaxies in local
clusters show strong evidence of being transformed now by ram-pressure
stripping \citep[e.g.][]{solanes,koopmann,vangorkom}. 

As discussed in $\S$~\ref{semi} the mechanism of suffocation was
proposed by \citet{larson} to explain the Butcher-Oemler effect, by
stripping the extended gas
reservoirs of recently accreted spiral galaxies in clusters at
\mbox{$z\!\sim\!0.4$}, slowly transforming them by the present day into
passively-evolving galaxies by
exhausting their remaining gas supplies through star-formation.  
However, suffocation should not change the morphology or radial
profiles of the spiral galaxies, turning them into anemic spirals
\citep{vandenbergh76} rather than the S0s whose numbers in clusters
have increased rapidly since \mbox{$z\!\sim\!0.4$} mirroring the decline of
cluster spirals \citep{dressler97,desai}.  

Instead, studies of spiral galaxies in the Virgo cluster find that a
much larger fraction of them have truncated H$\alpha$ and H\,{\sc i}
radial profiles (52\%) than appear simply anemic (6--13\%) with
globally reduced star-formation and gas densities
\citep{koopmann}. Similarly, \citet{vogt} find a significant fraction
of spirals with asymmetric  H$\alpha$ profiles and H\,{\sc i} rotation
curves in nearby rich clusters.
It is difficult for suffocation to produce these truncated or
asymmetric H$\alpha$
and H\,{\sc i} profiles, and instead it appears more likely that a
process such as ram-pressure stripping actively removes the gas from
the outside-in, and \citet{boselli06} show that ram-pressure stripping
is able to reproduce the observed radial profiles. This outside-in
removal of the gas should also produce inverted colour gradients in
which the inner regions of the galaxy are bluer than the outer
regions, as star-formation is truncated earlier in the periphery than
the core of the galaxy, an effect which has been observed for the
Virgo spiral NGC\,4569 \citep{boselli06}.

These truncated spiral galaxies are mostly confined to the inner
1--1.5\,Mpc of the Virgo cluster, that is within the virial radius,
although there are some well outside the virial radius
\citep{koopmann}. All of the asymmetric spiral galaxies are located
within 1\,$h^{-1}$Mpc of the cluster centres with the truncation of
the H\,{\sc i} and H$\alpha$ emission preferentially along the leading
edge \citep{vogt}. 
These results are more consistent with the effects of ram-pressure
stripping which should only affect massive galaxies within the cluster cores
\citep[$\la\!0.3\,R_{vir}$;][]{roediger}, although as the H\,{\sc i} deficient spiral
galaxies are observed to be on
highly-elliptical radial
orbits \citep{solanes}, those truncated or asymmetric galaxies at $0.3\!\la\!(r/{\rm
  R}_{vir})\!\la\!1.5$ could have recently passed through the cluster core
\citep{mamon}. 

The few anemic spirals with globally reduced H$\alpha$
and H\,{\sc i} emission are generally found further from the cluster
centres than truncated galaxies, half of them are outside the virial
radius \citep{koopmann}, while \citet{goto03} find that passive
spirals in the SDSS are generally found in the outskirts of clusters at
1--10$\,{\rm R}_{\rm vir}$. It could then be that these galaxies are the
natural results of suffocation, and may represent an early phase of
evolution from spirals to passively-evolving S0s.
 The finding of significant numbers of
truncated spiral galaxies in clusters, implies that the complete
removal of gas and transformation into passively-evolving galaxies
does not occur rapidly, at least in Virgo-like clusters. This appears
inconsistent with our finding of no environmental dependence in the
H$\alpha$ emission of bright star-forming galaxies, as that implies
the transformation is either rapid or occurs at high-redshift. One
plausible explanation could be due to our measuring the H$\alpha$
emission from a confined region in the galaxy centre rather than the
entire disk, and that star-formation within the truncation radius is
relatively unaffected, until the galaxy orbit brings it sufficiently
close to the cluster centre that the remaining gas is completely
stripped, rapidly terminating star-formation across the galaxy.

\subsection{Dwarf galaxies affected by their environment}
\label{dwarfs}

In Section~\ref{bimodality} we showed that the make-up of the dwarf galaxy
population varies strongly with environment. Whereas in galaxy groups
and clusters the bulk of dwarf galaxies are passively-evolving, as the
local density decreases the fraction of passively-evolving dwarfs
drops rapidly, reaching zero in the rarefied field. Indeed for the
lowest luminosity range covered \mbox{($-18<{\rm M}_{r}<-16$)}
none of the $\sim600$ galaxies in the lowest density quartile are
passively-evolving. In Section~\ref{neighbours} we examined in detail
the immediate environment of those few passively-evolving dwarf
galaxies in field regions, finding them very strongly clustered around
bright \mbox{(M$_{r}\!\la-20.5$)} galaxies. Almost without exception those
passively-evolving galaxies outside groups and clusters appear to be
satellites bound to massive galaxies. 

This association of passively-evolving dwarf galaxies as satellites
within more massive halos is consistent with the analysis of
\citet{zehavi} of the dependence of the galaxy two-point correlation
function on luminosity and colour. They find that whereas the
overall amplitude of clustering decreases monotonically with magnitude
over \mbox{$-23\la{\rm M}_{r}\!\la-18$}, the clustering of {\em red}
galaxies on $\la\!1$\,Mpc scales is strongest for \mbox{M$_{r}\!>-19$}. They
are able to describe these results using halo occupation distribution
(HOD) models in which the faint red galaxies are nearly all satellites
in high-mass halos.

These results confirm and significantly extend the long noted morphological
segregation of dwarf galaxies in the local \mbox{($\la\!3$0\,Mpc)}
neighbourhood, with dwarf ellipticals (dEs) confined to groups, clusters and
satellites to massive galaxies, while dwarf irregulars (dIrrs) tends to follow
the overall large-scale structure without being bound to any of the 
massive galaxies \citep{einasto,ferguson}. Here dwarf
ellipticals (dEs) are generally defined as galaxies with \mbox{$-18\la{\rm
  M}_{B}\la-14$} having smooth, symmetrical surface-brightness profiles
implying no spiral structures or star-forming regions, and typically
have very low H\,{\sc i} mass fractions, and hence we identify these
galaxies with our passively-evolving dwarf galaxies, although there
are galaxies \citep[e.g. NGC\,205 which are classed as dEs but have recent
star-formation in their central regions;][]{mateo}. 

The morphological segregation of dwarf galaxies was first noted by
\citet{einasto} who compared the spatial distributions of dwarf
companions to our Galaxy and the nearby massive spirals M\,31,
M\,81 and M\,101. They found a striking separation of the regions
populated by dE and dIrr galaxies with a well defined line of
segregation which had a strong luminosity dependence, with more
luminous dEs constrained to smaller regions around the primary
galaxy. They argued that tidal effects would be insufficient to produce
this segregation and that a dense corona of halo gas around massive
galaxies is necessary to strip the gas from the satellites as they
move through the corona \citep{gunn}.
In the Local Group
the only dwarf ellipticals are M\,32, NGC\,147, NGC\,185 and NGC\,205,
all of which are satellites of M\,31, while
three of the five dwarf irregulars of comparable brightness
(NGC\,6822, IC\,6822 and WLM) are free-floating within the Local Group
potential \citep[e.g.][]{mateo,vanderbergh}. 
 Beyond the Local Group, there are no known
 isolated dE galaxies within 8\,Mpc \citep{karachentsev}.

In a survey covering
900\,deg$^{2}$ \citet*{binggeli90} identify 179 dwarf galaxies, and
claim that ``{\em in the field there are virtually no isolated dEs, and that
  the few dEs outside of gravitationally bound systems are close
  satellites to massive giants.}'' They find just one good candidate
for an isolated dE, \#179 in their dwarf catalogue. The
reported location of this galaxy is covered by the SDSS, but no galaxy
is found there, and it appears most likely to correspond to a $z=0.08$
Sa galaxy 1 arcmin distant.

Dwarf ellipticals (including here dSphs) in contrast are
the most numerous galaxy type in clusters, although unlike the field
or within groups only a small fraction appear bound to massive galaxies
\citep{ferguson92}, the rest follow the general cluster potential. However, the ratio of dEs to giants is much
greater in clusters than in groups or the field, and so not all cluster dEs can be accounted for by the accretion of ``field'' dEs \citep*{conselice1}. 

This clear segregation of passively-evolving dwarf galaxies places
strong constraints on their formation and evolution, in particular as
to the physical mechanisms that could cause them to cease forming
stars.  
Most importantly,
unlike massive galaxies for which their build-up through mergers appears
fundamental in determining their star-formation history, we can rule
out merging as a formation mechanism for passively-evolving dwarfs.  

Galaxy merging is a stochastic process which occurs 
independently of the large-scale environment \mbox{($\ga\!1$\,Mpc)} of a
galaxy \citep{hopkins07a}. 
This means that mergers take place in all environments\footnote{except
in relaxed, rich clusters where encounter velocities are much higher
than the internal velocity dispersions of galaxies, preventing their
coalescence \citep{aarseth}}, even in voids, as has been observed for interacting
pairs \citep{alonso}. This results in merger remnants being
ubiquitous, as predicted by \citet{hopkins07a}, and observed in the
spatial distribution of post-starburst galaxies \citep{goto,hogg}. 
This is most clearly demonstrated with the presence in all
environments of passively-evolving, massive galaxies with old stellar
populations, which in field regions make up \mbox{$\ga\!5$0\%} of the total
population of massive galaxies, forming an equal interspersed mixture
with younger star-forming galaxies (Figs. 3 and 4 from Paper I). 

Hence
dwarf galaxies which have undergone major mergers should also be
ubiquitous, as \citet{conselice} shows that they have undergone on
average about the
same number of major mergers as massive galaxies since \mbox{$z\!\sim\!3$}
(albeit at later epochs), yet we find no passively-evolving galaxies
among the $\sim\!6$00\, \mbox{$-18\!<\!{\rm M}_{r}\!<\!-16$} dwarfs in the lowest density quartile. In addition, mergers cannot explain the
observed strong segregation of dwarf galaxies, in particular the
presence of a massive galaxy should not affect the
merging efficiency of a dwarf galaxy, or the observation that most dwarf
star-forming galaxies in clusters are in the process of being
transformed into passively-evolving galaxies, an environment where
mergers are {\em now} strongly inhibited.  

The ineffectiveness of mergers to permanently shut down star-formation in
dwarf galaxies can be understood in the context of the current
theoretical framework of galaxy evolution. \citet{springel} show
through hydrodynamical simulations of mergers that although both
models with and without black holes produce strong starbursts during
the mergers followed by a decline in star-formation, in those models
without black holes the remnants continue to form stars at a
non-negligible rate. Given that the growth of central black holes
during galaxy mergers are strongly regulated by the mass of the host
galaxy, in low-mass galaxies the resultant black holes (if indeed there
is one at all), based on the M$_{\rm BH}-\sigma$ relation
\citep{ferrarese05}, are too small to power the quasar winds which would
expel the remaining gas and shut down star-formation, as occurs in
more massive systems. Indeed for merging galaxies with
\mbox{$\sigma=8$0\,km\,s$^{-1}$} \citet{springel05b} show that the effects of black
hole growth are minimal on star-formation in the remnant galaxy, and
the galaxy does not become passive as a result of the
merger. Moreover, most dwarf galaxies \mbox{(M$_{B}\!\ga\!-18$)} appear to host
central compact stellar nuclei rather than a central supermassive
black hole \citep{cote}, consistent with our observation that the AGN
fraction of galaxies falls to close to zero by
\mbox{M$_{r}\!\sim\!-18$} ($\S$\ref{agn}). 
Even if a significant fraction of the
available gas in the remnant is used up during the star-burst, the
intrinsic low star-formation efficiency of low-mass galaxies and
regulatory effects of supernovae feedback, plus the continuous infall
and cooling of fresh gas from the halo along filaments \citep{keres},
ensures that the remaining gas is unlikely to be exhausted. Finally,
the quasi-continuous low-level AGN activity that \citet{croton}
suggest could inhibit cooling of gas from the diffuse hot gaseous halo
of massive galaxies and prevent subsequent star-formation in them, has in
contrast little effect against the clumpy nature of the gas infalling
along filamentary structures onto the dwarf merger remnants. 
 
In Section~\ref{sfr} we find a significant anti-correlation
(10$\sigma$) between the EW(H$\alpha$) of dwarf star-forming galaxies
and their local density, producing a systematic reduction of
\mbox{$\sim\!3$0\%} in the H$\alpha$ emission in high-density environments with
respect to field values. We argue that this implies that the bulk of
star-forming dwarf galaxies in groups and clusters are in the process
of being slowly transformed into passive galaxies over long
time-scales \mbox{($\ga\!1$\,Gyr)}, and is thus suggestive of
suffocation \citep{balogh00}.   

However, as discussed in Section~\ref{semi} the long gas consumption
time-scales predicted from Eq.~\ref{gasconsume} and observed by
\citet{vanzee} imply that even if deprived of further gas
accretion through suffocation, star-formation in dwarf galaxies
occurs at a sufficiently low rate that they are unlikely to have
consumed all their gas by the present day. 
In low-mass dwarf galaxies
 \mbox{($\mathcal{M}\!\la\!10^{7}{\rm M}_{\odot}$)}
as gas collapses and stars form, the resultant feedback from
supernovae is able to drive out the remaining gas, temporarily
shutting off star-formation, until the gas is able to cool and restart
star-formation, resulting in episodic bursts of star-formation every \mbox{$\sim\!300$\,Myr}
\citep{stinson}. In more massive dwarf galaxies, such as those studied
here,  SN feedback appears too inefficient to power galactic winds
 that could expel the remaining gas from dwarf galaxies, even during
 star-bursts \citep{maclow,marcolini06}. 
Instead it seems that star-formation in dwarf
galaxies appears strongly regulated and rather resilient over long
time-scales. 
This quasi-continuous star-formation in dwarf galaxies seems to have
been maintained since early epochs, as the global star-formation rates
of galaxies with \mbox{$\mathcal{M}\!<\!3\times10^{10}{\rm M}_{\odot}$} have
remained approximately constant since \mbox{$z\!\sim\!3$}
\citep{panter}.

A further difficulty for dEs being produced as the result of
suffocation is that
it does not affect the galaxy structurally, and so we should expect
the surface brightness and luminosity to decrease in parallel as the
stellar population evolves passively. However, the surface stellar
mass densities of passively-evolving dwarf galaxies are 0.5\,dex {\em
  higher} than their star-forming counterparts of the same stellar
mass \citep{k03b}. Hence, if star-formation in dIrr galaxies were to simply be
stopped, as in suffocation, the resultant surface densities would be
too low in comparison to present-day dEs \citep{grebel}. Equally, the
metallicity of dIrrs are too low for their luminosity as compared with dEs, for
them to be simply transformed by becoming passive \citep{grebel},
while suffocation provides no means for the rotationally-supported
dIrrs to lose their angular momentum to become the dEs where little or
no signs of rotation ($\la5$\,km\,s$^{-1}$) are seen. 

If gas cannot be exhausted through
star-formation, expelled by supernovae feedback during star-bursts, or
prevented from infalling and cooling through AGN feedback,  
 the end of star-formation and
gas removal in dwarf galaxies must come from external mechanisms,
such as ram-pressure or tidal interactions \citep{marcolini06}.

When galaxies pass through the dense ICM of
clusters or the gaseous halos of massive galaxies they feel an
effective ram-pressure which, if able to overcome the gravitational
attraction between the gas and the disk, 
 is able to effectively strip the gas from
the disk \citep[for a review see][]{vangorkom},
according to the \citet{gunn}
condition
\mbox{$\rho_{\rm ICM}\nu_{gal}^{2}>2\pi G \,\Sigma_{\star}
  \Sigma_{gas}$}, which has been shown to hold approximately from
hydrodynamical simulations \citep[e.g.][]{marcolini03}. Hence 
ram-pressure stripping should be more effective for lower mass,
low-surface-brightness galaxies, galaxies on more eccentric orbits,
and galaxies in richer environments \citep{hester}, such that while
for massive galaxies 
ram-pressure stripping is only effective in the cores of rich clusters, dwarf galaxies can be completely stripped of their gas
even in poor groups \citep{marcolini03}.

\citet{moore} proposed that cluster spirals could be disrupted by
``galaxy harassment'', whereby repeated close \mbox{($<\!5$0\,kpc)}
high-velocity \mbox{($>\!1$\,000\,km\,s$^{-1}$)} encounters with massive
galaxies and the cluster's tidal field cause impulsive gravitational
shocks that damage the fragile disks of late-type disks, transforming
them over a period of Gyr into spheroids. 
While \mbox{$\sim\! L^{*}$} spirals are 
relatively stable to the effects of harassment, suffering little or no
mass loss, low surface brightness dwarf spirals with
their shallower potentials may suffer significant mass losses (up to
90\%) of both their stellar and dark matter components during
harassment \citep{moore99}, resulting in remnants resembling dwarf
ellipticals \citep{mastropietro}.

Dwarf spiral galaxies orbiting as satellites to massive
galaxies may also be 
transformed into passively-evolving dEs through tidal
interactions with the primary galaxy and ram-pressure stripping as
they pass through its gaseous halo. 
\citet{mayer} show that dwarf spirals orbiting a Milky Way type galaxy
on eccentric orbits taking them within 50\,kpc of the primary 
experience tidal shocks during their pericentre passages, that can cause
significant mass loss (mostly of the outer gaseous halo and dark
matter, but also of the stellar disk), formation of bar instabilities
that channel gas inflows triggering nuclear star-bursts, and loss of angular
momentum, resulting in their transformation over a period of
$\sim5$\,Gyr an early-type dwarf. \citet{mayer06} indicate that while
tidal stirring of disky dwarf galaxies can transform them into
remnants that resemble dEs after a few orbits, ram-pressure stripping
is required to entirely remove their gas component. \citet{kravtsov}
show that many dwarf satellites of Milky Way type galaxies have
undergone significant mass loss through tidal stripping, and are able
to reproduce the observed morphological segregation of dE and dIrr
galaxies \citep{einasto}, in which the dEs are those that have suffered
significant tidal stripping.
 
As discussed above star-forming, late-type dwarf
galaxies in clusters, groups or bound to massive galaxies  
can be transformed into dEs through a combination of
ram-pressure stripping and tidal interactions. 
This is supported by the finding of significant populations of dEs in
the Virgo cluster having blue central regions caused by recent or ongoing
star-formation \citep{lisker}, significant amounts
\mbox{($\ga\!10^{7}{\rm M}_{\odot}$)} of H\,{\sc i} gas \citep{conselice4},   
or clear disk features
including spiral arms, bars or significant velocity gradients, with
rotational velocities similar to dwarf irregulars of the same
luminosity, and anisotropy parameters
($\nu/\sigma_{m}\ga1$) indicating stellar kinematics dominated by
rotation rather than random motions \citep{vanzee04}. These
populations are found predominately in the cluster outskirts and some
have flat distributions of radial velocities, suggesting that they
have been recently accreted by the cluster or are
on high angular momentum orbits and therefore never go through the
cluster core, while normal,
non-rotating dEs are concetrated in the cluster core or in
high-density clumps \citep{conselice4,vanzee04,lisker07}.
In a similar H\,{\sc i} study, this time of 11 dIrrs in the Virgo
cluster, \citet{lee} find that five of them are gas deficient by a
factor \mbox{$\ga\!1$0} with respect to field dIrrs at comparable oxygen
abundances, and this gas deficiency correlates approximately with the
X-ray surface brightness of the ICM. These gas-poor dIrrs have typical
colours and luminosities of normal field dIrrs, indicating that their
star-formation hasn't yet been affected, and suggesting that they have
only recently encountered the ICM for the first time. 
In the Coma cluster \citet{caldwell} and \citet{poggianti} find that whereas bright
post-starburst galaxies with k+a spectra are largely absent, there is a
significant population of faint galaxies \mbox{(M$_{V}\!>\!-18.5$)} with k+a
spectra, which appear associated with substructures in the ICM, the
galaxies lying close to the edges of two infalling structures. This
strongly suggests that the interaction with the ICM could be
responsible for rapidly quenching the star-formation in these
galaxies, possibly after a starburst. 
These observations all point towards the {\em ongoing} transformation of
late-type dwarf galaxies into passively-evolving dwarf ellipticals.

\section{Summary and Conclusions}
\label{summary}
We present an analysis of star-formation in galaxies as a function of both
luminosity and environment, in order to constrain the physical
mechanisms that drive the star-formation histories of galaxies of
different masses. In particular we wish to distinguish between
mechanisms that are internal to the galaxy such as AGN feedback, gas
consumption through star-formation and merging, and those related to
the direct interaction of the galaxy with its surroundings including
ram-pressure stripping and galaxy harassment. 

For this analysis we use the NYU-VAGC low-redshift galaxy
catalogues \citep{nyuvagc} taken from the SDSS DR4 spectroscopic
dataset. Using a sample of 27\,753 galaxies in the redshift range
\mbox{$0.005<z<0.037$} that is \mbox{$\ga\!9$0\%} complete to
\mbox{M$_{r}=-18.0$} we quantify the environment of each galaxy
using an adaptive kernel method that for galaxies in groups or
clusters resolves the local density on the scale of their host halo,
while in field regions smoothes over its 5-10 nearest neighbours.  
We use H$\alpha$-emission as a gauge of the current star-formation in
the galaxies and find that the 
EW(H$\alpha$) distribution of galaxies is strongly bimodal, allowing
them to be robustly separated into passively-evolving and
star-forming populations about a value \mbox{EW(H$\alpha)=2$\AA}.   
Aperture effects can strongly bias the estimates of star-formation
rates based on spectra obtained through fibers which cover less than
$\sim\!2$0\% of the integrated galaxy flux, as is the case for SDSS
spectra of galaxies in the redshift range of our sample
\citep{kewley05}. For the massive galaxies in our sample with
\mbox{M$_{r}\!<\!-20$} we confirm that aperture effects are important
finding a significant fraction of galaxies (mostly face-on spirals
with prominent bulges) which are classified as passive from their
spectra, but whose blue integrated ${\rm NUV}-r$ colours indicate
recent star-formation \citep{paper3}. Throughout the article we thus
quantify and take full account of the effects of aperture biases on our results for
the bright galaxies in our sample. For the dwarf galaxy population
which represents our primary interest in this article, we find that
aperture biases should not affect our results, in agreement with
\citet{brinchmann}, due primarily to the absence of faint early-type
spirals. 

In the case of massive galaxies \mbox{($\mathcal{M}\!\ga\!10^{10.5}{\rm
M}_{\odot}$;} \mbox{M$_{r}\!\la\!-20$)} we find only gradual trends of
star-formation with environment, the fraction of passively-evolving
galaxies increasing gradually with local density from \mbox{$\sim5$0\%} to
\mbox{$\sim7$0\%} in high-density regions, the trend extending to the lowest
density regions studied, well beyond the effects of cluster-related
processes. For these galaxies only the large-scale galaxy density
appears important for defining its star-formation history, and not
its immediate neighbours  
or even whether it is within a cluster or group, implying that 
cluster-related processes are not the primary mechanisms by which
massive galaxies become passive. The star-formation
histories of massive galaxies appear to be predefined by the initial
conditions in which they form, whereby in high-density regions they are
likely to form earlier and have more active merger histories than
those in low-density regions, resulting in the observed gradual
SF-density relations. 

In contrast, the star-formation histories of dwarf galaxies \mbox{($\mathcal{M}\!\sim\!10^{9.2}{\rm
 M}_{\odot}$, M$_{r}\!\sim\!-18$)} are strongly dependent on their local environment, the fraction of
 passively-evolving galaxies dropping from \mbox{$\sim\!7$0\%} in dense
 environments, to \mbox{$\sim\!0$\%} in the rarefied field.
Indeed for the
lowest luminosity range covered \mbox{($-18<M_r<-16$)} none of the
\mbox{$\sim\!6$00} galaxies in the lowest density quartile are passively-evolving.
The few passively-evolving dwarfs in field regions
are strongly clustered around bright \mbox{($\ga\!L^{*}$)} galaxies,
and throughout the SDSS sample we find no passively-evolving dwarf
galaxies more than \mbox{$\sim\!2$} virial radii from a massive halo,
whether that be a cluster, group or massive galaxy. This limit of
around 2--2.5 virial radii corresponds to the maximum distance from a
cluster or massive galaxy that a galaxy can rebound to having been
previously subhalos within massive halos \citep{mamon,diemand}, and so
it seems reasonable to believe that all passively-evolving dwarf
galaxies are or have been satellites within a massive DM halo.

Our finding that passively-evolving dwarf galaxies are only found
within cluster, groups or as satellites to massive galaxies indicates
that internal processes or merging are not responsible for terminating
star-formation in these galaxies.
Instead the evolution of dwarf
galaxies is primarily driven by the mass of their host halo, probably
through the combined effects of tidal forces and ram-pressure stripping. 
We find a significant anti-correlation ($10\sigma$) between the
EW(H$\alpha$) of dwarf \mbox{($-19<{\rm M}_{r}<-18$)} star-forming
galaxies and local density, producing a  
a systematic reduction of \mbox{$\sim\!3$0\%} in the H$\alpha$ emission in
high-density regions with respect to field values. 
We argue that this implies that the bulk of star-forming 
dwarf galaxies in groups and clusters are currently in the process of
being slowly transformed into passive galaxies. The transformation of
dwarf galaxies solely through environmental processes results in the
wide variety of star-formation histories observed in the local dwarf
galaxy population \citep{mateo}.

Examining the fraction of passively-evolving galaxies as a function of
both luminosity and local environment, we find that in high-density
regions \mbox{$\sim7$0\%} of galaxies are passively-evolving
independent of luminosity. In the rarefied field, where environmental
related processes are unlikely to be effective, however the fraction
of passively-evolving galaxies is a strong function of luminosity,
dropping from 50\% for \mbox{M$_{r}\!\la\!-21$} to zero by
\mbox{M$_{r}=-18$}. 
This strong luminosity dependence of the fraction of passive galaxies
in field regions reflects with the systematic trend with increasing
galaxy mass for star-formation histories to be driven by internal
mechanisms as opposed to environmental processes. This transition from
 environmentally to internally-driven star-formation histories is likely due to a combination of factors:
\begin{enumerate}
\item the increasing efficiency and rapidity with which gas is
  converted into stars for more massive galaxies, resulting in
  gas-consumption time-scales for massive galaxies that are much shorter than the Hubble time;
\item the mode for the accretion and cooling of fresh gas onto the galaxy
  from its surroundings. When galaxy halos reach a mass
  \mbox{$\sim\!10^{12}{\rm M}_{\odot}$} expanding shocks heat
  infalling gas to the virial temperature of the halo, producing a
  stable atmosphere of hot, diffuse gas which is vulnerable to
  feedback effects that can prevent the gas from cooling onto
  the galaxy \citep{keres,dekel};
\item AGN feedback in the form of quasar winds which can expel the
  remaining gas from a galaxy and/or the quasi-continuous low-level AGN
  activity that prevents cooling of the diffuse atmosphere of hot gas
  in massive galaxies. The observed paralleled increasing fractions of
  AGN and passive field galaxies with luminosity supports the
  importance of AGN feedback in shutting down star-formation in
  galaxies, a process which should become increasingly efficient with galaxy mass as the result of the tight M$_{\rm BH}-\sigma$
  relation; 
\item the increased susceptability of low-mass galaxies to disruption
  by environmental effects such as tidal shocks and ram-pressure
  stripping due to their shallow potential wells.
\end{enumerate}
The combined effect of these processes is to produce the observed
bimodality in the global properties of galaxies about a characteristic
mass of \mbox{$\sim\!3\times10^{10}{\rm M}_{\odot}$} \citep{k03a}.
 However given that several models that treat each of these processes in
diverse ways are qualitatively able to reproduce this bimodality \citep[e.g.][]{menci,bower,croton,cattaneo,birnboim}, it will
be difficult to quantify the relative importance of the mechanisms for
driving the star-formation histories of galaxies, although
\citet{hopkins07b} show some observational tests that could distinguish
between these models.

\section*{acknowledgements}

\noindent
The authors thank Ivan Baldry, Agatino Rifatto and Eelco van Kampen
for reading a draft version of the article, and the referee for useful
comments regarding this work.
CPH acknowledges the financial supports provided
  through the European Community's Human Potential Program, under
  contract HPRN-CT-2002-0031 SISCO. AM acknowledges the
  INAF-Osservatorio Astronomico di Capodimonte for a grant. AG thanks
  her PhD supervisor Massimo Capaccioli. 

Funding for the SDSS and SDSS-II has been provided by the Alfred
P. Sloan Foundation, the Participating Institutions, the National
Science Foundation, the U.S. Department of Energy, the National
Aeronautics and Space Administration, the Japanese Monbukagakusho, the
Max Planck Society, and the Higher Education Funding Council for
England. The SDSS Web Site is http://www.sdss.org/. 

The SDSS is managed by the Astrophysical Research Consortium for the
Participating Institutions. The Participating Institutions are the
American Museum of Natural History, Astrophysical Institute Potsdam,
University of Basel, University of Cambridge, Case Western Reserve
University, University of Chicago, Drexel University, Fermilab, the
Institute for Advanced Study, the Japan Participation Group, Johns
Hopkins University, the Joint Institute for Nuclear Astrophysics, the
Kavli Institute for Particle Astrophysics and Cosmology, the Korean
Scientist Group, the Chinese Academy of Sciences (LAMOST), Los Alamos
National Laboratory, the Max-Planck-Institute for Astronomy (MPIA),
the Max-Planck-Institute for Astrophysics (MPA), New Mexico State
University, Ohio State University, University of Pittsburgh,
University of Portsmouth, Princeton University, the United States
Naval Observatory, and the University of Washington. 

This research has made use of the NASA/IPAC Extragalactic Database
(NED) which is operated by the Jet Propulsion Laboratory, California
Institute of Technology, under contract with the National Aeronautics
and Space Administration.

The Millennium Run simulation used in this paper was carried out by the Virgo Supercomputing Consortium at the Computing Centre of the Max-Planck Society in Garching. The semi-analytic galaxy catalogue is publicly available at
http://www.mpa-garching.mpg.de/galform/agnpaper

\appendix
\section{Testing of the density estimator}
\label{tests}

\begin{figure*}
\includegraphics[width=150mm]{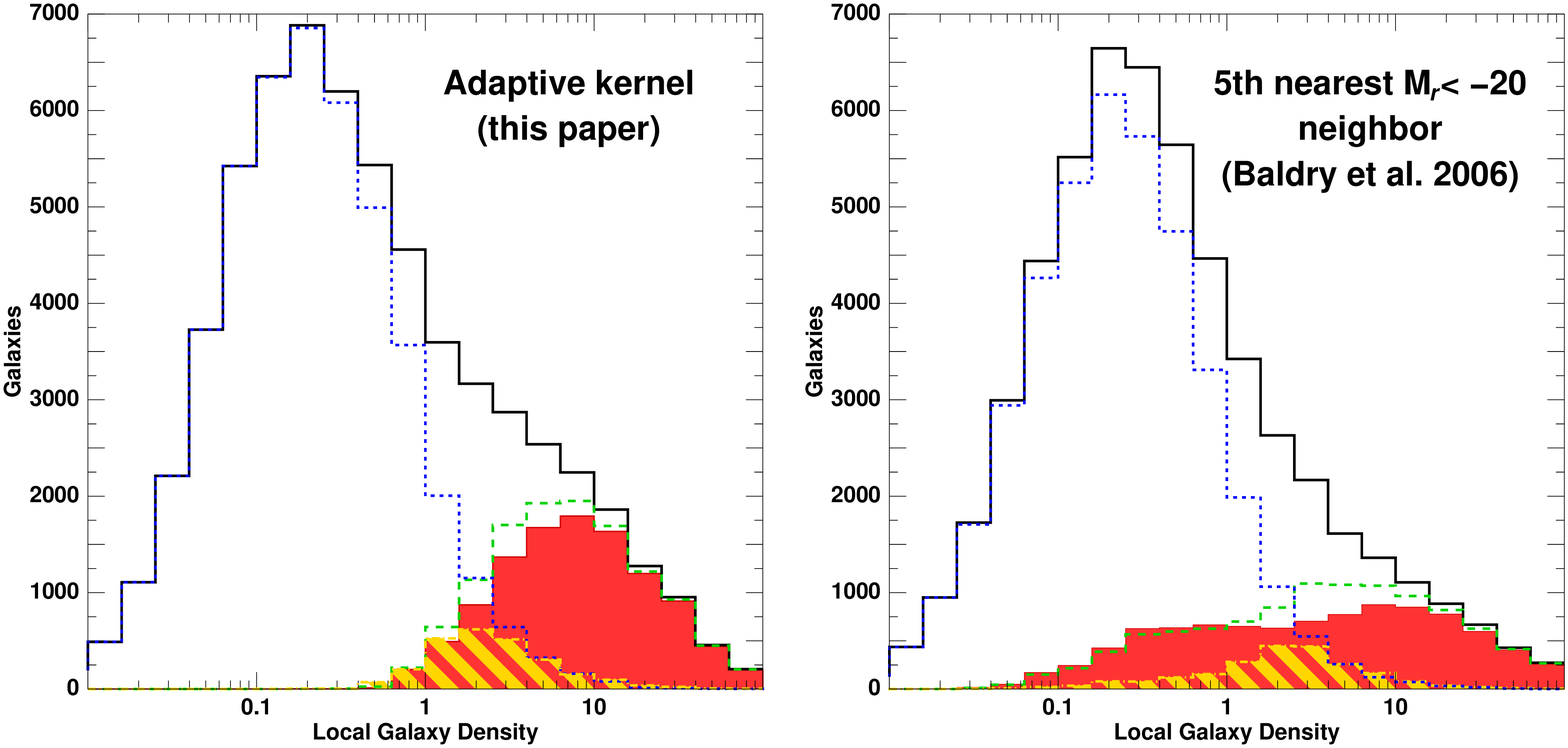}
\caption{Distribution of ``observed'' local galaxy densities from Millennium
  simulation as calculated using the adaptive kernel algorithm (left)
  and based on the distance of the 5$^{th}$ nearest neighbour with
  M$_r<-20$ \citep{balogh04,baldry06}. The histograms show the local density distributions of
  all \mbox{M$_r<-18$} galaxies (black solid lines), those galaxies within groups
  containing 4 or more members (red filled histogram), those galaxies
  within \mbox{1\,R$_{\rm vir}$} (green dashed lines),
  \mbox{$0.8<(r/{\rm R}_{\rm vir})<1.2$}  (yellow hatched histogram
  with dot-dashed lines), and more than \mbox{2\,R$_{\rm vir}$} (blue short-dashed lines) from the nearest group or cluster.}
\label{frac_rho}
\end{figure*}

We test the efficiency of this density estimation algorithm, by
applying it to the public galaxy catalogues from the Millennium
simulation \citep{springelsim,croton}. These simulations cover a \mbox{(500\,$h^{-1}$Mpc)$^{3}$}
volume producing a galaxy catalogue complete to \mbox{M$_r=-17.4$} containing
some 9 million galaxies, for which positions, peculiar velocities,
absolute magnitudes in each of the SDSS filters, stellar masses are
all provided. 
From this, we consider 61\,981 \mbox{M$_r<-18$} galaxies from a
subregion of volume \mbox{$200\times200\times100\,{\rm Mpc}^3$}. 
We estimate
the local density of each galaxy in physical space, $\rho(3D)$, by applying the
adaptive kernel method, representing each galaxy by a spherically-symmetric
Gaussian kernel whose width is equal to the distance to its 5$^{th}$
nearest neighbour. Galaxy groups and clusters are identified through
a percolation analysis using a friends-of-friends algorithm with a
linking length equal to 0.2 times the mean interparticle separation.  
For each group containing four or more members, the velocity
dispersion \mbox{($\sigma_{\nu}$)} and virial radius \mbox{(R$_{\rm vir}$)} are calculated
following \citet{girardi}.
For each
galaxy we combine the position and peculiar velocity in the
$z$-direction to create a redshift, producing a volume-limited
redshift catalogue, to which we apply exactly the same
adaptive kernel method as for the SDSS data, representing each galaxy
by a Gaussian kernel of transverse width defined by the distance to
its 3$^{rd}$ nearest neighbour within \mbox{500\,km\,s$^{-1}$} and of width
\mbox{500\,km\,s$^{-1}$} in the radial/redshift direction. The resultant
distribution of local galaxy densities, $\rho$, is shown in the left panel
Fig.\ref{frac_rho} by the solid black lines. It shows a peak at
\mbox{$\sim0.2$}, which is close to the global mean \mbox{M$_r<-18$} galaxy
density of \mbox{0.11\,Mpc$^{-2}$\,(500\,km\,s$^{-1}$)$^{-1}$}, and a long
tail extending to high densities. 

To examine the efficiency
of this density estimator in identifying galaxies within
groups the red filled histogram shows the density distribution of
galaxies that were identified by the percolation analysis to belong to
a group containing four or more members. Alternatively, the
green-dashed histogram shows the density distribution of those
galaxies within the virial radius of a group containing four or more galaxies.
In both cases, the observed local galaxy density of group galaxies are strongly
biased to high densities with \mbox{$\rho>1$}, while {\em no} group galaxy
has \mbox{$\rho<0.5$}. The yellow hatched histogram shows the density
distribution of those galaxies in the transition regions between group and
field environments having \mbox{$0.8<(r/{\rm R}_{\rm
    vir})<1.2$}. Galaxies in these transition regions have local
densities in the range \mbox{$1\la\rho\la4$}. Finally, the blue short-dashed
histogram shows those galaxies in isolated field regions more than two
virial radii from the nearest group, and hence are unlikely to have
encountered the group environment during their evolution or be
affected by group/cluster-related physical processes. As expected,
they are observed to have low local densities with \mbox{$\rho\la1$}. 

The left panel of Fig.~\ref{frac_rho} demonstrates the efficiency of this density
estimator in identifying group and isolated field galaxies, in
particular the latter. By selecting galaxies with \mbox{$\rho<0.5$} we can be
sure of obtaining a sample of field galaxies, the vast majority
\mbox{($>97$\%)} at \mbox{$>2\,{\rm R}_{\rm vir}$}, and with no contamination from any galaxies
belonging to groups or clusters, even those containing as few as four
galaxies. Considering instead those galaxies with \mbox{$\rho<1.0$}, $>9$5\%
still have \mbox{$>2\,{\rm R}_{\rm vir}$}, while only 0.7\% lie within the
virial radius of a group. In contrast, \mbox{$\sim6$0\%} of \mbox{$\rho>1.0$} and
\mbox{$\sim9$0\%} of \mbox{$\rho>4$} galaxies are at \mbox{$<1\,{\rm R}_{\rm vir}$}. 

For comparison the right panel of Fig.~\ref{frac_rho} shows the same
density distributions of field and group galaxies brighter than
\mbox{M$_r=-18$} using the nearest neighbour algorithm as applied by
BB04 and BB06 
who estimate the local density on the basis of the
projected distance to the 5$^{th}$ nearest neighbour that is brighter
than \mbox{M$_{r}=-20$} and has a radial velocity within
\mbox{1\,000\,km\,s$^{-1}$} of each galaxy. The overall density
distributions are quite similar, as although there are three times
fewer \mbox{M$_r\!<\!-20$} galaxies than \mbox{M$_r\!<\!-18$} galaxies, the recession velocity range
over which the projected density is estimated is quadruple that used for
the adaptive kernel estimator (2\,000\,km\,s$^{-1}$ instead of 500\,km\,s$^{-1}$). The most important difference between
the two estimators is the much broader density distribution of group
galaxies (red filled histogram) for the nearest neighbour algorithm,
which extends to much lower ``observed'' densities than our
approach. Even at the lowest densities studied by BB04
and BB06, corresponding to \mbox{$\Sigma_{5}\simeq0.1$}, \mbox{$\sim5$\%} of
the galaxies are group members, and hence could have been affected by
group-related environmental processes. Equally, $\Sigma_{5}$ does not
appear very sensitive to the position of a galaxy within a halo, as is
apparent by the densities of those galaxies at and around the virial
radius of groups being covering a wide range of values of $\Sigma_{5}$
and peaking at a mid-range value for galaxies within groups, rather
than that obtained from the adaptive kernel approach where galaxies near
the virial radius have generally the lowest densities of those
galaxies in groups. The value of $\Sigma_{5}$ is instead most
sensitive to the {\em mass} of the nearest massive halo, rather
than whether a galaxy is inside or outside that halo. 

To fairly compare the efficiency of the adaptive kernel estimator
against nearest neighbour approaches, we retested the nearest
neighbour algorithm using the same \mbox{M$_{r}\!<\!-18$} datasets,
varying the number of neighbours used over the range 3--10, and the
velocity range used to select neighbours from 400 to
1000\,km\,s$^{-1}$. We define the efficiency of the density estimators
in two ways: firstly in terms of the rank correlation between the
observed projected density and the actual physical 3-dimensional
galaxy density $\rho(3D)$; and secondly its sensitivity to the
position of the galaxy within a halo, as measured by the correlation
with the distance $r$ do the nearest massive halo (containing four or
more galaxies) scaled by the virial radius R$_{vir}$. We measure the
strengths of these correlations through the Spearman rank correlation
test as presented in Table~\ref{estimators}. Firstly, varying the
numbers of neighbours used to estimate the local density, we find that
$\Sigma_{5}$ is the most sensitive to the actual physical local
density $\rho(3D)$, while $\Sigma_{10}$ is the most sensitive to the
position of the galaxy within the halo (for neighbours within
500\,km\,s$^{-1}$.  Secondly, varying the velocity range over which
neighbours are selected, we find that a range of 400\,km\,s$^{-1}$ is
most sensitive to the physical density, while a range of
600\,km\,s$^{-1}$ is the most sensitive to the position of the galaxy
within the halo. These confirm that using a range of 500\,km\,s$^{-1}$
is the optimal general purpose value for estimating the local density
of galaxies (at least for samples extending to M$_{r}\sim-18$, while
the commonly used value of 1\,000\,km\,s$^{-1}$ is much less efficient
due to contamination of projected background galaxies. Comparing our
adaptive kernel estimator to the nearest neighbour algorithm, we find
that it is always more sensitive to the position of the galaxy within
a halo, and is only marginally less efficient than the optimal choice
of parameters for the nearest neighbour algorithm
($\Sigma_{5}$). The adaptive kernel estimator we have adopted is at
least as efficient as any comparable nearest neighbour algorithms, and
is particularly sensitive to the position of a galaxy within a halo,
which is likely to be the most important aspect of the environment in
terms of affecting its evolution \citep{lemson}. 
    
\begin{table}
\begin{center}
\caption{Comparison of efficiencies of density estimators}
\label{estimators}
\begin{tabular}{cccccc} \hline
Method&No. of& Magnitude &Velocity& \multicolumn{2}{c}{Spearman rank} \\
used &neigh-& limit & range & \multicolumn{2}{c}{correlation}\\
&bours& & (km\,s$^{-1}$) & $\rho$(3D) &
$r/R_{\rm vir}$ \\
\hline 
AK & 3 & M$_{r}\!<\!-18$ & 500 &  {\bf 0.891} & {\bf -0.800} \\ \hline 
NN & 3 &  M$_{r}\!<\!-18$ & 500 & 0.849 & -0.744 \\
NN & 5 & M$_{r}\!<\!-18$ & 500 & {\bf 0.896} & -0.774 \\
NN & 8 & M$_{r}\!<\!-18$ & 500 & 0.890 & -0.786 \\
NN & 10 & M$_{r}\!<\!-18$ & 500 & 0.877 & {\bf -0.789} \\ \hline
NN & 5 & M$_{r}\!<\!-18$ & 1000 & 0.856 & -0.766 \\
NN & 5 & M$_{r}\!<\!-18$ & 800 & 0.872 & -0.772 \\
NN & 5 & M$_{r}\!<\!-18$ & 600 & 0.889 & {\bf -0.775} \\
NN & 5 & M$_{r}\!<\!-18$ & 500 & 0.896 & -0.774 \\ 
NN & 5 & M$_{r}\!<\!-18$ & 400 & {\bf 0.898} & -0.766 \\ \hline
NN & 5 & M$_{r}\!<\!-20$ & 1000 & 0.785 & -0.741 \\ \hline
\end{tabular}
\end{center}
\end{table}

In comparison we note that the particular algorithm used by BB04, BB06
is significantly less sensitive to both the $\rho(3D)$ and the
position of the galaxy within the host halo (last row of Table~\ref{estimators}).
These differences reflect the practical issue of the
density of information used to characterize the environment of a
galaxy. By measuring the local environment using just \mbox{M$_{r}<-20$}
galaxies, the density of information is much sparser, making it that
much less sensitive to effects on the scales of poor
groups, which may contain three or less \mbox{M$_{r}<-20$} galaxies. However,
a volume of the universe can be covered that is a factor ten larger
than is possible using our approach which is limited by the necessity
of being complete to \mbox{M$_{r}=-18$}, and indeed for galaxies brighter
than \mbox{$\sim{\rm M}^{\star}$} our analysis of their environmental
trends are strongly limited by the small sample size. In addition, in
the volume covered by our analysis, the only rich structures (with
$\sigma_{\nu}\ga5$00\,km\,s$^{-1}$) are the
supercluster associated with the rich cluster Abell 2199 studied in
Paper I, and Abell 1314, limiting our ability to follow the environmental trends to
the highest densities.

The adaptive kernel method has the added
advantage of being able to be used as a group-finder
\citep[e.g.][]{bardelli,haines04a}, by identifying groups and clusters as local maxima in the galaxy
density function $\rho(\textbf{x}, z)$. A comparison of the groups identified in the
Millennium simulation from a percolation analysis, finds that all such
groups containing four or more members are marked as clear local
maxima in the luminosity-weighted density maps.
Unlike group finding
algorithms based on a percolation analysis, the adaptive kernel
approach is also able to efficiently define substructures that are the
natural consequence of the hierarchical merging of galaxy groups and
clusters.

\label{lastpage}

\begin{thebibliography}{99}
\bibitem[\protect\citeauthoryear{Aarseth \& Fall}{1980}]{aarseth}
 Aarseth S. J., Fall S. M., 1980, ApJ, 236, 43
\bibitem[\protect\citeauthoryear{Adelman-McCarthy et al.}{2006}]{sdssdr4}
 Adelman-McCarthy J. K., et al., 2006, ApJS, 163, 38
\bibitem[\protect\citeauthoryear{Alonso et al.}{2006}]{alonso}
 Alonso S. M., Lambas D. G., Tissera P., Coldwell G., 2006,
 MNRAS, 367, 1029
\bibitem[\protect\citeauthoryear{Baldry et al.}{2004}]{baldry04}
 Baldry I. K., Glazebrook K., Brinkmann J., Ivezi\'{c} \u{Z},
 Lupton R. H., Nichol R. C., Szalay A. S., 2004, ApJ, 600, 681
\bibitem[\protect\citeauthoryear{Baldry et al.}{2006}]{baldry06}
 Baldry I. K., Balogh M. L., Bower R. G., Glazebrook K., Nichol
 R. C., Bamford S. P., Budavari T., 2006, MNRAS, 373, 469 (BB06)
\bibitem[\protect\citeauthoryear{Baldwin, Phillips, \&
    Terlevich}{Baldwin et al.}{1981}]{baldwin}
 Baldwin J. A., Phillips M. M., Terlevich R., 1981, PASP, 93, 5
\bibitem[\protect\citeauthoryear{Balogh et al.}{1999}]{balogh99}
 Balogh M. L., Morris S. L., Yee H. K. C., Carlberg R. G.,
 Ellingson E., 1999, ApJ, 527, 54
\bibitem[\protect\citeauthoryear{Balogh, Navarro \& Morris}{Balogh et al.}{2000}]{balogh00}
 Balogh M. L., Navarro J. F. Morris S. L., 2000, ApJ, 540, 113
\bibitem[\protect\citeauthoryear{Balogh et al.}{2004a}]{balogh04}
 Balogh M. L., et al., 2004a, MNRAS, 348, 1355 (B04)
\bibitem[\protect\citeauthoryear{Balogh et al.}{2004b}]{balogh04b}
 Balogh M. L., Baldry I. K., Nichol R., Miller C., Bower R.,
 Glazebrook K., 2004b, ApJL, 615, 101 (BB04)
\bibitem[\protect\citeauthoryear{Bardelli et al.}{1998}]{bardelli}
 Bardelli S., Pisani A., Ramella M., Zucca E., Zamorani
 G., 1998, MNRAS, 300, 589
\bibitem[\protect\citeauthoryear{Bekki, Couch \& Shioya}{Bekki et al.}{2001}]{bekki01}
 Bekki K., Couch W. J., Shioya Y., 2001, PASJ, 53, 395
\bibitem[\protect\citeauthoryear{Bekki, Couch \& Shioya}{Bekki et al.}{2002}]{bekki02}
 Bekki K., Couch W. J., Shioya Y., 2002, ApJ, 577, 651
\bibitem[\protect\citeauthoryear{Bell \& de Jong}{2001}]{belldejong}
 Bell E. F., de Jong R., 2001, ApJ, 550, 212
\bibitem[\protect\citeauthoryear{Bell et al.}{2004}]{bell}
 Bell E. F., et al., 2004, ApJ, 608, 752
\bibitem[\protect\citeauthoryear{Binggeli, Tarenghi \&
 Sandage}{Binggeli et al.}{1990}]{binggeli90}
 Binggeli B., Tarenghi M., Sandage A., 1990, A\&A, 228, 42
\bibitem[\protect\citeauthoryear{Birnboim, Dekel \&
 Neistein}{Birnboim et al.}{2007}]{birnboim}
 Birnboim Y., Dekel A., Neistein E., 2007, preprint (astro-ph/0703435)
\bibitem[\protect\citeauthoryear{Blanton, Berlind \& Hogg}{Blanton et al.}{2006}]{blanton06}
 Blanton M. R., Berlind A. A., Hogg D. W., 2006, preprint (astro-ph/0608353)
\bibitem[\protect\citeauthoryear{Blanton et al.}{2003a}]{blanton03a}
 Blanton M. R., et al., 2003a, AJ, 125, 2276
\bibitem[\protect\citeauthoryear{Blanton et al.}{2003b}]{blanton03}
 Blanton M. R., et al., 2003b, ApJ, 594, 186
\bibitem[\protect\citeauthoryear{Blanton et al.}{2003c}]{kcorrect}
 Blanton M. R., et al., 2003c, AJ, 125, 2348
\bibitem[\protect\citeauthoryear{Blanton et al.}{2005a}]{blanton05}
 Blanton M. R., et al., 2005a, ApJ, 629, 143
\bibitem[\protect\citeauthoryear{Blanton et al.}{2005b}]{blanton05a}
 Blanton M. R., et al., 2005b, ApJ, 631, 208
\bibitem[\protect\citeauthoryear{Blanton et al.}{2005c}]{nyuvagc}
 Blanton M. R., et al., 2005c, AJ, 129, 2562
\bibitem[\protect\citeauthoryear{Boselli \& Gavazzi}{2006}]{boselli}
 Boselli A., Gavazzi G., 2006, PASP, 118, 517
\bibitem[\protect\citeauthoryear{Boselli et al.}{2006}]{boselli06}
 Boselli A., et al., 2006, ApJ, 651, 811
\bibitem[\protect\citeauthoryear{Bower et al.}{2006}]{bower}
 Bower R. G., et al., 2006, MNRAS, 370, 645
\bibitem[\protect\citeauthoryear{Brinchmann et al.}{2004}]{brinchmann}
 Brinchmann J., Charlot S., White S. D. M., Tremonti C.,
 Kauffmann G., Heckman T., Brinkmann J., 2004, MNRAS, 351, 1151
\bibitem[\protect\citeauthoryear{Bruzual \& Charlot}{2003}]{bc03}
 Bruzual A. G., Charlot S., 2003, MNRAS, 344, 1000
\bibitem[\protect\citeauthoryear{Caldwell et al.}{1993}]{caldwell}
 Caldwell N., Rose J. A., Sharples R. M., Ellis R. S., Bower
 R. G., 1993, AJ, 106, 473
\bibitem[\protect\citeauthoryear{Cardelli, Clayton \& Mathis}{Cardelli
 et al.}{1989}]{cardelli}
 Cardelli J. A., Clayton G. C., Mathis J. S., 1989, ApJ, 345, 245
\bibitem[\protect\citeauthoryear{Cattaneo et al.}{2006}]{cattaneo}
 Cattaneo A., Dekel A., Devriendt J., Guideroni B., Blaizot J., 2006,
 MNRAS, 370, 1651
\bibitem[\protect\citeauthoryear{Charlot \& Fall}{2000}]{charlot}
 Charlot S., Fall S. M., 2000, ApJ, 539, 718
\bibitem[\protect\citeauthoryear{Chiosi \& Cararro}{2002}]{chiosi}
 Chiosi C., Carraro G., 2002, MNRAS, 335, 335
\bibitem[\protect\citeauthoryear{Christlein \& Zabludoff}{2005}]{christlein}
 Christlein D., Zabludoff A. I., 2005, ApJ, 616, 192 
\bibitem[\protect\citeauthoryear{Cirasuolo et al.}{2006}]{cirasulo}
 Cirasuolo M., et al., 2006, preprint(astro-ph/0609287)
\bibitem[\protect\citeauthoryear{Conselice}{2006}]{conselice}
 Conselice C. J., 2006, ApJ, 638, 686
\bibitem[\protect\citeauthoryear{Conselice, Gallagher \& Wyse}{Conselice et al.}{2001}]{conselice1}
 Conselice C. J., Gallagher J. S. III, Wyse R. F. G., 2001, ApJ, 559, 791
\bibitem[\protect\citeauthoryear{Conselice et al.}{2003}]{conselice4}
 Conselice C. J., O'Neill K., Gallagher J. S. III,  Wyse
 R. F. G., 2003, ApJ, 591, 167
\bibitem[\protect\citeauthoryear{Cooper et al.}{2006}]{cooper}
 Cooper M. C., et al., 2006, MNRAS, 370, 198
\bibitem[\protect\citeauthoryear{C\^{o}t\'{e} et al.}{2006}]{cote}
 C\^{o}t\'{e} P., et al., 2006, ApJS, 165, 57
\bibitem[\protect\citeauthoryear{Cox et al.}{2006}]{cox}
 Cox T. J., Di Matteo T., Hernquist L., Hopkins P. F., Robertson
 B., Springel V., 2006, ApJ, 643, 692
\bibitem[\protect\citeauthoryear{Croton et al.}{2006}]{croton}
 Croton D. J., et al., 2006, MNRAS, 368, 11 (C06)
\bibitem[\protect\citeauthoryear{Davoodi et al.}{2006}]{davoodi}
 Davoodi P., et al., 2006, MNRAS, 371, 1113
\bibitem[\protect\citeauthoryear{Dekel \& Birnboim}{2006}]{dekel}
 Dekel A., Birnboim Y., 2006, MNRAS, 368, 2
\bibitem[\protect\citeauthoryear{de Lapparent}{2003}]{delapparent}
 de Lapparent V., 2003, A\&A, 408, 845
\bibitem[\protect\citeauthoryear{Desai et al.}{2007}]{desai}
 Desai V., et al., 2007, ApJ, 660, 1151
\bibitem[\protect\citeauthoryear{de Vaucouleurs}{1961}]{devaucouleurs}
 de Vaucouleurs G., 1961, ApJS, 5, 233
\bibitem[\protect\citeauthoryear{Diemand, Kuhlen \& Madau}{Diemand et
    al.}{2007}]{diemand}
 Diemand J., Kuhlen M., Madau P., 2007, preprint (astro-ph/0703337)
\bibitem[\protect\citeauthoryear{di Matteo, Springel \& Hernquist}{di
    Matteo et al.}{2005}]{dimatteo}
 di Matteo T., Springel V., Hernquist L., 2005, Nature, 433, 604
\bibitem[\protect\citeauthoryear{Dressler}{1980}]{dressler80}
 Dressler A., 1980, ApJ, 236, 772
\bibitem[\protect\citeauthoryear{Dressler et al.}{1997}]{dressler97}
 Dressler A., et al., 1997, ApJ, 490, 577
\bibitem[\protect\citeauthoryear{Driver et al.}{2006}]{driver}
 Driver S. P., et al., 2006, MNRAS, 368, 414 
\bibitem[\protect\citeauthoryear{Einasto et al.}{1974}]{einasto}
 Einasto J., Saar E., Kaasik A., Chernin A. D., 1974, Nature,
 252, 111 
\bibitem[\protect\citeauthoryear{Ferguson}{1992}]{ferguson92}
 Ferguson H. C., 1992, MNRAS, 255, 389
\bibitem[\protect\citeauthoryear{Ferguson \& Binggeli}{1994}]{ferguson}
 Ferguson H. C., Binggeli B., 1994, A\&ARv, 6, 67 
\bibitem[\protect\citeauthoryear{Ferrarese \& Ford}{2005}]{ferrarese05}
 Ferrarese L., Ford, H., 2005, Space Science Reviews, 116, 523
\bibitem[\protect\citeauthoryear{Ferrarese et al.}{2006}]{ferrarese}
 Ferrarese L., et al., 2006, ApJS, 164, 334
\bibitem[\protect\citeauthoryear{Fujita \& Nagashima}{1999}]{fujita}
 Fujita Y., Nagashima M., 1999, ApJ, 516, 619
\bibitem[\protect\citeauthoryear{Gebhardt et al.}{2000}]{gebhardt}
Gebhardt K., et al., 2000, ApJL, 539, 13
\bibitem[\protect\citeauthoryear{Girardi et al.}{1998}]{girardi}
 Girardi M., Giuricin G., Mardirossian F., Mezzetti M., 
 Boschin W., 1998, ApJ, 505, 74
\bibitem[\protect\citeauthoryear{Glazebrook et al.}{2004}]{glazebrook}
 Glazebrook K., et al., 2004, Nature, 430, 181
\bibitem[\protect\citeauthoryear{G\'{o}mez et al.}{2003}]{gomez}
 G\'{o}mez P., et al., 2003, ApJ, 584, 210
\bibitem[\protect\citeauthoryear{Goto}{2005}]{goto}
 Goto T., 2005, MNRAS, 357, 937
\bibitem[\protect\citeauthoryear{Goto et al.}{2003}]{goto03}
 Goto T.,  et al., 2003, PASJ, 55, 757
\bibitem[\protect\citeauthoryear{Gottl\"{o}ber, Klypin \&
    Kravtsov}{Gottl\"{o}ber et al.}{2001}]{gottlober}
Gottl\"{o}ber S., Klypin A., Kravtsov A. V., 2001, ApJ, 546, 223
\bibitem[\protect\citeauthoryear{Graham \& Driver}{2007}]{graham07}
 Graham A. W., Driver S., 2007, ApJ, 655, 77
\bibitem[\protect\citeauthoryear{Gray et al.}{2004}]{gray}
 Gray M. E., Wolf C., Meisenheimer K., Taylor A., Dye S., Borch
 A., Kleinheinrich M. 2004, MNRAS, 357, L73
\bibitem[\protect\citeauthoryear{Grebel, Gallagher \& Harbeck}{Grebel
 et al.}{2003}]{grebel}
 Grebel E. K., Gallagher J. S., Harbeck D., 2003, AJ, 125, 1926
\bibitem[\protect\citeauthoryear{Gunn \& Gott}{1972}]{gunn}
 Gunn J. E., Gott J. R., 1972, ApJ, 176, 1
\bibitem[\protect\citeauthoryear{Haehnelt, Natarajan \& Rees}{1998}]{haehnelt}
 Haehnelt M. G., Natarajan P., Rees M. J., 1998, MNRAS, 300, 817  
\bibitem[\protect\citeauthoryear{Haines et al.}{2004a}]{haines04a}
 Haines C. P., Campusano L. E., Clowes R. G., 2004a, A\&A, 421, 157
\bibitem[\protect\citeauthoryear{Haines et al.}{2004b}]{haines04b}
 Haines C. P., Mercurio A., Merluzzi P., La Barbera F.,
 Massarotti M., Busarello G., Girardi M., 2004b, A\&A, 425, 783
\bibitem[\protect\citeauthoryear{Haines et al.}{2006a}]{haines}
 Haines C. P., Merluzzi P., Mercurio A., Gargiulo A., Krusanova
 N., Busarello G., La Barbera F., Capaccioli M., 2006a, MNRAS, 371, 55
\bibitem[\protect\citeauthoryear{Haines et al.}{2006b}]{paper1}
 Haines C. P., La Barbera F., Mercurio A., Merluzzi P.,
 Busarello G., 2006b, ApJL, 647, 21 (Paper I)
\bibitem[\protect\citeauthoryear{Haines et al.}{2007}]{paper3}
 Haines C. P., Gargiulo A., Merluzzi P., 2007, in preparation (Paper III)
\bibitem[\protect\citeauthoryear{H\"{a}ring \& Rix}{2004}]{haring}
 H\"{a}ring N., Rix H.-W., 2004, ApJL, 604, 89
\bibitem[\protect\citeauthoryear{Hao et al.}{2005}]{hao}
 Hao L., et al., 2005, AJ, 129, 1783 
\bibitem[\protect\citeauthoryear{Heavens et al.}{2004}]{heavens}
 Heavens A., Panter B., Jimenez R., Dunlop J., 2004, Nature,
 428, 625
\bibitem[\protect\citeauthoryear{Hester}{2006}]{hester}
 Hester J. A., 2006, ApJ, 647, 910
\bibitem[\protect\citeauthoryear{Hogg et al.}{2006}]{hogg}
 Hogg D. W., Masjedi M., Berlind A. A., Blanton M. R., Quintero A. D.,
 Brinkmann J., 2006, ApJ, 650, 763
\bibitem[\protect\citeauthoryear{Hopkins et al.}{2003}]{hopkins}
 Hopkins A. M., et al., 2003, ApJ, 599, 971
\bibitem[\protect\citeauthoryear{Hopkins et al.}{2006a}]{hopkins06a}
 Hopkins P. F., Hernquist L., Cox T. J., Di Matteo T., Robertson
 B., Springel V., 2006a, ApJS, 163, 1 
\bibitem[\protect\citeauthoryear{Hopkins et al.}{2006b}]{hopkins06b}
 Hopkins P. F., Hernquist L., Cox T. J., Robertson
 B., Springel V., 2006b, ApJS, 163, 50 
\bibitem[\protect\citeauthoryear{Hopkins et al.}{2007a}]{hopkins07a}
 Hopkins P. F., Hernquist L., Cox T. J., Kere\v{s} D., 2007a, preprint (astro-ph/0706.1243) 
\bibitem[\protect\citeauthoryear{Hopkins et al.}{2007b}]{hopkins07b}
 Hopkins P. F., Cox T. J., Kere\v{s} D., Hernquist L., 2007b, preprint (astro-ph/0706.1246) 
\bibitem[\protect\citeauthoryear{Hubble}{1926}]{hubble1}
 Hubble E. P., 1926, ApJ, 64, 321
\bibitem[\protect\citeauthoryear{Hubble}{1936}]{hubble2}
 Hubble E. P., 1936, The Realm of the Nebulae. Yale University Press, New Haven
\bibitem[\protect\citeauthoryear{Hubble \& Humason}{1931}]{hubble3}
 Hubble E. P., Humason M. L., 1931, ApJ, 71, 43
\bibitem[\protect\citeauthoryear{Jungwiert, Combes \& Palous}{2001}]{jungwiert}
 Jungwiert B., Combes F., Palous J., 2001, A\&A, 376, 85
\bibitem[\protect\citeauthoryear{Kapferer et al.}{2007}]{kapferer}
 Kapferer W., et al., 2007, A\&A, 466, 813
\bibitem[\protect\citeauthoryear{Karachentsev}{2005}]{karachentsev05}
 Karachentsev I. D., 2005, AJ, 129, 178
\bibitem[\protect\citeauthoryear{Karachentsev et al.}{2004}]{karachentsev}
 Karachentsev I. D., Karachentseva V. E., Huchtmeier W. K.,
 Makarov D. I., 2004, AJ, 127, 2031
\bibitem[\protect\citeauthoryear{Kauffmann, White \&
    Guideroni}{1993}]{kauffmann93}
 Kauffmann G., White S. D. M., Guideroni B., 1993, MNRAS, 264, 201
\bibitem[\protect\citeauthoryear{Kauffmann et al.}{2003a}]{k03a}
 Kauffmann G., et al., 2003a, MNRAS, 341, 33 (K03)
\bibitem[\protect\citeauthoryear{Kauffmann et al.}{2003b}]{k03b}
 Kauffmann G., et al., 2003b, MNRAS, 341, 54 
\bibitem[\protect\citeauthoryear{Kauffmann et al.}{2003c}]{k03c}
 Kauffmann G., et al., 2003c, MNRAS, 346, 1055
\bibitem[\protect\citeauthoryear{Kauffmann et al.}{2004}]{kauffmann04}
 Kauffmann G., et al., 2004, MNRAS, 353, 713
\bibitem[\protect\citeauthoryear{Kauffmann et al.}{2006}]{kauffmann06}
 Kauffmann G., et al., 2006, MNRAS, 367, 1408 
\bibitem[\protect\citeauthoryear{Kennicutt}{1998}]{kennicutt}
 Kennicutt R. C. Jr., 1998, ARA\&A, 36, 189
\bibitem[\protect\citeauthoryear{Kere\v{s} et al.}{2005}]{keres}
 Kere\v{s} D., Katz N., Weinberg D. H., Dav\'{e} R., 2005, MNRAS, 363, 2
\bibitem[\protect\citeauthoryear{Kewley et al.}{2001}]{kewley}
 Kewley L. J., Dopita M. A., Sutherland R. S., Heisler C. A.,
 Trevena J., 2001, ApJ, 556, 121 
\bibitem[\protect\citeauthoryear{Kewley, Jansen \& Geller}{Kewley et al.}{2005}]{kewley05}
 Kewley L. J., Jansen R. A., Geller M. J., 2005, PASP, 117, 227
\bibitem[\protect\citeauthoryear{Kewley et al.}{2006}]{kewley06}
 Kewley L. J., Groves B., Kauffmann G., Heckman T., 2006, MNRAS,
 372, 961
\bibitem[\protect\citeauthoryear{Koopmann \& Kenney}{2004}]{koopmann}
 Koopmann R. A., Kenney J. D. P., 2004, ApJ, 613, 851
\bibitem[\protect\citeauthoryear{Kravtsov, Gnedin \& Klypin}{2004}]{kravtsov}
 Kravtsov A. V., Gnedin O. Y., Klypin A. A., 2004, ApJ, 609, 482
\bibitem[\protect\citeauthoryear{Kriek et al.}{2006}]{kriek}
 Kriek M., et al., 2006, ApJL, 649, 71
\bibitem[\protect\citeauthoryear{Lacey \& Cole}{1993}]{lacey}
 Lacey C., Cole S., 1993, MNRAS, 262, 627
\bibitem[\protect\citeauthoryear{Larson, Tinsley \& Caldwell}{Larson et al.}{1980}]{larson}
 Larson R. B., Tinsley B. M., Caldwell C. N., 1980, MNRAS, 237, 692
\bibitem[\protect\citeauthoryear{Lee, McCall \& Richer}{2003}]{lee}
 Lee H., McCall M. L., Richer M. G., 2003, AJ, 125, 2975 
\bibitem[\protect\citeauthoryear{Lemson \& Kauffmann}{1999}]{lemson}
 Lemson G., Kauffmann G., 1999, 302, 111
\bibitem[\protect\citeauthoryear{Lewis et al.}{2002}]{lewis}
 Lewis I., et al., 2002, MNRAS, 334, 673
\bibitem[\protect\citeauthoryear{Lisker et al.}{2006}]{lisker}
 Lisker T., Glatt K., Westera P., Grebel E. K., 2006, AJ, 132, 2432
\bibitem[\protect\citeauthoryear{Lisker et al.}{2007}]{lisker07}
 Lisker T., Grebel E. K., Binggeli B., Glatt K., 2007, ApJ, 660, 1186
\bibitem[\protect\citeauthoryear{Mac Low \& Ferrara}{1999}]{maclow}
 Mac Low M.-M., Ferrara A., 1999, ApJ, 513, 142
\bibitem[\protect\citeauthoryear{Mahdavi \& Geller}{2004}]{mahdavi}
 Mahdavi A., Geller M. J., 2004, ApJ, 607, 202
\bibitem[\protect\citeauthoryear{Mamon et al.}{2004}]{mamon}
 Mamon G. A., Sanchis T., Salvador-Sol\'{e} E., Solanes
 J. M., 2004, A\&A, 414, 445
\bibitem[\protect\citeauthoryear{Marcolini et al.}{2003}]{marcolini03}
 Marcolini A., Brighenti F., D'Ercole A., 2003, MNRAS, 345, 1329
\bibitem[\protect\citeauthoryear{Marcolini et al.}{2006}]{marcolini06}
 Marcolini A., D'Ercole A., Brighenti F., Recchi S., 2006,
 MNRAS, 371, 643 
\bibitem[\protect\citeauthoryear{Martin \& Kennicutt}{2001}]{martin}
 Martin C. L., Kennicutt R. C., 2001, ApJ, 555, 301
\bibitem[\protect\citeauthoryear{Martini et al.}{2006}]{martini}
 Martini P., Kelson D. D., Kim E., Mulchaey J. S., Athey A. A., 2006, ApJ, 644, 116
\bibitem[\protect\citeauthoryear{Mastropietro et al.}{2005}]{mastropietro}
 Mastropietro C., Moore B., Mayer L., Debattista V. P.,
 Piffaretti R., Stadel J. 2005, MNRAS, 364, 607
\bibitem[\protect\citeauthoryear{Mateo}{1998}]{mateo}
 Mateo M., 1998, ARA\&A, 36, 435
\bibitem[\protect\citeauthoryear{Mateus et al.}{2006}]{mateus2}
 Mateus A., Sodr\'{e} L., Cid Fernandes R., Stasi\'{n}ska G.,
 Schoenell W., Gomes J. M., 2006, MNRAS, 370, 721
\bibitem[\protect\citeauthoryear{Mateus et al.}{2007}]{mateus4}
 Mateus A., Sodr\'{e} L., Cid Fernandes R., Stasi\'{n}ska G.,
 2007, MNRAS, 374, 1457
\bibitem[\protect\citeauthoryear{Mathews \& Brighenti}{2003}]{mathews}
 Mathews W. G., Brighenti F., 2003, ARA\&A, 41, 191 
\bibitem[\protect\citeauthoryear{Maulbetsch et al.}{2007}]{maulbetsch}
 Maulbetsch C., Avila-Reese V., Colin P., Gottl\"{o}ber S., Khalatyan
 A., Steinmetz M., 2007, ApJ, 654, 53
\bibitem[\protect\citeauthoryear{Mayer et al.}{2001}]{mayer}
 Mayer L., Governato F., Colpi M., Moore B., Quinn T., Wadsley
 J., Stadel J., Lake G., 2001, ApJ, 559, 754
\bibitem[\protect\citeauthoryear{Mayer et al.}{2006}]{mayer06}
 Mayer L., Mastropietro C., Wadsley J., Stadel J., Moore
 B., 2006, MNRAS, 369, 1021
\bibitem[\protect\citeauthoryear{Menci et al.}{2005}]{menci}
 Menci N., Fontana A., Giallongo E., Salimbeni S., 2005, ApJ, 632, 49
\bibitem[\protect\citeauthoryear{Mercurio et al.}{2006}]{mercurio}
 Mercurio A., Merluzzi P., Haines C. P., Gargiulo A., Krusanova
 N., La Barbera F., Busarello G., Capaccioli M., Covone G., 2006, MNRAS, 368, 109
\bibitem[\protect\citeauthoryear{Merlin \& Chiosi}{2006}]{merlin}
 Merlin E., Chiosi C., 2006, A\&A, 457, 437
\bibitem[\protect\citeauthoryear{Miller et al.}{2003}]{miller}
 Miller C. J., et al., 2003, ApJ, 597, 142
\bibitem[\protect\citeauthoryear{Moore et al.}{1996}]{moore}
 Moore B., Katz N., Lake G., Dressler A., Oemler A. Jr., 1996,
 Nature, 379, 613
\bibitem[\protect\citeauthoryear{Moore et al.}{1999}]{moore99}
 Moore B., Lake G., Quinn T., Stadel J., 1999, MNRAS, 364, 465
\bibitem[\protect\citeauthoryear{Morgan}{1958}]{morgan}
 Morgan W. W., 1958, PASP, 70, 415
\bibitem[\protect\citeauthoryear{Moustakas, Kennicutt \&
 Tremonti}{Moustakas et al.}{2006}]{moustakas}
 Moustakas J., Kennicutt R. C. Jr., Tremonti, C. A., 2006, ApJ,
 642, 775
\bibitem[\protect\citeauthoryear{Naab et al.}{2007}]{naab}
 Naab T., Johansson P. H., Ostriker J. P., Efstathiou G., 2007, ApJ,
 658, 710
\bibitem[\protect\citeauthoryear{Nelan et al.}{2005}]{nelan}
 Nelan J. E., Smith R. J., Hudson M. J., Wegner G. A., Lucey
 J. R., Moore S. A. W., Quinney S. J., Suntzeef N. B., 2005, ApJ,
 632, 137
\bibitem[\protect\citeauthoryear{Noeske et al.}{2007a}]{noeske07a}
 Noeske K. G., et al. 2007a, ApJL, 660, L43
\bibitem[\protect\citeauthoryear{Noeske et al.}{2007b}]{noeske07b}
 Noeske K. G., et al. 2007b, ApJL, 660, L47
\bibitem[\protect\citeauthoryear{Nolan, Raychaudhury, \& Kab\'{a}n}{Nolan et al.}{2007}]{nolan}
 Nolan L. A., Raychaudhury S., Kab\'{a}n A., 2007, MNRAS, 375, 381
\bibitem[\protect\citeauthoryear{Panter et al.}{2007}]{panter}
 Panter B., Jimenez R., Heavens A. F., Charlot S., 2007, MNRAS, 378, 1550
\bibitem[\protect\citeauthoryear{Panuzzo et al.}{2007}]{panuzzo}
Panuzzo P., et al., 2007, ApJ, 656, 206
\bibitem[\protect\citeauthoryear{Pisani}{1993}]{pisani93}
 Pisani A., 1993, MNRAS, 265, 706
\bibitem[\protect\citeauthoryear{Pisani}{1996}]{pisani96}
 Pisani A., 1996, MNRAS, 278, 697
\bibitem[\protect\citeauthoryear{Poggianti et al.}{2004}]{poggianti}
 Poggianti B. M., Bridges T. J., Komiyama Y., Yagi M., Carter D.,
 Mobasher B., Okamura S., Kashikawa N., 2004, ApJ, 601, 197
\bibitem[\protect\citeauthoryear{Popesso et al.}{2004}]{popesso}
 Popesso P., B\"{o}hringer H., Brinkmann J., Voges W., York
 D. G., 2004, A\&A, 423, 449
\bibitem[\protect\citeauthoryear{Quadri et al.}{2007}]{quadri}
 Quadri R., et al., 2007, ApJ, 654, 138
\bibitem[\protect\citeauthoryear{Renzini}{2006}]{renzini}
 Renzini A., 2006, ARA\&A, 44, 141
\bibitem[\protect\citeauthoryear{Rigby et al.}{2006}]{rigby}
 Rigby J. R., Rieke G. H., Donley J. L., Alonso-Herrero A.,  P\'{e}rez-Gonz\'{a}lez P. G., 2006, ApJ, 645, 11
\bibitem[\protect\citeauthoryear{Rines \& Diaferio}{2006}]{rines}
 Rines K., Diaferio A., 2006, AJ, 132, 1275
\bibitem[\protect\citeauthoryear{Rines et al.}{2005}]{rines05}
 Rines K., Geller M. J., Kurtz M. J., Diaferio A., 2005, AJ, 130, 1482 (R05)
\bibitem[\protect\citeauthoryear{Robertson et al.}{2006}]{robertson}
 Robertson B., Cox T. J., Hernquist L., Franx M., Hopkins P. F.,
 Martini P., Springel V., 2006, ApJ, 641, 21
\bibitem[\protect\citeauthoryear{Roediger \& Hensler}{2005}]{roediger}
 Roediger E., Hensler G., 2005, A\&A, 433, 875
\bibitem[\protect\citeauthoryear{Roettiger, Burns \&
    Loken}{1996}]{roettiger}
 Roettiger K., Burns J. O., Loken C., 1996, ApJ, 673, 451 
\bibitem[\protect\citeauthoryear{Rossa et al.}{2006}]{rossa}
Rossa J., et al., 2006, AJ, 132, 1074
\bibitem[\protect\citeauthoryear{Ruderman \& Ebeling}{2005}]{ruderman}
Ruderman J. T., Ebeling H., 2005, ApJL, 623, 81
\bibitem[\protect\citeauthoryear{Sales et al.}{2007}]{sales}
 Sales L. V., Navarro J. F., Abadi M. G., Steinmetz M., 2007, preprint (astro-ph/0704.1770)
\bibitem[\protect\citeauthoryear{Scannapieco et al.}{2006}]{scannapieco}
 Scannapieco C., Tissera P. B., White S. D. M., Springel
 V., 2006, MNRAS, 371, 1125
\bibitem[\protect\citeauthoryear{Schlegel, Finkbeiner \&
    Davis}{Schlegel et al.}{1998}]{schlegel}
 Schlegel D. J., Finkbeiner D. P., Davis M., 1998, ApJ, 500, 525
\bibitem[\protect\citeauthoryear{Schmidt}{1959}]{schmidt}
 Schmidt M., 1959, ApJ, 129, 243
\bibitem[\protect\citeauthoryear{Silk \& Rees}{1998}]{silkrees}
 Silk J., Rees M., 1998, A\&A, 334, L1
\bibitem[\protect\citeauthoryear{Silverman}{1986}]{silverman}
 Silverman B. W., 1986, Density Estimation for Statistics and Data
 Analysis. Chapman and Hall, London
\bibitem[\protect\citeauthoryear{Smith et al.}{2005}]{smithg}
 Smith G. P., Treu T., Ellis R. S., Moran S. M., Dressler A. 2005, ApJ, 620, 78
\bibitem[\protect\citeauthoryear{Smith et al.}{2006}]{smith}
 Smith R. J., Hudson M. J., Lucey J. R., Nelan J. E., Wegner
 G. A., 2006, MNRAS, 369, 1419
\bibitem[\protect\citeauthoryear{S\"{o}chting, Clowes \&
    Campusano}{S\"{o}chting et al.}{2004}]{sochting}
 S\"{o}chting I. K., Clowes R. G., Campusano L. E., 2004, MNRAS,
    1241, 1254
\bibitem[\protect\citeauthoryear{Solanes et al.}{2001}]{solanes}
 Solanes J. M., Manrique A., Garc\'{i}a-G\'{o}mez C.,
 Gonz\'{a}lez-Casado G., Giovanelli R., Haynes M. P., 2001, ApJ, 548, 97
\bibitem[\protect\citeauthoryear{Sorrentino, Antonuccio-Delogo \& Rifatto}{Sorrentino et al.}{2006}]{sorrentinovoids}
 Sorrentino G., Antonuccio-Delogu V., Rifatto A., 2006, A\&A, 460, 673
\bibitem[\protect\citeauthoryear{Sorrentino, Radovich \& Rifatto}{Sorrentino et al.}{2006}]{sorrentino}
 Sorrentino G., Radovich M., Rifatto A., 2006, A\&A, 451, 809
\bibitem[\protect\citeauthoryear{Springel, di Matteo \&
    Hernquist}{Springel et al.}{2005a}]{springel}
 Springel V., di Matteo T., Hernquist L., 2005a, MNRAS, 361, 776
\bibitem[\protect\citeauthoryear{Springel, di Matteo \&
    Hernquist}{Springel et al.}{2005b}]{springel05b}
 Springel V., di Matteo T., Hernquist L., 2005b, ApJL, 620, 79
\bibitem[\protect\citeauthoryear{Springel et al.}{2005c}]{springelsim}
 Springel V., et al., 2005c, Nat, 435, 629
\bibitem[\protect\citeauthoryear{Stoughton et al.}{2002}]{stoughton}
 Stoughton C., et al., 2002, AJ, 123, 485 
\bibitem[\protect\citeauthoryear{Stinson et al.}{2007}]{stinson}
 Stinson G. S., Dalcanton J. J., Quinn T., Kaufmann T., Wadsley J.,
 2007, preprint (astro-ph/0705.4494)
\bibitem[\protect\citeauthoryear{Strateva et al.}{2001}]{strateva}
 Strateva I., et al., 2001, AJ, 122, 1861
\bibitem[\protect\citeauthoryear{Strauss et al.}{2002}]{strauss}
 Strauss M. A., et al., 2002, AJ, 124, 1810
\bibitem[\protect\citeauthoryear{Struck}{2006}]{struck}
 Struck C., 2006, in ed. J. W. Mason, Astrophysics Update 2. Springer
 Verlag, Heidelberg, p. 115
\bibitem[\protect\citeauthoryear{Tanaka et al.}{2004}]{tanaka}
 Tanaka M., Goto T., Okamura S., Shimasaku K., Brinkmann,
 J., 2004, AJ, 128, 2677 (T04)
\bibitem[\protect\citeauthoryear{Tanaka et al.}{2005}]{tanaka05}
 Tanaka M., et al., 2005, MNRAS, 362, 268
\bibitem[\protect\citeauthoryear{Tassis, Kravtsov, \& Gnedin}{2006}]{tassis}
 Tassis K., Kravtsov A. V., Gnedin N. Y., 2006, preprint (astro-ph/0609763)
\bibitem[\protect\citeauthoryear{Thomas et al.}{2005}]{thomas}
 Thomas D., Maraston C., Bender R., Mendes de Oliveira C., 2005,
 ApJ, 621, 673
\bibitem[\protect\citeauthoryear{Toomre}{1977}]{toomre}
 Toomre A., 1977, in Tinsley B. M., Larson R. B., eds, Evolution of
 Galaxies and Stellar Populations. Yale Univ. Obs., New Haven, p. 401
\bibitem[\protect\citeauthoryear{Tremonti et al.}{2004}]{tremonti}
 Tremonti C. A., et al., 2004, ApJ, 613, 898
\bibitem[\protect\citeauthoryear{Treu et al.}{2003}]{treu}
 Treu T., Ellis R. S., Kneib J.-P., Dressler A., Smail I.,
 Czoske O., Oemler A., Natarajan P., 2003, ApJ, 591, 53
\bibitem[\protect\citeauthoryear{van den Bergh}{1976}]{vandenbergh76}
 van den Bergh S., 1976, ApJ, 206, 883
\bibitem[\protect\citeauthoryear{van den Bergh}{1999}]{vanderbergh}
 van den Bergh S., 1999, A\&A Review, 9, 273
\bibitem[\protect\citeauthoryear{van Dokkum \& van der Marel}{2007}]{vandokkum}
 van Dokkum P., van der Marel R. P., 2007, ApJ, 655, 30
\bibitem[\protect\citeauthoryear{van Gorkom}{2004}]{vangorkom}
 van Gorkom J., 2004, in eds. Mulchaey J. S., Dressler A., Oemler A., Clusters of Galaxies: Probes of Cosmological
 Structure and Galaxy Evolution. Cambridge University Press, New York, p. 306
\bibitem[\protect\citeauthoryear{van Zee}{2001}]{vanzee}
 van Zee L., 2001, AJ, 121, 2003
\bibitem[\protect\citeauthoryear{van Zee, Skillman \& Haynes}{2004}]{vanzee04}
 van Zee L., Skillman E. D., Haynes M. P., 2004, AJ, 128, 121
\bibitem[\protect\citeauthoryear{Vogt et al.}{2004}]{vogt}
 Vogt N., Haynes M. P., Giovanelli R., Herter T., 2004, AJ, 127, 3300
\bibitem[\protect\citeauthoryear{Vollmer et al.}{2001}]{vollmer}
 Vollmer B., Cayatte V., Balkowski C., Duschl W. J., 2001, ApJ,
 561, 708
\bibitem[\protect\citeauthoryear{Wehner \& Harris}{2006}]{wehner}
 Wehner E. H., Harris W. E., 2006, ApJL, 644, 17
\bibitem[\protect\citeauthoryear{Weinmann et al.}{2006a}]{weinmann06a}
 Weinmann S. M., van den Bosch F. C., Yang X., Mo H. J., 2006a, MNRAS, 366, 2
\bibitem[\protect\citeauthoryear{Weinmann et al.}{2006b}]{weinmann06b}
 Weinmann S. M., van den Bosch F. C., Yang X., Mo H. J., Croton
 D. J., Moore B., 2006b, MNRAS, 372, 1161
\bibitem[\protect\citeauthoryear{White \& Rees}{1978}]{white}
 White S. D. M., Rees M. R., 1978, MNRAS, 183, 341
\bibitem[\protect\citeauthoryear{White et al.}{1999}]{white99}
 White R. A., Bliton M., Bhavsar S. P., Bornmann P., Burns J. O.,
 Ledlow M. J., Loken C., 1999, AJ, 118, 2014
\bibitem[\protect\citeauthoryear{Willmer et al.}{2006}]{willmer}
 Willmer C. N. A. et al., 2006, 647, 853
\bibitem[\protect\citeauthoryear{Wolf, Gray, \&
    Meisenheimer}{Wolf et al.}{2005}]{wolf05}
 Wolf C., Gray M. E., Meisenheimer K., 2005, A\&A, 443, 435
\bibitem[\protect\citeauthoryear{Yang et al.}{2005}]{yang}
 Yang X., Mo H. J., van den Bosch F. C., Jing Y. P., 2005, MNRAS, 356, 1293
\bibitem[\protect\citeauthoryear{York et al.}{2000}]{york}
 York D. G. et al., 2000, AJ, 120, 1579 
\bibitem[\protect\citeauthoryear{Zehavi et al.}{2005}]{zehavi}
 Zehavi I. et al., 2005, ApJ, 630, 1 
\end{thebibliography}
\end{document}